\documentclass[usenatbib]{mnras}

\usepackage[T1]{fontenc}
\usepackage{ae,aecompl}

\usepackage{graphicx}	
\usepackage{amsmath}	
\usepackage{amssymb}	
\usepackage{deluxetable}
\usepackage{amsmath,amssymb,amsfonts}
\usepackage{color}
\usepackage{epsfig}
\usepackage{natbib}

\newcommand \beq {\begin{equation}}
\newcommand \enq {\end{equation}}
\newcommand \Sigmag {\Sigma_{\rm g}}

\newcommand \tstop    {t_{\rm{stop}}}
\newcommand \tstopi   {t_{\rm{stop}}{_{,i}}}
\newcommand \ar  {\alpha_{\rm R}}

\newcommand \ag  {\alpha_{\rm g}}
\newcommand \atot  {\alpha_{\rm tot}}
\newcommand \qloss {\mathcal{L}}
\newcommand \tcool {t_{\rm {cool}}}
\newcommand \Dp    {D_{\rm {p,x}}}
\newcommand \Dg    {D_{\rm {g,x}}}
\newcommand \ux    {u_{\rm {x}}}

\newcommand \vx    {v_{\rm {x}}}
\newcommand \vy    {v_{\rm {y}}}

\definecolor{grey}{gray}{0.5}

\newcommand \lx  {L_{\rm {x}}}
\newcommand \ly  {L_{\rm {y}}}

\newcommand \cs  {c_{\rm {s}}}

\newcommand \drangle {\rangle_{\scriptscriptstyle \Sigma}}

\title[dust dynamics in 2D gravito-turbulent disks]{Dust dynamics in 2D
gravito-turbulent disks}
\author[J.-M. Shi et al.]{
Ji-Ming Shi,$^{1}$\thanks{E-mail:~jmshi@astro.princeton.edu}
Zhaohuan Zhu,$^{1}$
James M. Stone,$^{1}$
Eugene Chiang $^{2}$
\\
$^{1}$ {Department of Astrophysical Sciences, Princeton University, 4 Ivy Ln,
    Princeton, NJ 08544} \\
$^{2}$ {Department of Astronomy, UC Berkeley, 501 Campbell Hall,
    Berkeley, CA 94720-3411}
}

\date{Accepted XXX. Received YYY; in original form ZZZ}

\pubyear{2016}

\begin{document}
\label{firstpage}
\pagerange{\pageref{firstpage}--\pageref{lastpage}}
\maketitle

\begin{abstract}
\noindent 
The dynamics of solid bodies in protoplanetary disks
are subject to the properties of any underlying gas turbulence.
Turbulence driven by disk self-gravity shows features
distinct from those driven by 
the magnetorotational instability (MRI).
We study the dynamics of
solids in gravito-turbulent disks with two-dimensional (in
the disk plane), hybrid (particle and gas) 
simulations.  Gravito-turbulent disks
can exhibit stronger gravitational stirring
than MRI-active disks,
resulting in greater radial diffusion
and larger eccentricities and relative
speeds for large particles
(those with dimensionless
stopping times $\tstop\Omega > 1$, where
$\Omega$ is the orbital frequency). The
agglomeration of large particles
into planetesimals by pairwise collisions
is therefore disfavored in gravito-turbulent
disks.
However, the relative speeds of intermediate-size particles
($\tstop\Omega \sim 1$) are significantly reduced as such particles are
collected by gas drag and gas gravity 
into coherent filament-like structures with
densities high enough to trigger
gravitational collapse.
First-generation planetesimals may
form via gravitational instability of dust
in marginally gravitationally unstable gas 
disks.
\end{abstract}


\section{INTRODUCTION \label{sec:introduction}}

As protoplanetary disks are often thought to be turbulent 
(\citealt{armitage2011araa}; but see \citealt{flahertyetal2015}),
understanding how disk solids interact with
turbulent gas is crucial to modelling
the formation of planetesimals and planets
\citep{weidenschillingcuzzi93,chiangyoudin10,johansenetal2014prpl,testietal2014prpl}
and to explaining observations of disks \citep{william_cieza2011araa,andrews2015PASP}.
Turbulence determines the spatial distribution of solid
particles and their relative collision velocities  
\citep[``turbulent stirring"; ][]{voelketal1980,cuzzietal1993, YL2007,OC2007}.
For example, dust particles can be
concentrated in local pressure maxima
at the interstices of turbulent eddies;
pressure bumps can also be found in
spiral density waves, anti-cyclonic vortices, or zonal
flows associated with whatever mechanism drives
disk transport
\citep{maxey87,Klahr2003,riceetal2004,barrancomarcus05,mamarice2009,Johansen2009,
johansenetal2011,panetal2011}.
Turbulent density fluctuations can also exert stochastic gravitational torques on solid objects and alter their orbital dynamics 
\citep[``gravitational stirring";][]{laughlinetal2004,nelson2005,ogiharaetal2007,IGM2008}.

Disk self-gravity can drive turbulence, provided
disks are sufficiently massive and provided
their cooling times are longer than their dynamical times
\citep{Paczynski78,gammie01,Forgan2012,ShiChiang2014}. 
``Gravito-turbulence'' may characterize
the early phase of star formation, when disks are
still massive
\citep{rodriguez2005,eisneretal2005,andrews_williams2007}. Observations show signs of early grain
growth in some very young stellar systems \citep{riccietal2010,tobinetal2013}. Millimeter-sized chondrules in primitive meteorites indicate they
might once have been melted via strong shock waves in
self-gravitating disks \citep{cuzzi2006nature,alexanderetal2008sci}. Simulations
suggest large dust concentrations via spiral waves and vortices present in gravito-turbulent disks,
possibly leading to
planetesimal formation via gravitational instability in
the dust itself
\citep{riceetal2004,gibbonsetal2015}. 

~~~
Many studies of dust dynamics to date focus on particles in disks made turbulent by the magneto-rotational
instability (MRI; \citealt{BH91,BH98}).
Both
analytical and numerical works have been carried out to obtain radial/vertical diffusivities and particle
relative velocities due to turbulent stirring \citep{cuzzietal1993, YL2007,OC2007,carb2010,carb2011,
zhuetal2015}. 
For large particles, gravitational forces by MRI-turbulent
density fluctuations exceed aerodynamic drag forces
by gas and generate relative particle velocities
too high to be conducive to planetesimal formation
\citep{nelson2005,johnsonetal2006,
yangetal2009,yangetal2012,NG2010,gnt2012mnras}.
A few useful metrics common to many of these papers include: (1)\,the diffusion coefficient, which 
characterizes how quickly solids
random walk through the disk, (2)\, the particle eccentricity or velocity dispersion, 
and (3) the pairwise relative velocity,
which is crucial for determining collision outcomes.
Quantity (2) usually serves as a good proxy
for quantity (3).\footnote{We will show in section~\ref{sec:vel} that this
approximation actually breaks down for $\Omega\tstop\sim 1$ particles in gravito-turbulent disks.} 

Relatively fewer groups investigated the dynamics of dust in turbulence driven by disk self-gravity.
\citet{gibbonsetal2012,gibbonsetal2014,gibbonsetal2015} study particles with a 
range of sizes (with stopping times $\tstop = 10^{-2}$--$10^2\Omega^{-1}$, where $\Omega$
is the local orbital frequency) accumulate in
local, two-dimensional (in the disk plane) simulations.
They find that intermediate-sized dust can concentrate by
up to two orders of magnitude, and that the dispersion of
particle velocites can approach the gas sound speed,
consistent with
the results of 2D global simulations
\citep{riceetal2004,riceetal2006}.
\citet{britschetal2008} and \citet{walmswelletal2013}
find strong eccentricity growth 
for large-sized planetesimals forced more by
gravitational stirring than by gas drag.
\citet{boss2015} investigates the radial diffusion process for particles $1$\,cm--$10$\,m in size
($\tstop \sim 10^{-2}$--$1\Omega^{-1}$ in their model),
finding enhanced diffusion for $m$-sized or
larger bodies. 

However, no systematic study has yet been performed to
directly measure the dynamical properties
(diffusivities, eccentricities, and relative speeds as
listed above) of solids in gravito-turbulent disks as has
been done for MRI-active disks.
Gravito-turbulence tends to produce relatively
stronger density fluctuations ($\delta\rho/\rho\sim 1$ for a typical Shakura-Sunyaev turbulence parameter $\alpha\sim 10^{-2}$; see \citealt{ShiChiang2014})
than are seen in MRI turbulence ($\delta\rho/\rho\sim 0.1$ for $\alpha\sim 10^{-2}$).
The prominent spiral density features that characterize self-gravitating disks
and that help trap dust particles are
also absent in MRI-turbulent disks. 

It is the goal of this paper to study the dynamics and
spatial distribution of dust in gravito-turbulent
disks in a systematic manner, placing our measurements
into direct comparison with analogous measurements
made for MRI-turbulent disks.
We first describe our simulation setup in section~2. Results are given in section~3, where we describe the radial diffusion, eccentricity
growth, and relative velocities of particles,
and how these quantities are affected by gravitational
stirring, particle stopping time, gas cooling rate, numerical resolution, and simulation domain
size. In section~4, we put our results into physical context, discuss their astrophysical
implications, and make comparison with MRI-active disks.
We conclude in section~5.

\section{METHODS\label{sec:methods}}
\subsection{Equations solved and code description \label{sec:eqn}}
We study the diffusion of solids in 
gravito-turbulent disks using hybrid (particle+fluid) hydro simulations in the disk plane.

For the gas, we solve the hydrodynamic equations governing 2D,
self-gravitating accretion disks, including the effects of secular cooling. The disk is modeled in
the local shearing sheet approximation assuming the disk aspect ratio $H/r\ll 1$. In a Cartesian
reference frame corotating with the disk at fixed orbital frequency $\Omega$, the equations solved
are similar to those \citet{ShiChiang2014}, but restricted
to be in the disk plane:
\begin{align}
  \frac{\partial \Sigmag}{\partial t} + \nabla\cdot (\Sigmag \mathbf{u}) = 0 \,, 
  \label{eq:continuity} \\
  \frac{\partial \Sigmag\mathbf{u}}{\partial t} + \nabla\cdot\left(\rho\mathbf{u}\mathbf{u}
  +P\mathbf{I} + \mathbf{T_{\rm g}} \right) =  \nonumber \\
  2q\Sigmag\Omega^2 x \hat{\mathbf{x}} -2\Sigmag\Omega\hat{\mathbf{z}}\times\mathbf{u}\,,
  \label{eq:eom} \\
  \frac{\partial E}{\partial t} + \nabla\cdot (E+P)\mathbf{u} =
  -\Sigmag\mathbf{u}\cdot\nabla\Phi \nonumber \\
  +\rho\Omega^2\mathbf{u}\cdot\left(2 q x \hat{\mathbf{x}} - z\hat{\mathbf{z}}\right) -\Sigmag\qloss \,,
  \label{eq:eoe} \\
  \nabla^2\Phi = 4\pi G (\Sigmag-\Sigma_0)\delta(z)\,, 
  \label{eq:poisson}
\end{align}
where $\hat{\mathbf{x}}$ points in the radial direction,
$\rho$ is the gas mass density, $\mathbf{u} = (u_{\rm x}, u_{\rm y}) $ is the gas velocity relative
to the background Keplerian flow $\mathbf{u_0} = (0, -q\Omega x)$,
$P$ is the gas pressure, $\Phi$ is the self-gravitational potential of a razor-thin disk,
$q = 3/2$ is the Keplerian shear parameter, 
\beq
  E = {U} + {K} = \frac{P}{\Gamma -1} + \frac{1}{2}\Sigmag u^2
  \label{eq:eos}
\enq
is the sum of the internal energy density $U$ and bulk kinetic energy
density $K$ for an ideal gas with 2D specific heat ratio $\Gamma = 2$, and
\beq
  \mathbf{T_{\rm g}} = \frac{1}{4\pi G}\left[\nabla\Phi\nabla\Phi -
  \frac{1}{2}\left(\nabla\Phi\right)\cdot\left(\nabla\Phi\right)\mathbf{I}\right] 
  \label{eq:grav_tensor}
\enq
is the gravitational stress tensor with identity tensor $\mathbf{I}$.

We choose a very simple cooling function, 
\beq
\Sigmag\qloss = U/\tcool = \Omega U / \beta \,\,\,\,
{\rm (constant \,\, cooling \,\, time)}
\enq
with $\beta \equiv \Omega \tcool$ constant everywhere.
The assumption of constant cooling time $\tcool$ is adopted
by many 2D
\citep[e.g.,][]{gammie01,JG2003,Paardekooper2012} and three-dimensional (3D)
\citep[e.g.,][]{Rice2003,LR2004,LR2005,Mejia2005,Cossins2009,MB2011}
simulations of self-gravitating disks.
This prescription enables direct experimental control over the
rate of energy loss.

For the solids, 
we assume the dust particles only passively respond to the GT turbulence via the aerodynamical drag
and also the gravitational acceleration from the gas. No particle feedback is included in this study. In the 2D local
approximation, the equation of particle motion reads
\beq
\frac{d\mathbf{v_i}}{dt} = \frac{\mathbf{u}-\mathbf{v_i}}{\tstopi} - \nabla\Phi
        -2\Omega\hat{\mathbf{z}}\times \mathbf{v_i} + 2q\Omega^2 x\hat{\mathbf{x}} \,,
\label{eq:eom_par}
\enq
in which the first term is the drag force per unit mass,
the second term $-\nabla\Phi$ is the gravitational pull from the self-gravitating gas.
The particle velocity $\mathbf{v_i} = (v_{\rm x},v_{\rm y})$ represents the $i$--th particle specie relative
to the background shear.

Our simulations are run with \texttt{Athena} \citep{stoneetal08} with the built-in particle module
\citep{baistone10a}.  We adopt
the van Leer integrator \citep{vl06,sg09}, a piecewise linear spatial
reconstruction in the primitive variables, and the HLLC
(Harten-Lax-van Leer-Contact) Riemann solver.  We solve Poisson's
equation of a razor-thin disk using fast Fourier transforms
\citep{gammie01,Paardekooper2012}\footnote{{In this
paper, we do not smooth the potential over a length $\delta$ in the $z$-direction to mimic the finite
thickness of the disk, as we find the density and velocity dispersions in our unsmoothed 2D simulations to be
similar to those in 3D  \citep{ShiChiang2014}. The effect of
smoothing on the perturbations is modest; e.g., $\delta\Sigma/\Sigma$ decreases by at most a factor of two when $\delta=\cs/\Omega$ is used.}}
Boundary conditions for our physical variables ($\rho$,
$\mathbf{v}$, $U$, and $\Phi$)
are shearing-periodic in radius ($x$) and periodic in azimuth ($y$).
We also use orbital advection algorithms to shorten the timestep
and improve conservation \citep{Masset2000, Johnson2008, sg10}.  

\subsection{Initial conditions and run setup \label{sec:ic}}
\begin{figure*}
	\includegraphics[width=8cm]{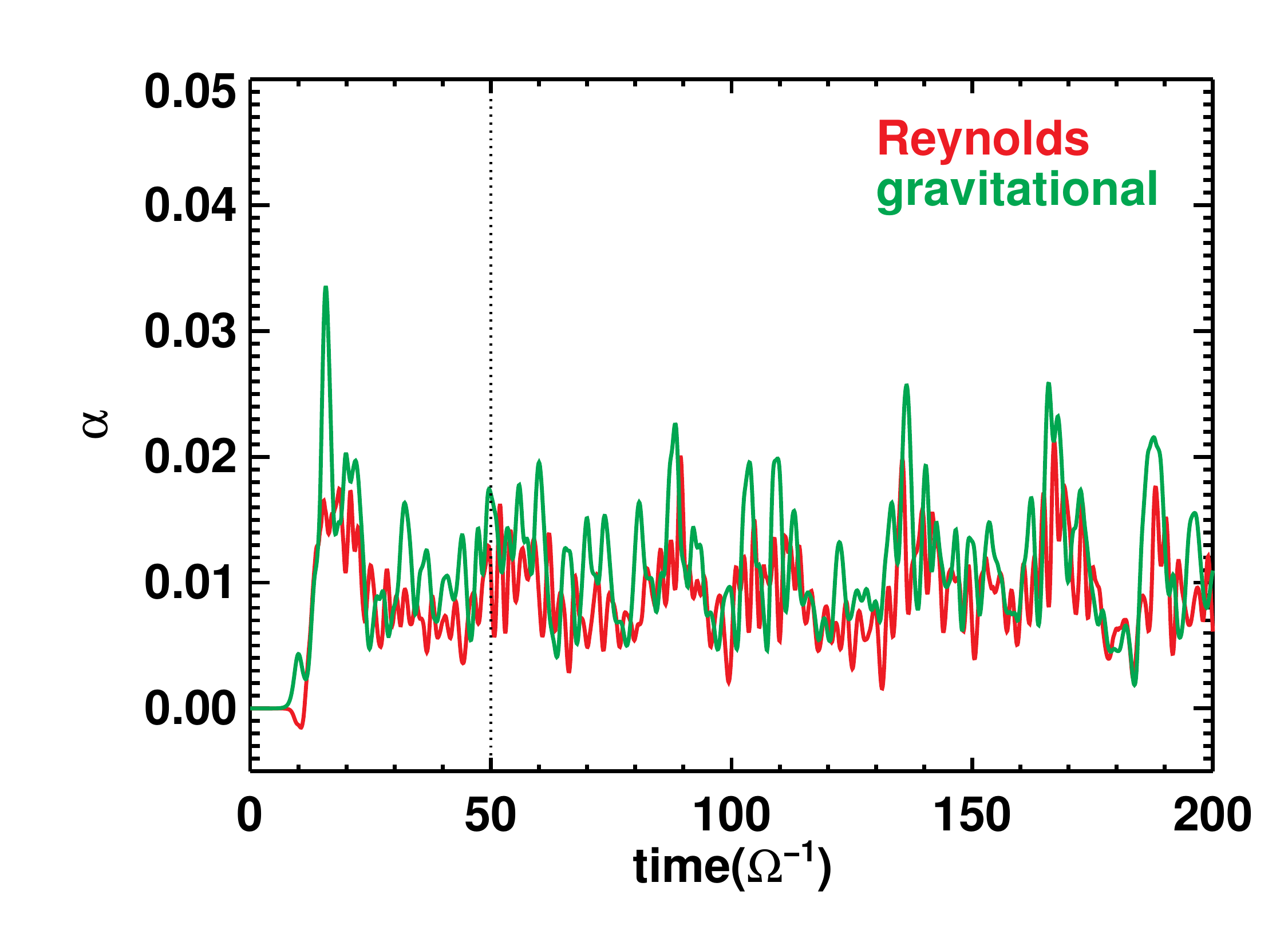}
	\includegraphics[width=8cm]{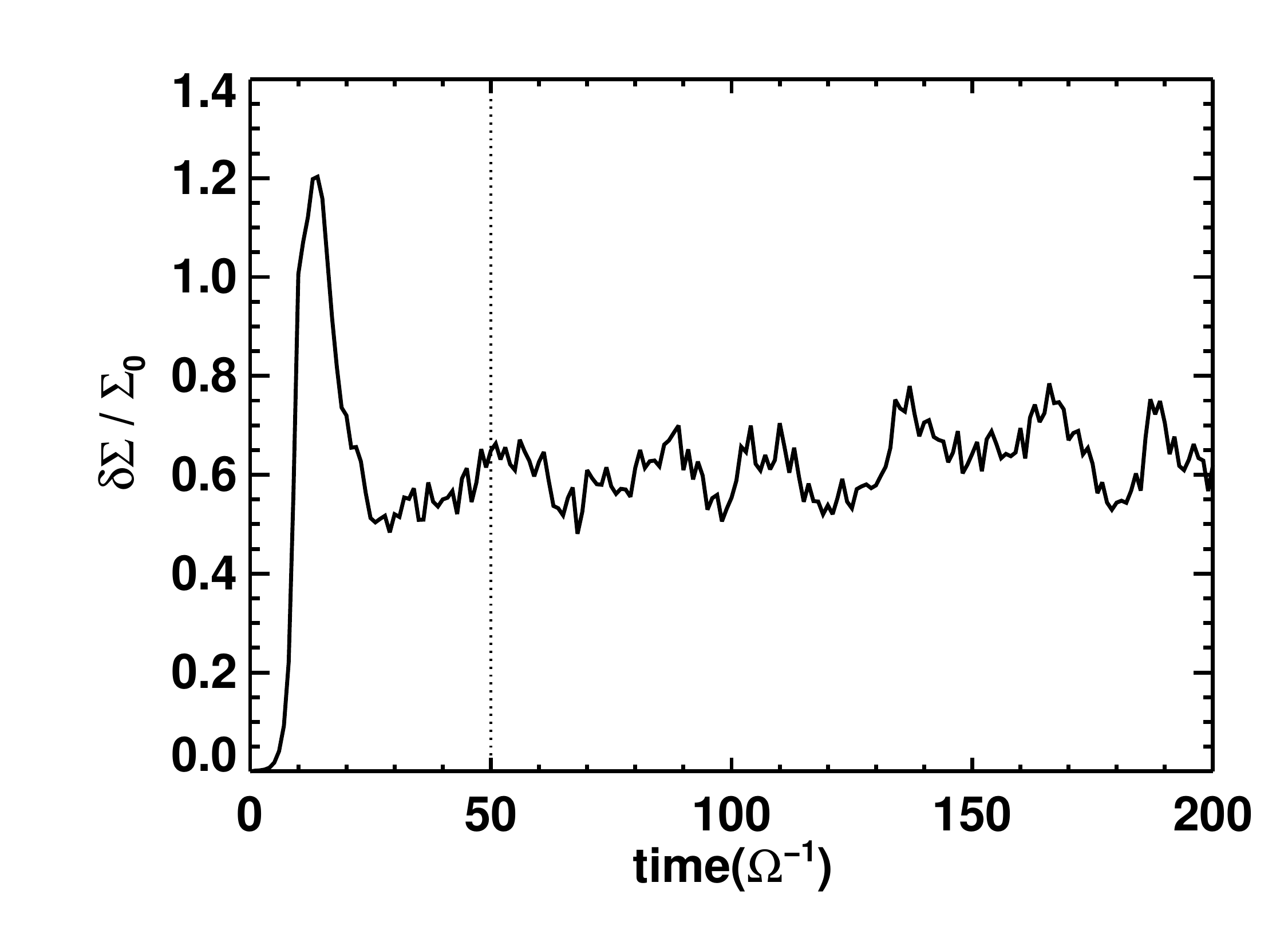} \\
	\includegraphics[width=8cm]{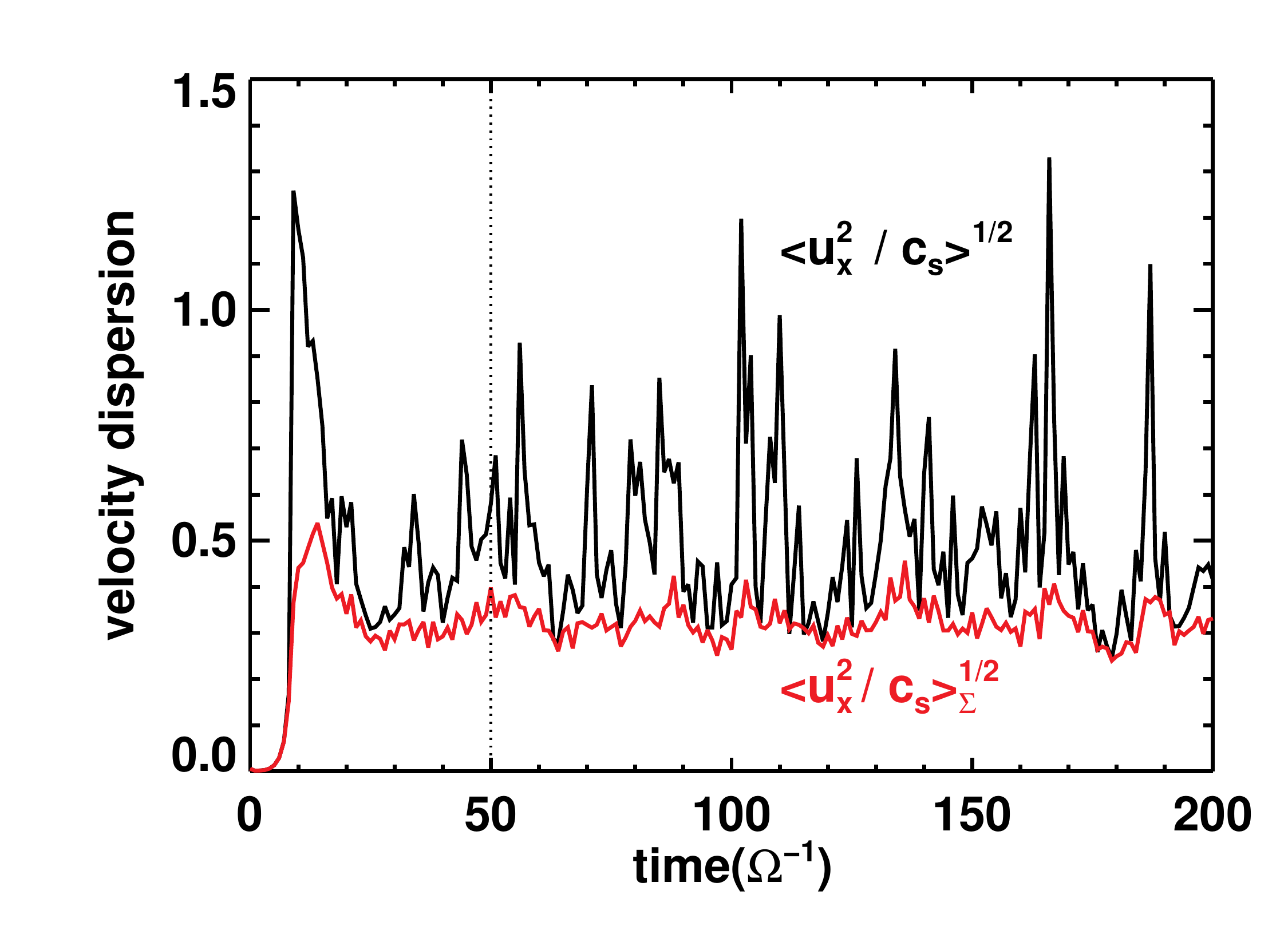}
	\includegraphics[width=8cm]{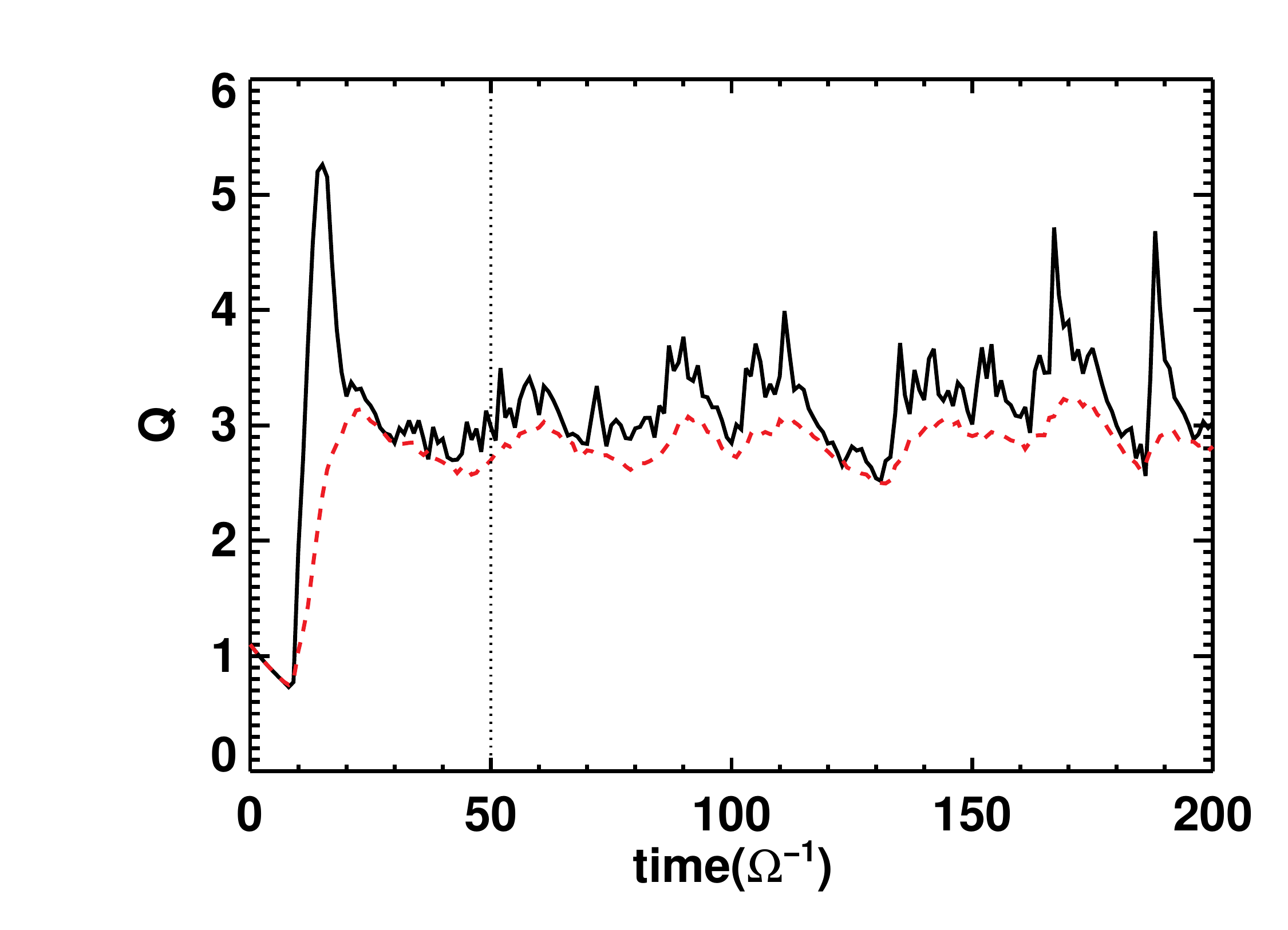}
	\caption{\small{Time history of the gravito-turbulent disk of
pure gas case with $\tcool\Omega=10$. Top left: The gravitational (green) and Reynolds
(red) stresses normalized with averaged pressure as a function of time. Top Right: the density
dispersion versus time. 
Bottom left: the velocity dispersion with (red) and without (black) density weight. Bottom right: The averaged
Toomre Q-parameter using sound speed with (red dashed) and without (black solid)density weight. 
In our dust+gas hybrid simulations, we choose
$t=50\Omega^{-1}$ (indicated by vertical dotted lines) as our initial state and distribute dust particles randomly in space.}}
\label{fig:sst}
\end{figure*}
We start with pure gas simulations. At $t=0$ we initialize a uniformly distributed gas disk and set
$\Sigma_0 = \Omega = G =1$. The thermal energy is such that $Q = \cs\Omega/(\pi G\Sigma_0) = 1.1$
close to the critical Toomre $Q$-parameter for a razor-thin disk \citep[$\simeq
1$;][]{toomre64,goldreichbell65}, where $\cs = \sqrt{\Gamma(\Gamma-1)U/\Sigma}$ is the sound speed. 
The velocity is $\mathbf{u_0} = (\delta_x,-q\Omega x + \delta_y)$, where $\delta_x$ and $\delta_y$ are
randomized perturbations at $\sim$$1\%$ of the initial sound speed $\cs{_0=1.1\pi}$. Our simulation domain is a box which covers $[-80, 80]\,G\Sigma_0/\Omega^2$ in
both radial ($x$) and azimuthal ($y$) direction with $512^2$ grid points. This amounts to
a spatial span of $160 H/(\pi Q)\simeq 51 Q^{-1} H$ and a resolution of $\simeq 10 Q/H$, where
$H\equiv \cs/\Omega = \pi Q (G\Sigma_0/\Omega^2)$ is the disk scale height. 

We allow the disk to cool off immediately with a fixed cooling time $\tcool\Omega \in
\{5,10,20,40\}$. After a short transient phase which normally takes about twice the cooling time,
the disk settles to a quasi-steady gravito-turbulent state in which the heating from compression and
shocks is balanced by the imposed cell-by-cell cooling. For example, we show the time evolution of Reynolds and
gravitational stresses, surface density and velocity dispersions and Toomre's Q parameter
from the $\tcool\Omega=10$ case in Figure~\ref{fig:sst}.

All quantities saturate after $\geq
20\Omega^{-1}$, and well established turbulence sustains to the end of the simulations (hundreds of
dynamical times). The time averaged ($t>50\Omega^{-1}$) nominal $\alpha$, i.e., stresses normalized
with pressure, is $\alpha\simeq 0.020$ for the sum of the Reynolds and gravitational stress. 
Toomre's Q is hovering $\sim 3$ \citep{gammie01} which also sets the spatial averaged sound speed $\langle
\cs\rangle = \pi Q (G\Sigma_0/\Omega)$. The Q-parameter using density weighted sound speed
$\langle \cs\drangle$ (red dashed curve in bottom right panel) shows less fluctuation and slightly diminished ($\sim
10\%$) mean. We choose the density weighted measure hereafter and simply use $\langle\cs\rangle$ to
represent the density weighted sound speed. 
But we do note that the spatial and temporal averaged velocity dispersion $\langle \ux \rangle$
(black) is about twice the density weighted value $\langle\ux\drangle$ (red) as shown in the bottom left
panel of Figure~\ref{fig:sst}. The cause and effects will be discussed in
section~\ref{sec:dp}.
We also emphasize that the density dispersion
$\delta\Sigma_g/\Sigma_0\simeq
0.6$ for
$\alpha\sim 0.02$, much stronger than found in the MRI-driven turbulence case where
$\delta\rho/\rho\sim \sqrt{0.5\alpha}\sim 0.1$ for
similar $\alpha$ values with or without net weak magnetic field \citep[][]{NG2010,OH2011,ssh2016}. We will discuss it further in section~\ref{sec:tcool}.

We then randomly distribute dust particles in space at $t =
50\Omega^{-1}$ (for $\tcool\Omega \leq 20$) or $100\Omega^{-1}$ (for $\tcool\Omega = 40$), and
evolve the particle+fluid system for another $200\Omega^{-1}$. We
implemented seven types of particles with constant stopping time such that the Stokes number
$\tau_s\equiv \Omega\tstop \in [10^{-3},10^{-2},0.1,1,10,10^2,10^3]$, evenly spaced in logarithmic 
scale. For each type, we use $2^{19}$ ($\sim$$5\times
10^5$) particles, or $\sim$ 2 particles per grid cell on average. Their velocities follow the background shear
initially. 
Since gas densities vary in time and space,
it would be more physical to fix
the size of each particle rather than its
stopping time; nevertheless, our default
simulations fix stopping times to
more easily compare with previous
simulations that do likewise. We also verify
explicitly that our simulations with
fixed stopping time agree well with a
simulation that uses fixed particle
sizes (see Section \ref{sec:fixed_size}).
For this latter run,
we employ a stopping time based on
the Epstein drag law:
\beq
\tstop{_{,i}} = \frac{f a^*_i}{\Sigma \cs} \,,
\label{eq:epstein}
\enq
where $a^*_i = 10^{i-4}$ for $i=1$--$7$ is the dimensionless
size of the $i$-th particle species.  The converting factor $f \simeq 3\pi (G\Sigma_0^2/\Omega^2)$ is
chosen such that $f a^*_i/\langle\!\langle\Sigma\cs\rangle\!\rangle_{\rm
t}\sim a^*_i$, matching $\tau_s$ used in the fixed stopping time runs.

\section{RESULTS\label{sec:results}}
\subsection{Standard run \label{sec:std}}
\begin{figure}
	\includegraphics[width=\columnwidth]{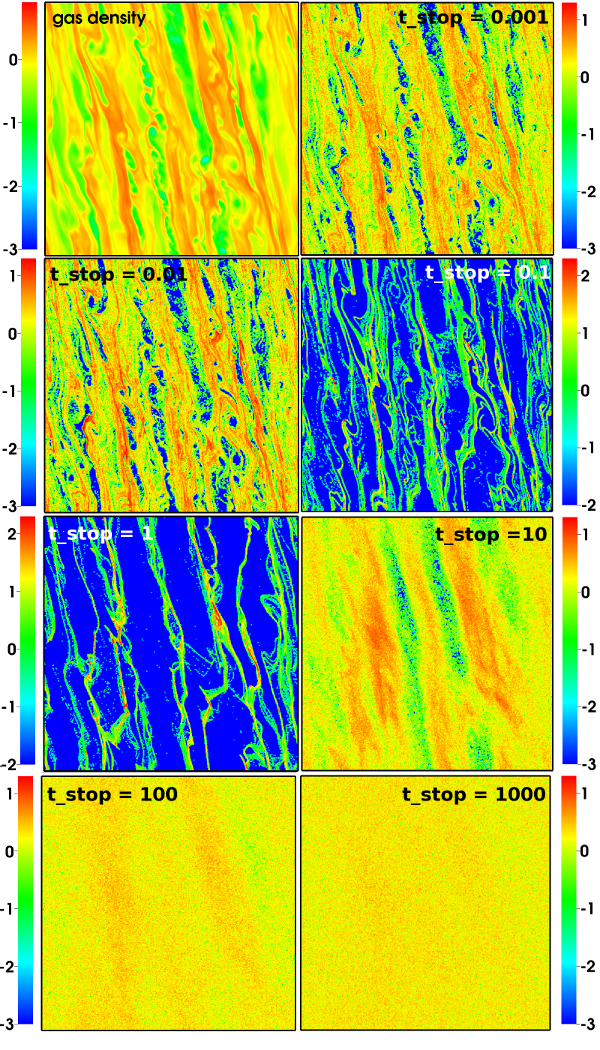}
\caption{\small{Snapshots of gas and dust surface density distributions at the end of simulation.
		The domain size is $\lx = \ly = 160 G\Sigma_0/\Omega^2 \simeq 17 H$, and cooling time is
		$10\Omega^{-1}$.
Values are normalized against initial values and color coded in log scale. We see small $\tstop$ particles
tracing the gas; particles with intermediate $\tau_s = 0.1$-$1$ strongly clustering; and 
larger particles diffused across the domain. }}
\label{fig:8panel}
\end{figure}
We first present our standard run tc=10, i.e., $\tcool\Omega = 10$ run (see Table~\ref{tab:tab1} for
the gas properties). 
After distributing the particles randomly on the grid, the particles quickly adjust in
response to the dynamical
gas flows. After $\sim 20\Omega^{-1}$, the distribution of particles reaches a steady state. As an illustration, we have shown the gas and dust density in Figure~\ref{fig:8panel}
at $t=180\Omega^{-1}$. 
Clearly, the small particles ($\tau_s\leq 10^{-2}$) are nearly
perfectly coupled to the gas and therefore share the same density structures of the gas. 
Particles with intermediate stopping time, for both $\tau_s = 0.1$ and $1$, appear to
concentrate
along the dense gas filaments and cause dust density enhancement of two orders of magnitude
relative to the mean (note the color bars for $\tau_s = 0.1$ and $1$ now extend to higher
values). {Transient vortices are also observable in the snapshots for gas and $\tau_s < 1$ particles, but are probably under-resolved; see \citet{gibbonsetal2015} for the effects
of vortices on particle concentration.}
For large particles ($\tau_s\ge 10$), they are strongly disturbed by the
gravitational stirring from the fluctuating gas (see further discussion of the effects of gravity in
section~\ref{sec:selfg}). After $\sim 20 \Omega^{-1}$, they are completely redistributed and their
end status recovers a random distribution similar to the initial. 

\begin{figure}
\includegraphics[width=\columnwidth]{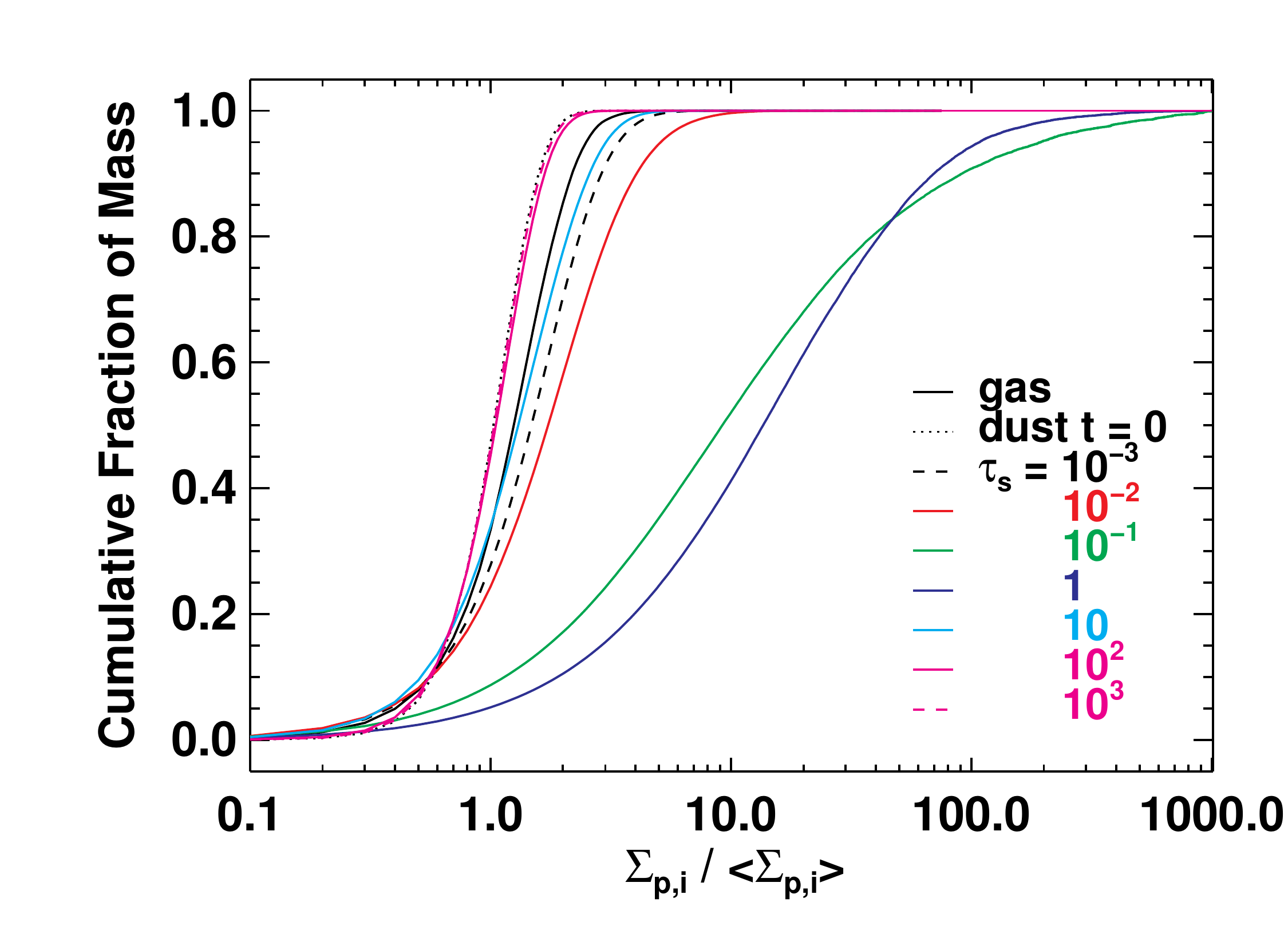}
\caption{\small{The time-averaged cumulative fraction of the total particle mass as a function of  particle
surface density. The surface density is normalized by the averaged value (equal to the initial value) and
thus represents the concentration factor of the dust particles. We also show the time-averaged gas distribution (black solid) and the initial dust distribution (black dotted) for comparison. }}
\label{fig:mass}
\end{figure}

In Figure~\ref{fig:mass}, we show the time-averaged fraction of cumulative mass for each type of
particle, binned
in local surface density (number of particles in each cell) normalized to the mean
($2$/cell). 
The running profiles of small sized
particles resemble that of the time-averaged gas (solid black curve). Those with large stopping times track the
initial random distribution (dotted black). In contrast, the blue ($\tau_s=1$)
and green ($\tau_s=0.1$) curves show that intermediate-sized particles exhibit relatively higher densities,
$\Sigma_p/\langle\Sigma_p\rangle \sim 10$--$100$.

After reaching quasi-steady state, we further evolve the dust+gas mixture for a total duration of $200\Omega^{-1}$. We
then measure the radial diffusion coefficients, particle eccentricity and relative velocities averaged over all particles of the same type. The
results are presented in the following subsections.

\subsection{Radial diffusion\label{sec:dp}}
\begin{figure*}
	\includegraphics[width=8cm]{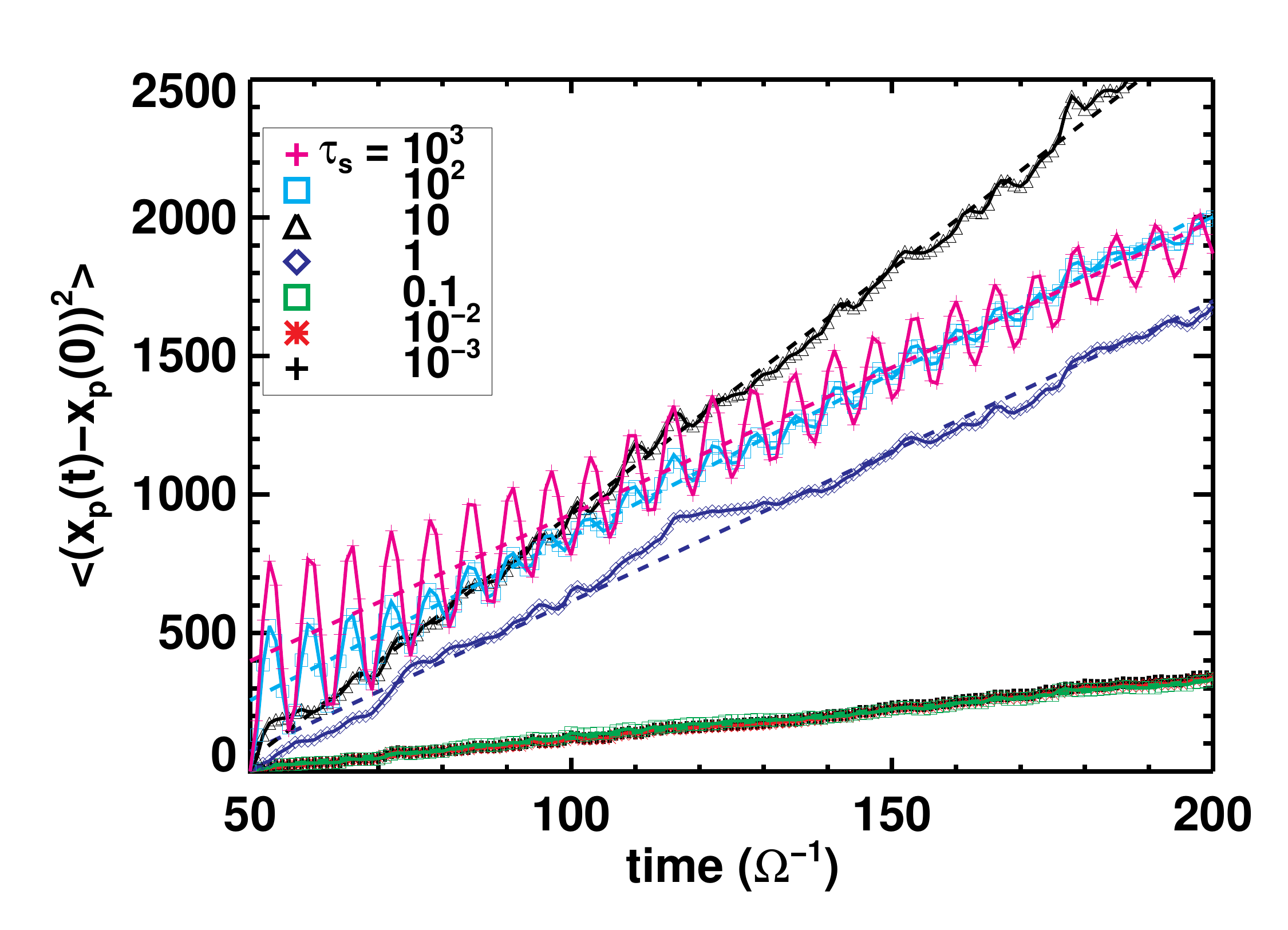}\hfill
	\includegraphics[width=8cm]{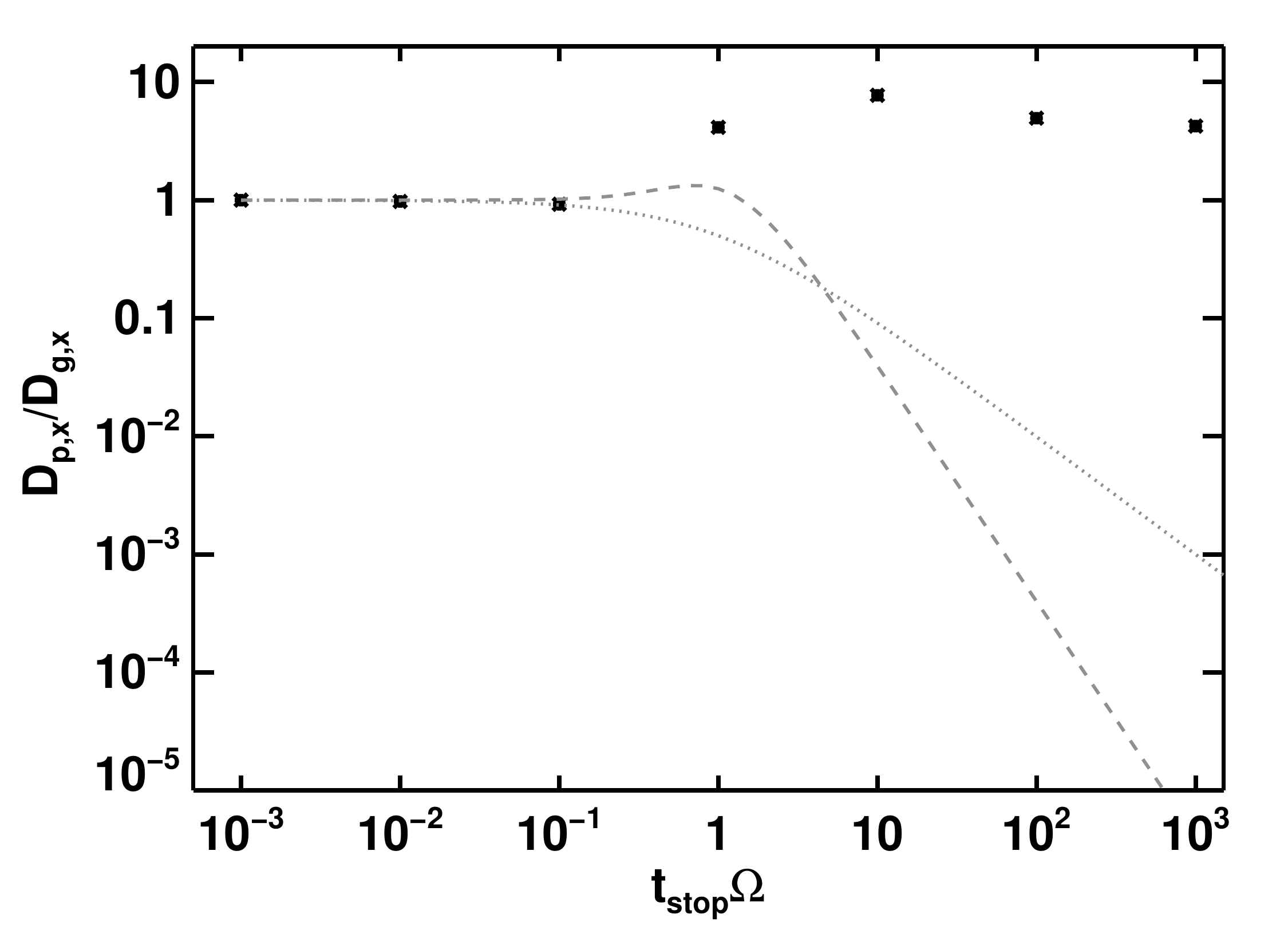}
\caption{\small{Left: The squared radial displacement versus time for all types of 
        particles ($\tau_s \equiv \tstop\Omega$) in run tc=10 ($\tcool\Omega = 10$. 
		The color symbols (with solid curves) are measurement, the dashed curves 
		are the linear fits to the data. Their slopes are then twice the diffusion 
		coefficients based on Equation~\ref{eq:def_dp}.
		Right: Symbols mark the radial diffusion coefficients for particles with 
		different stopping time. At larger stopping time, the measured coefficients 
		are largely different than models of either \citet{YL2007} (dashed gray 
		curve) or \citet{cuzzietal1993} (dotted) due to extra stirring from the 
		gas self-gravity.}}
\label{fig:fixed_ts}
\end{figure*}
We utilize the following formula to derive the radial diffusion coefficients $\Dp$ of different
types of particles in our
simulations \citep{YL2007,carb2011}: 
\beq
\Dp \equiv \frac{1}{2}\frac{d\langle |x_p(t)-x_p(0)|^2\rangle}{dt} \,,
\label{eq:def_dp}
\enq
where $\Dp$ is the diffusion coefficient for given particle stopping time, $x_p(t)$ and $x_p(0)$ are
the radial position at time $t$ and initial (taken to be $50\Omega^{-1}$ after
injecting the particles to the gas disk). The radial coordinate is extended beyond the edges of the sheet so that particles moves 
on radially without periodic boundary conditions.
The measurements are made at
every time interval $\delta t=0.5\Omega^{-1}$, and are performed for a duration of $150\Omega^{-1}$
long. The average $\langle\,\rangle$ here is taken for all particles within the same type.  

The squared displacements as a function of time on the right hand side of Equation~(\ref{eq:def_dp})
are shown in the left panel of Figure~\ref{fig:fixed_ts}. Each curve represents the displacement of
one distinctive type of particles. For particles of $\tau_s\leq 0.1$, the curves overlap. As they
are well-coupled with the turbulent gas flows, their displacements reflect the properties of gas
diffusion.
For larger particles, the gravitational stirring dominates the drag force which introduces some
extra effective diffusion. As a result, the displacement curves of those particles
have bigger slopes. For the largest two dust species we implemented, the curves show periodic
oscillations which are due to the epicyclic motion of individual particles. The large amplitude of the
epicyclic motion is a result of the strong gravitational forcing of the background gas.  

Now with the help of Equation~\ref{eq:def_dp} and Figure~\ref{fig:fixed_ts}, we can derive the
radial diffusion coefficients based on the slope of the squared displacements using linear fitting.
The results are shown in the right panel of Figure~\ref{fig:fixed_ts} and also recorded in
Table~\ref{tab:tab2}. The radial diffusion
coefficients $\Dp$ are normalized against the gas diffusion coefficient $D_{g,x}$; the latter is
approximated using the $\Dp$ of particles with $\tau_s = 10^{-3}$. We find $\Dp\simeq
\Dg$ for $\tau_s\leq 0.1$. However, $\Dp$ exceeds $\Dg$ for particles with
longer $\tstop$, and stay roughly constant, $\Dp \sim 6$--$8 \Dg$. For comparison, we also
plot the diffusion coefficient predicted by models in \citet[][dotted gray
curve]{cuzzietal1993} and \citet[][dashed gray]{YL2007}. Both predict small and decreasing values
for larger particles based on homogeneous turbulence without extra forcing like self-gravity,
in clear contrast with what we obtain in our gravito-turbulent disk. When gravitational stirring is artificially suppressed (see section~\ref{sec:selfg}), we do recover similar relationship as they predicted. 
\begin{figure}
	\includegraphics[width=\columnwidth]{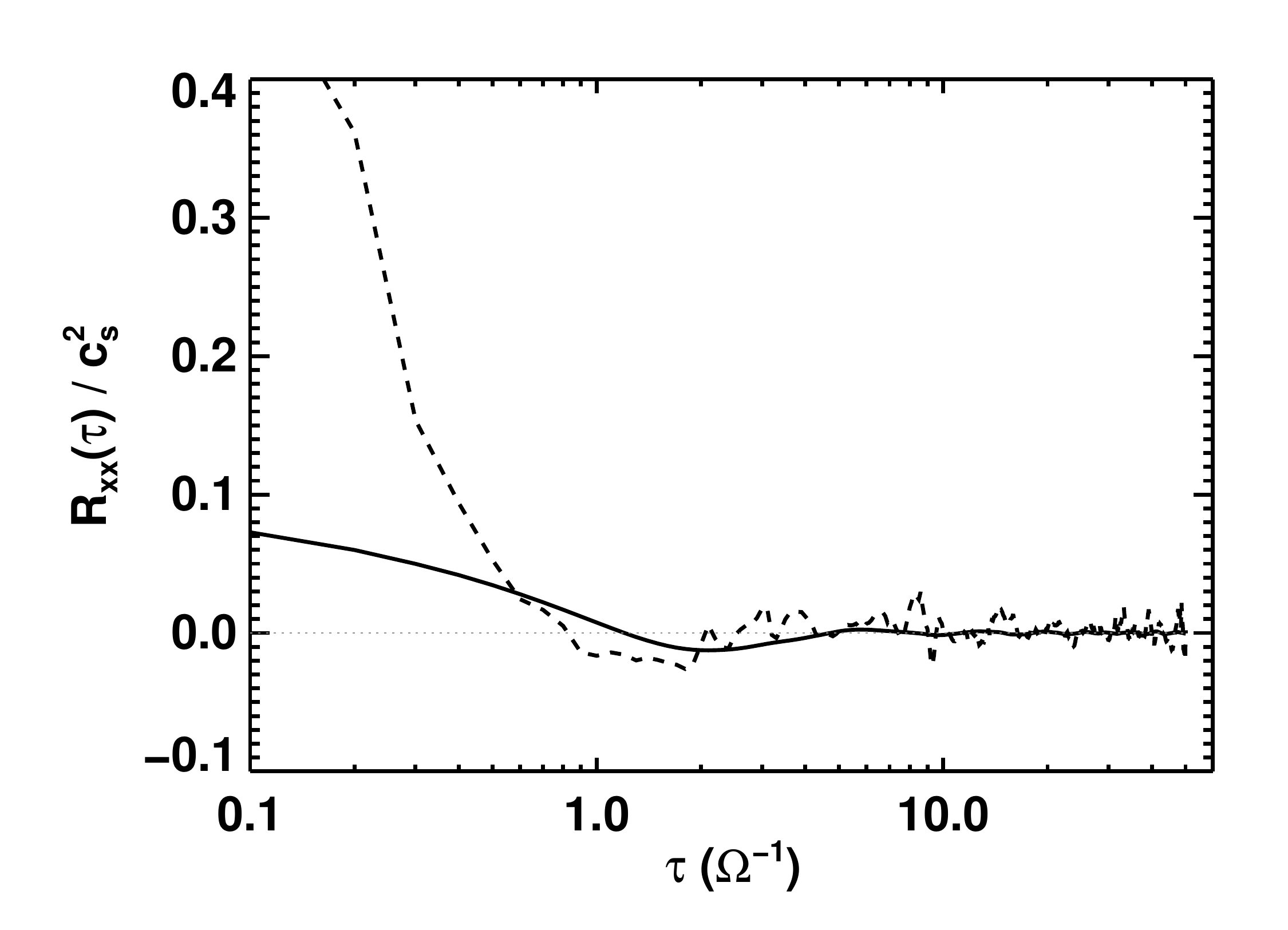} 
\caption{\small{ The radial velocity auto-correlation
functions of the gravito-turbulent gas disk calculated with (solid) and without (dashed) density weight.  }}
\label{fig:rxx}
\end{figure}

We also check the validity of approximating $\Dg$ with $\Dp$ by measuring the auto-correlation
function of the turbulent velocity field directly. In general, 
\beq
\Dg = \int_0^{\infty}R_{\rm xx}(\tau) d\tau \,,
\label{eq:def_dg}
\enq
where $R_{\rm xx}(\tau) = \langle \ux(\tau) \ux(0) \rangle$
is the auto-correlation of the gas velocity at time $\tau$. However, in gravito-turbulent disk, the
spiral density shock waves cause low density ($\Sigma/\Sigma_0\sim O(10^{-2})$) valleys between high density ($\Sigma/\Sigma_0\sim
O(1)$) ridges. Most of the matter in the high density regions has small velocity, but the gas
in low density region has very high velocity ($\sim \cs$). The auto-correlation function
defined above would strongly bias toward the low density instead of the high density region where
most of the small dust particles reside. One way to remove the bias is to calculate the
gas or dust\footnote{We use the surface density of the dust in our calculation; using the gas density would
	change the estimated diffusion coefficient by $\sim$10\%.}
density weighted auto-correlation function
\beq
R_{\rm xx}(\tau) = \frac{\int\! \Sigma(\tau)\ux(\tau)\Sigma(0)\ux(0) dxdy}{\int\! \Sigma(\tau)\Sigma(0)
dxdy}  = \langle \ux(\tau)\ux(0)\drangle \,,
\label{eq:rxx}
\enq
in which $\ux(0)$ and $\Sigma(0)$ are velocity and density at a reference time, $\ux(\tau)$ and
$\Sigma(\tau)$ are measured at time $\tau$ from the reference point and are both sheared back to
that point in order to calculate the correlation. Shown in Figure~\ref{fig:rxx} is the velocity
auto-correlation calculated with (solid curve) and without (dashed curve) density weight. Integrating
the density weighted $R_{\rm xx}$, we get $\Dg\simeq 0.014 \langle\!\langle \cs
\rangle\rangle\Omega^{-1}$ close to the $\Dp\simeq 0.012 \langle\!\langle \cs
\rangle\rangle\Omega^{-1}$ measured with particles of $\tau_s = 10^{-3}$. However, using $R_{xx}$
without density weight would overestimate the diffusion coefficient by a factor of $15$.

\subsection{Eccentricity growth \label{sec:ecc}}
\begin{figure}
	\includegraphics[width=\columnwidth]{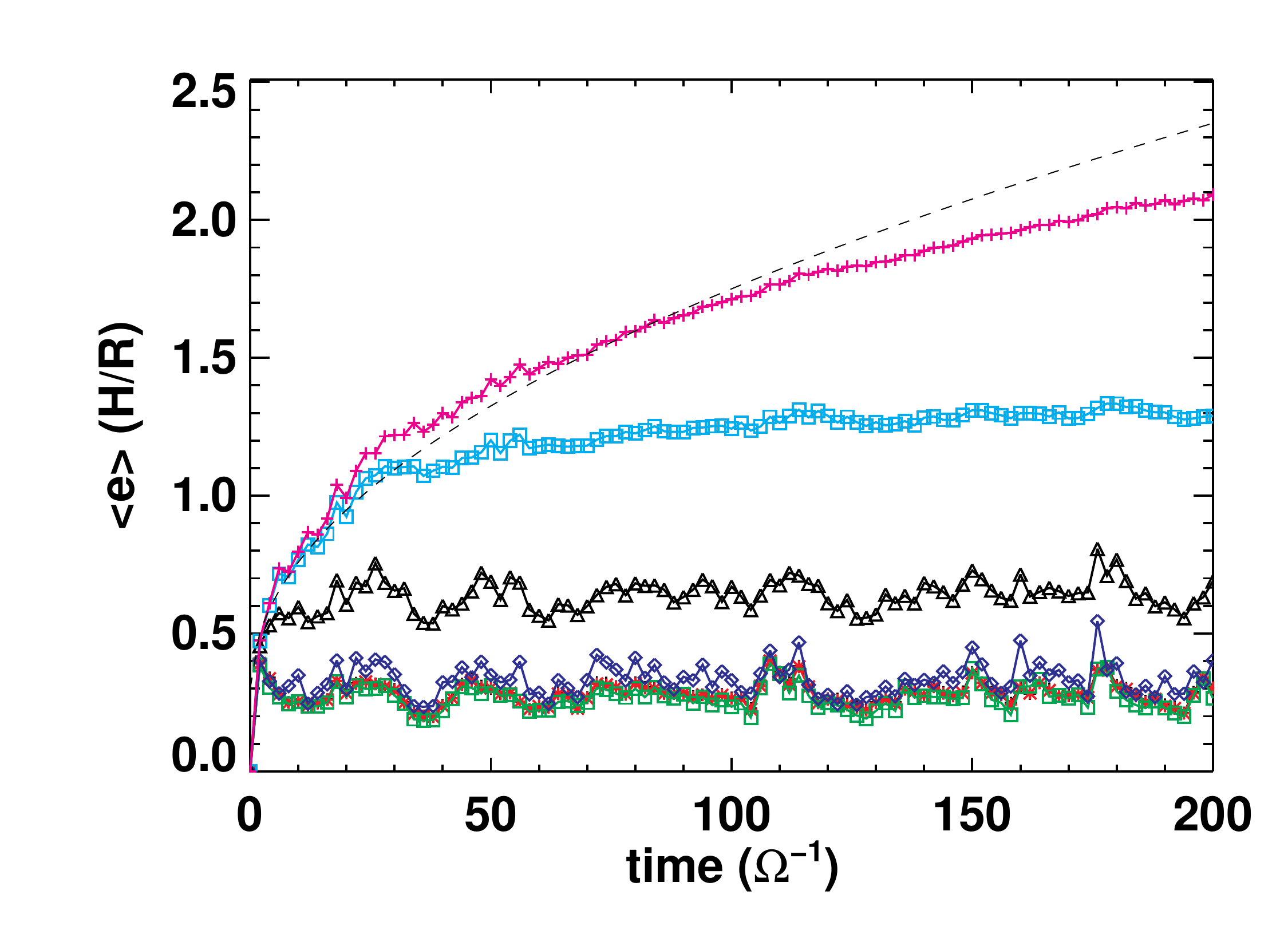}\\
	\includegraphics[width=\columnwidth]{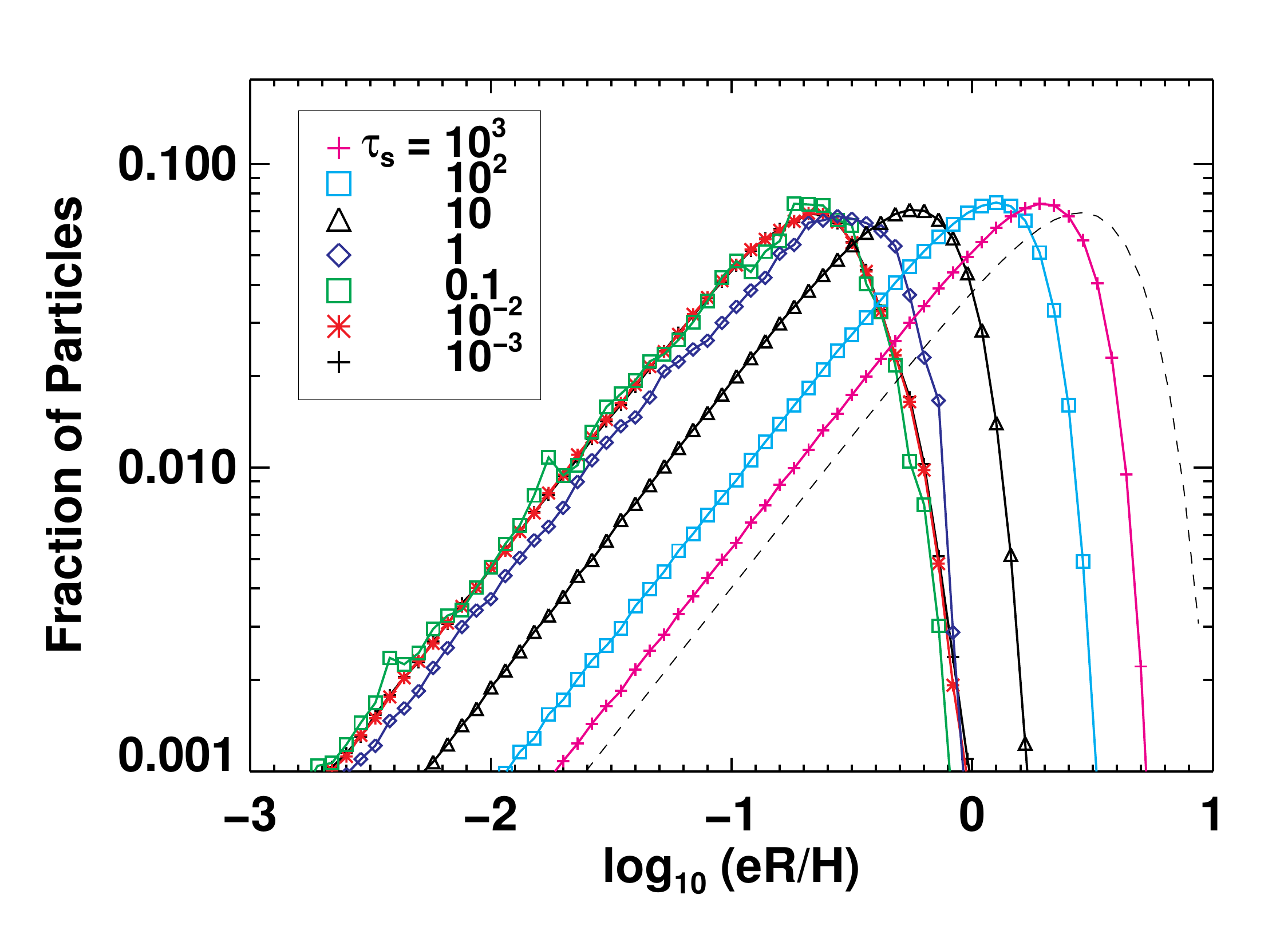}
\caption{\small{Top: The time evolution of orbital eccentricities averaged over all 
        particles of the same type. 
		The dashed line is $0.15 (\Omega t)^{1/2}$, which characterize the early
		growth of the eccentricity for large particles. 
		Bottom: The relative number distribution of particle eccentricity at late time of
		simulation. We choose
		100 bins evenly divided between $\log_{10} (eR/H) = -3 $ and $1$ for the distribution plot.  
		The dashed line shows a typical Rayleigh distribution for comparison. 
 }}
\label{fig:ecc}
\end{figure}

\begin{figure}
	\includegraphics[width=\columnwidth]{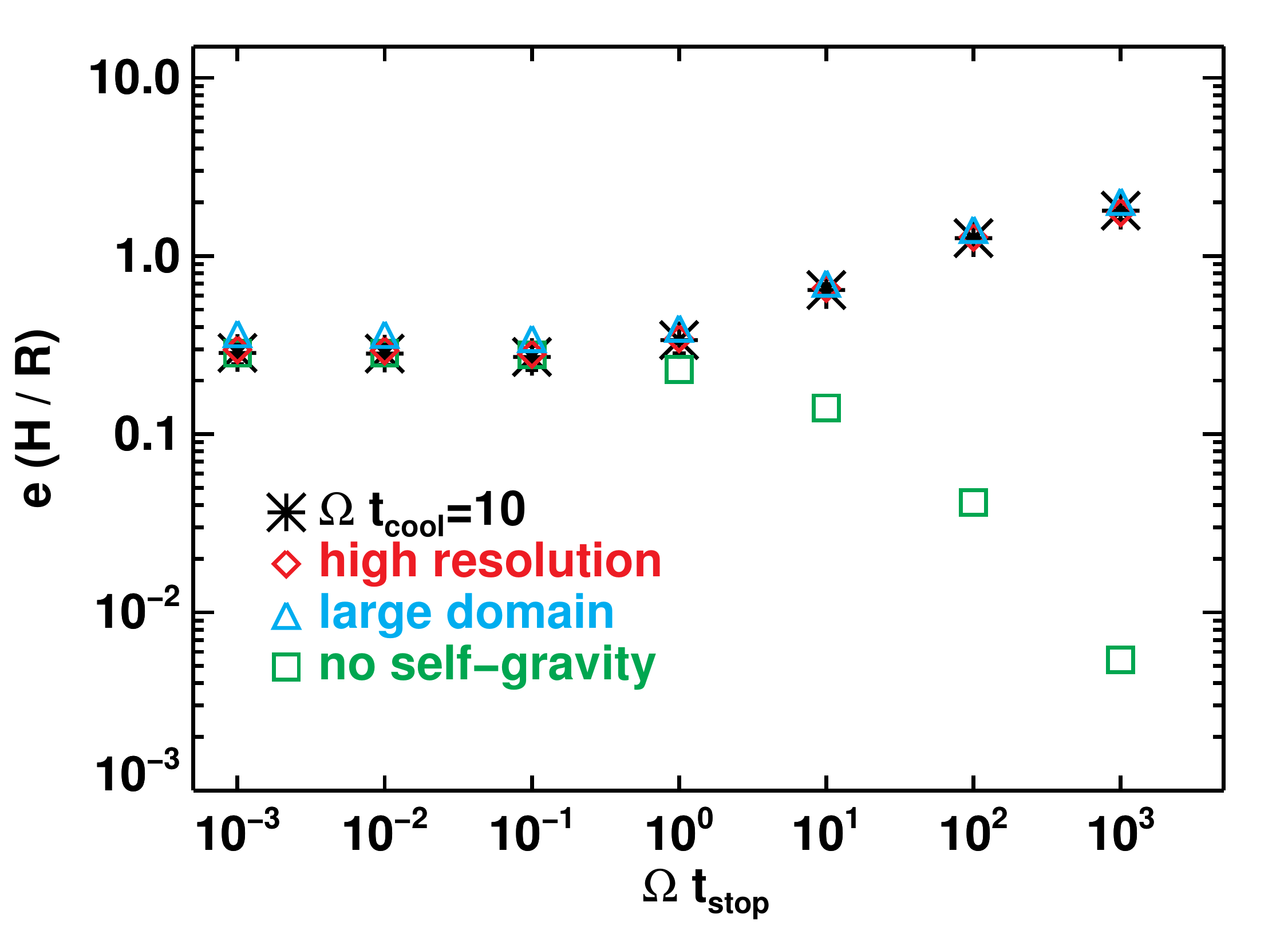} 
\caption{\small{The time averaged (last $15$ orbits) particle eccentricity (or equivalently the velocity
        dispersion $\delta v/\cs$) varies with stopping time. 
        The standard run using $\tcool = 10\Omega^{-1}$ are shown in asterisk.
		The label `high resolution', `large domain', and `no
		self-gravity' represent results from  tc=10.hires using doubled resolution,
		tc=10.dble with doubled sheet size, and tc=10.wosg without gravitational forcing
		from the gas respectively, and are discussed in section~\ref{sec:selfg} and~\ref{sec:box}.
 }}
\label{fig:ecc_ts}
\end{figure}

When particle eccentricity is small, we have the relation 
\beq
\vx^2 \simeq 2 \vy^2 \simeq e^2\Omega^2 r^2 \,. 
\label{eq:def_ecc}
\enq
We can therefore measure the orbital eccentricity of each individual particle according to this
relation, and obtain the evolution of the eccentricity as shown in Figure~\ref{fig:ecc}. Although
the initial $e = 0$, we find the
mean eccentricity quickly saturates at $e\simeq 0.2$-$0.3 (H/R)$ for particles of $\tau_s
\leq 1$. It then saturates slower and levels off at greater value for increasing $\tau_s > 1$. The
saturated
values are  $e \simeq 0.7 (H/R)$ and $1.3 (H/R)$ for $\tau_s = 10$ and $10^2$ particles (see
Figure~\ref{fig:ecc_ts}). The eccentricity
of $\tau_s = 10^3$ particle keeps rising gradually through the end of the 
simulation, suggesting that a saturation level of $\gtrsim 2 (H/r)$ might be achieved for a longer
simulation which is consistent with previous simulations of planetesimals in gravito-turbulent disks
\citep{britschetal2008,walmswelletal2013}. 
We also find that the particle eccentricity obtained in gravito-turbulent disk is in
general much greater
than that could be excited in the MRI-driven turbulent disk with similar turbulent stress-to-pressure
ratio $\alpha$. The latter usually gives $e \sim 10^{-2}$--$10^{-1} (H/R) Q^{-1}$
\citep{yangetal2009,yangetal2012,NG2010}, orders of
magnitude smaller than what we have observed here in our simulations.

Since the particle eccentricity is excited by nearly random gravity field, the evolution will follow the
general $\propto \sqrt{t}$ law \citep{ogiharaetal2007},
\beq
e = C \left(\frac{H}{R}\right) (\Omega t)^{1/2} \,,
\label{eq:ecc}
\enq
where the dimensionless coefficient $C$ determines the growth rate and can be measured with our
simulation. We fit the early growing phase of the both $\tau_s=10^2$ (cyan square) and $10^3$
(magenta cross) with the above relation and obtain $C\simeq 0.15$ as the best fit coefficient (see
the black dashed curve in Figure~\ref{fig:ecc})
The
excitation time scale of eccentricity could be estimated as $t_{\rm exc}\sim e/(de/dt)\sim 2e^2(R/H)^2
C^{-2}\Omega^{-1}$. We can therefore predict the saturated eccentricity by equating $t_{\rm exc}$
with the damping time scale, in our case, is simply the stopping time $\tstop$. The
eccentricity at saturation is therefore
\beq
e  \sim C (H / R) (\tau_s/2)^{1/2} \,.
\enq
For $\tau_s=10^2$ particles, this gives $e \sim 1.1 (H/R)$
that matches what we observe in Figure~\ref{fig:ecc} very well. It also predicts a saturation level
of $e\sim 4.7 (H/R)$ if we extend the simulation for another $\sim 300\Omega^{-1}$. 

Equation~\ref{eq:ecc} also allows us to measure the dimensionless parameter $\gamma$ which
characterizes the amplitude of the fluctuating gravity field as defined in
\citet{ogiharaetal2007}. Following \citet{IGM2008} and \citet{OO2013a}, we write the eccentricity
growth as  
\beq
e \simeq 1.6\gamma \left(\frac{M_{\odot}}{M_*}\right) 
\left(\frac{\Sigma}{10\,{\rm g}\,{\rm cm^{-3}}}\right)
\left(\frac{R}{100\,{\rm AU}}\right)^2
\left(\Omega t\right)^{1/2} \,.
\label{eq:ecc_IGM}
\enq
After comparing with Equation~\ref{eq:ecc}, we find the dimensionless turbulent strength
$\gamma\simeq 0.01$ in our gravito-turbulent
disk with $\Omega\tcool=10$ or $\alpha\simeq 0.02$, considerably larger than that in MRI disks with
similar $\alpha$ \citep[e.g., $\gamma\sim 10^{-4}$ in][]{yangetal2012}.

We also calculate the saturated eccentricity distributions of particles. In Figure~\ref{fig:ecc}, we find that particles of all types obey a Rayleigh-like distribution
\citep{yangetal2009,yangetal2012}. 
The probability rises toward greater eccentricity, and then drops at roughly the
mean eccentricity measured in the top
panel of Figure~\ref{fig:ecc}. Particles having $\tau_s < 1$ share nearly the same properties as gas. As $\tau_s$ increases above unity, increasingly many particles obtain higher eccentricities owing to stochastic gravitational forcing by gas. {This behavior contrasts with that shown in Figure $6$ of \citet{gibbonsetal2012}, in which the particle velocity distribution narrows as $\tau_s$ exceeds unity; their simulations
omit gravitational stirring.}

\subsection{Relative velocity \label{sec:vel}}
\begin{figure*}
\includegraphics[width=8cm]{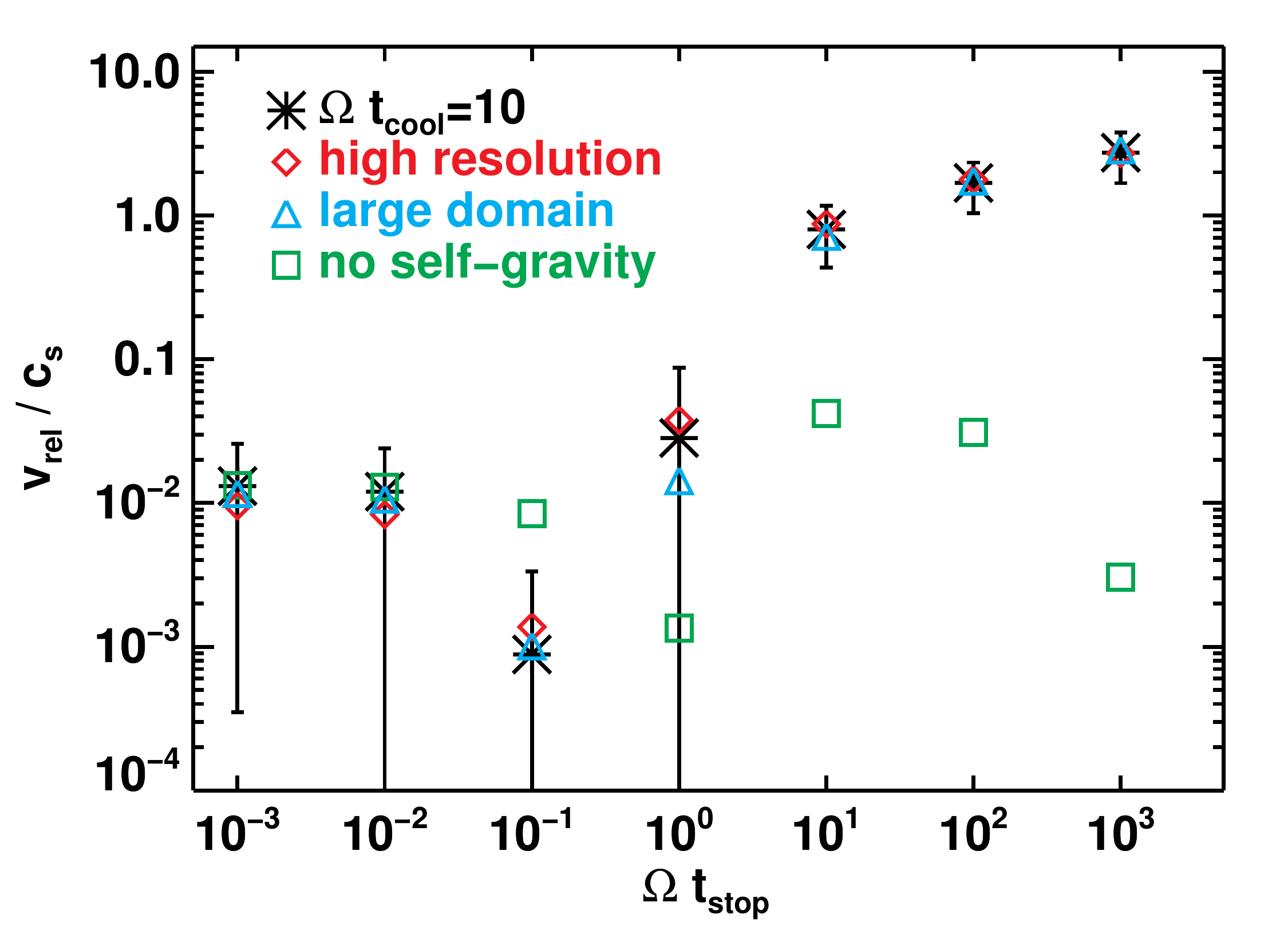}\hfill
\includegraphics[width=8cm]{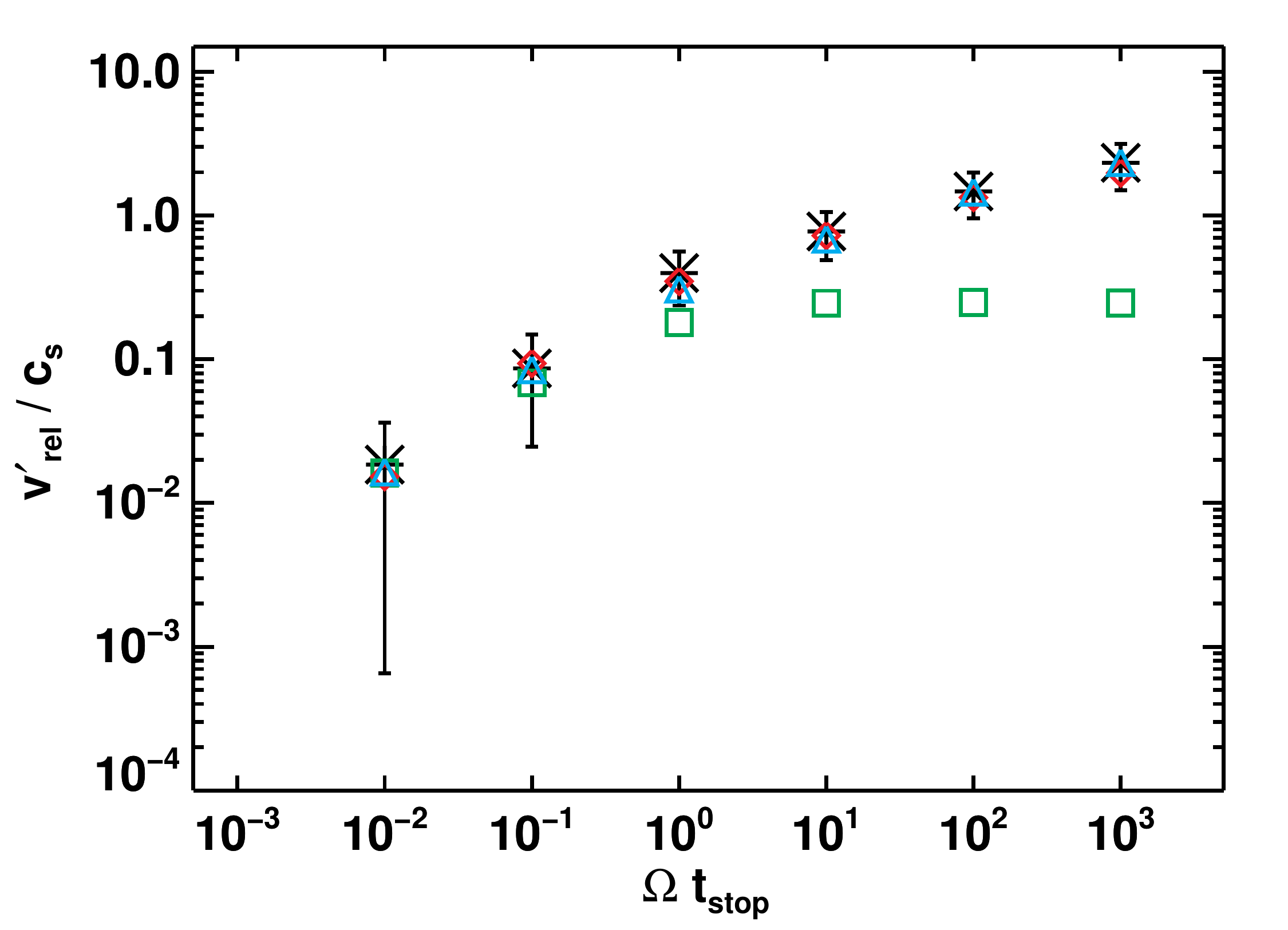}\hfill
\includegraphics[width=8cm]{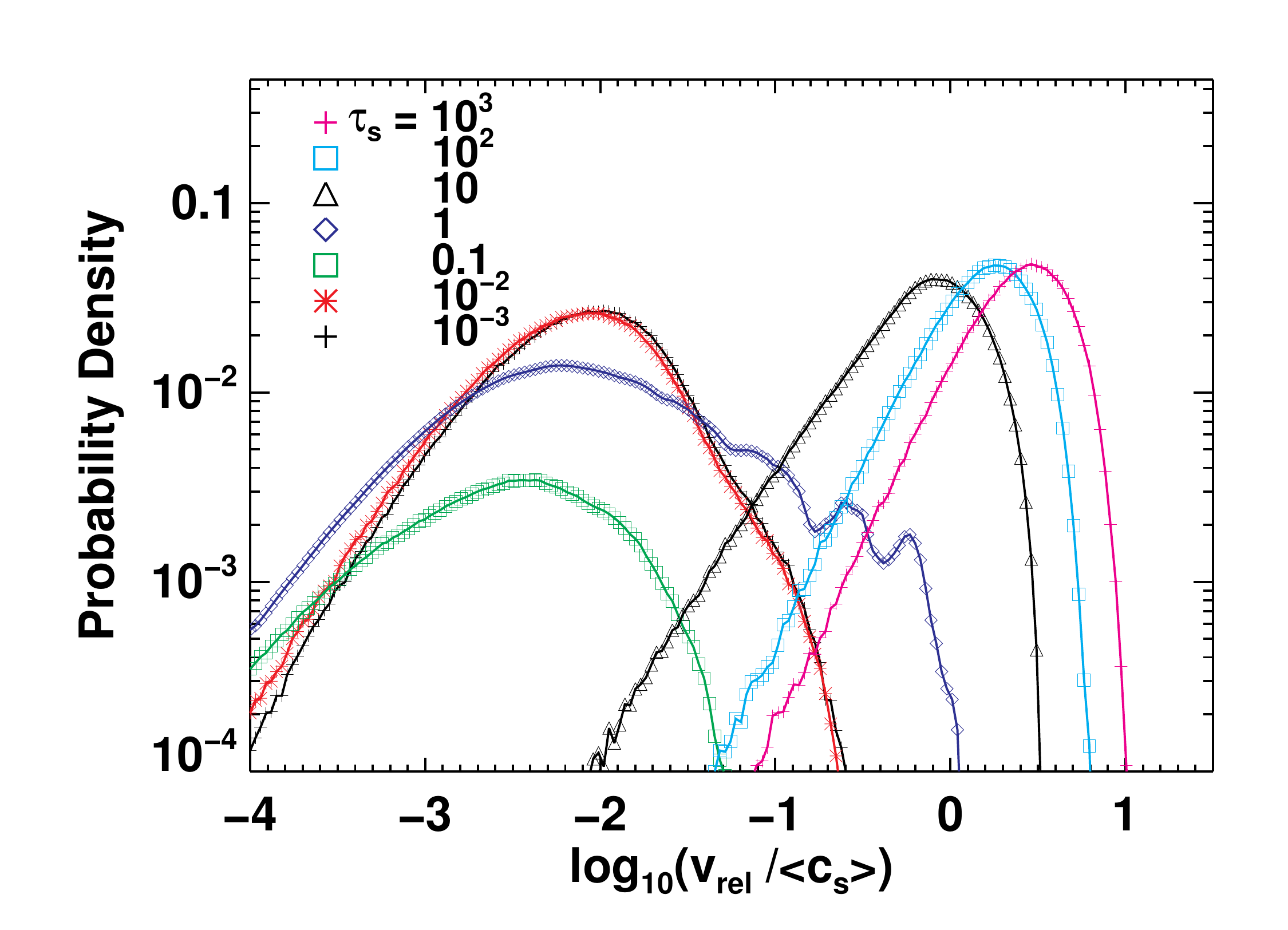}\hfill
\includegraphics[width=8cm]{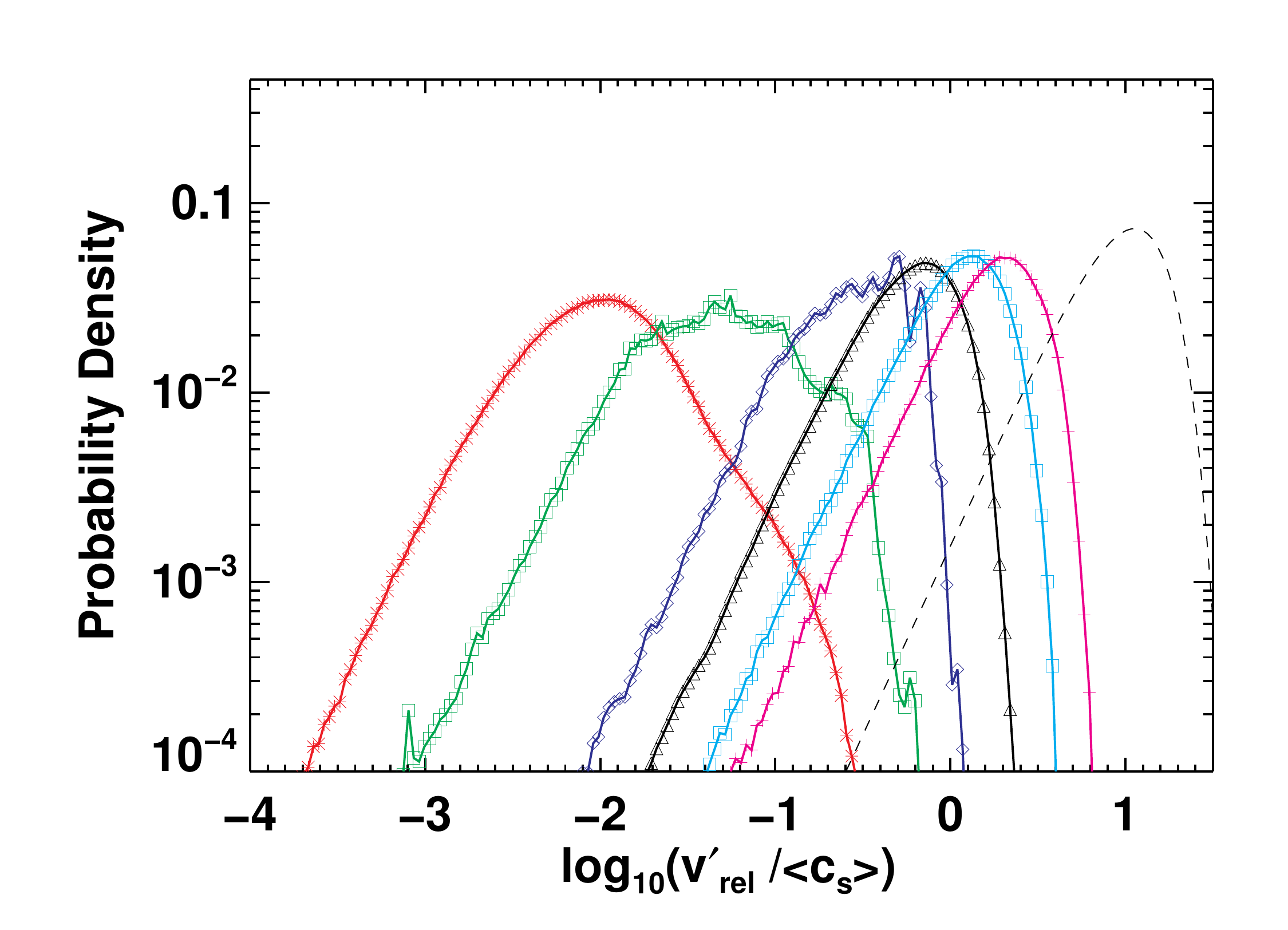}
	\caption{\small{Average relative velocities versus particle stopping time (top row),
			and underlying probability density distributions of relative velocity (bottom row). 
			Averages are taken over the final orbit of 
			the simulations. The relative velocity $v_{\rm rel}$ is evaluated between particles of the same type (stopping time), while $v^{\prime}_{\rm rel}$
			is measured with respect to $\tau_s=10^{-3}$ particles. Error bars show 
			the standard deviation for run tc=10 alone (asterisks). 
			The probability density curve for $v_{\rm rel}$ between $\tau_s = 0.1$ particles (green squares
			in bottom left panel) has a diminished amplitude because there is significant probability
			at $v_{\rm rel}/\langle\cs\rangle < 10^{-4}$, which
			is off the scale of the plot. For comparison, a Maxwellian distribution is shown in the bottom right panel as a dashed curve.
 }}
\label{fig:vrel}
\end{figure*}
The high eccentricities obtained in these
particles indicate large velocity dispersion, which leads to the question of what is the relative
velocity between pair collision.  
We thus measure the relative velocity of the same type particle or the mono-disperse
case $v_{\rm rel}$, and also the relative
velocity with respect to the smallest grains ($\tau_s=10^{-3}$) or a bi-disperse velocity
$v_{\rm rel}^{\prime}$, in each grid cell assuming
there is no sub-structure below this scale. The results are shown in Figure~\ref{fig:vrel}. 

In general, the relative velocity (see the asterisk symbols in top two panels) $v_{\rm rel}$ and
$v_{\rm rel}^{\prime}$ increase from
$\sim 10^{-2}\cs$ to $\gtrsim \cs$ as the $\tau_s$ increases from $10^{-3}$ to $10^3$ which is
indicated by the increasing velocity dispersion or eccentricity as discussed in previous section. 
Exceptions occur at $\tau_s= 0.1\rm{-}1$, where particles show the strongest clustering effects
(coherent and therefore less relative motions in the filaments), the
relative velocity of the same type particles is largely reduced. It even drops below 
$10^{-3}\cs$ for $\tau_s=0.1$ case. The general approach of approximating particle relative velocity
via the measurement of velocity dispersion fails here. 
Our $v_{\rm rel}$ and $v_{\rm rel}^{\prime}$ for smaller particles ($\tau_s \leq 0.1$) seem similar to what are found in MRI-driven turbulent disks \citep[][]{carb2010}. 
However, our $v_{\rm rel}$ and $v^{\prime}_{\rm rel}$ increases with $\tstop$
for larger particles that have $\tau_s>0.1$, in contrast to MRI-driven turbulence results \citep[c.f. Figure~2 and 3
in][]{carb2010} where turbulent torquing due to the fluctuating gravity field of the gas is not
considered. The latter show $v_{\rm rel}$ falls continuously, and
$v^{\prime}_{\rm rel}$ stays roughly constant with $\tstop$, in agreement with the
theoretical prediction in isotropic turbulence \citep{OC2007}. We will discuss the effects of 
gravitational stirring in section~\ref{sec:selfg}. 

\begin{figure}
\includegraphics[width=\columnwidth]{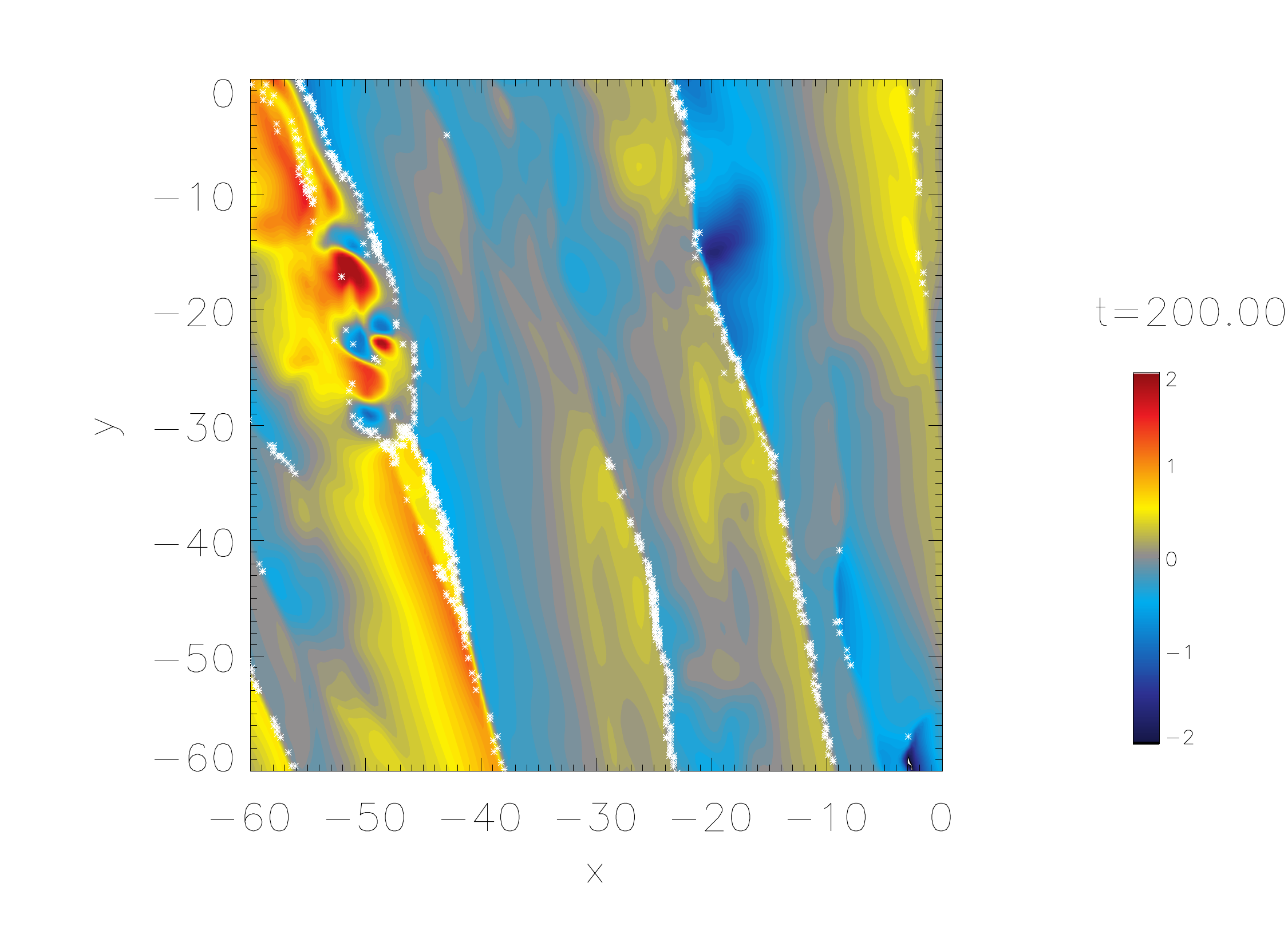}
\caption{\small{Snapshot of gas radial velocities (normalized by the volume-averaged sound speed) at $t=200\Omega^{-1}$. Large relative velocities ($>0.1\cs$) between $\tau_s=10^{-2}$ and $10^{-3}$ particles are overlaid as white symbols.
Evidently, large relative velocities between
particles characterize regions just behind shock fronts; in post-shock regions, particles decelerate at different rates according to their different stopping times \citep{NM2004,JT2014}.
 }}
\label{fig:postshock}
\end{figure}

We also show the probability density functions (PDFs) for $v_{\rm rel}$ and 
$v_{\rm rel}^{\prime}$ in Figure~\ref{fig:vrel}. 
The PDFs of larger particles ($\tau_s > 1$) are similar to 
Maxwellian distribution (see the dashed curve in the bottom right panel as 
an example of Maxwellian distribution) \citep{windmark2012b,garaudetal2013}. 
Particles smaller than $\tau_s\sim 1$, however, show broader distributions and 
smaller mean values than the bigger particles. They deviate significantly from a Maxwellian and resemble more of a log-normal distribution \citep{mitraetal2013,PPS2014b}. 
Relative velocities with respect to the smallest particles, i.e., $v^{\prime}_{\rm rel}$, have
distributions that shift gradually to higher values as
the size of the larger particle increases. {Between $\tau_s = 0.01$ and $\tau_s = 0.001$ particles, the largest
relative velocities $v^{\prime}_{\rm rel} > 0.1 \cs$, which might lead to collisional destruction, are mostly due to their different 
deceleration rates in post-shock regions \citep{NM2004,JT2014}. 
As plotted in Figure~\ref{fig:postshock}, the locations where $v^{\prime}_{\rm rel} >0.1 \cs$ (white 
symbols) are found just behind shock fronts,
where gas radial velocities are discontinuous.}

Relative velocities $v_{\rm rel}$ between
particles of the same type can be separated
into two groups. 
One group corresponds to the small-size particles ($\tau_s\leq 1$), for which
relative velocities peak 
at $v_{\rm rel} \simeq 0.01 \cs$.
The other group corresponds to large-sized particles ($\tau_s > 1$), for which peak velocities are sonic. We note that $\tau_s = 0.1$ (green squares) particles 
show diminished amplitude and a significant shift toward smaller relative velocity 
due to the clustering effect: $\sim 80\%$ of contribution from 
$v_{\rm rel}/\cs \ll 10^{-4}$ is not shown in this plot. While for $\tau_s = 1$ 
particles, there is also a second, near sonic, contribution, which shows the 
transitional behavior from small size (friction-dominated) to large size (gravity-dominated)
particles.

\subsection{Effects of gravitational stirring\label{sec:selfg}}
The gravitational stirring effect has been studied in MRI-driven turbulent
disks and shown to induce high velocity dispersion for solid bodies bigger than 
$\sim 100\,\rm{m}$ (or $\tau_s \gtrsim 10^3$ at radius $R\sim$\,AU)
\citep{NG2010,yangetal2009,yangetal2012,OO2013a,OO2013b}. But we will show the dominating
gravitational forcing kicks in for even smaller particles owning to its much stronger density
fluctuation, $\delta\Sigma/\Sigma\sim 5\sqrt{\alpha}$ in gravito-turbulent disks (see
section~\ref{sec:tcool}) rather than
$\sim \sqrt{0.5\alpha}$ in MRI-driven turbulent disks \citep{NG2010,yangetal2012}. 

As the stopping time increases, the first term in
Equation~(\ref{eq:eom_par}), the drag force,  becomes less important compared to the second term, the
gravitational acceleration. Assuming the characteristic length scale of the spiral density features
is $l$, then $|\nabla \Phi |\sim \Phi / l \sim G\delta\Sigma$, where $\delta\Sigma\sim \Sigma$ in
our gravito-turbulent disk is the density fluctuation.
The length scale can be approximated with the most unstable wavelength for axisymmetric
disturbances, or simply the disk scaleheight, 
$l \sim \cs^2/G\Sigma \sim Q^2 (G\Sigma/\Omega^2) \sim H$. This is also supported by our
simulations; the density waves have typical radial wavelength $\sim H$ in the top left panel of
Figure~\ref{fig:8panel}.
The drag force $|\ux - \vx|/\tstop \lesssim \Omega l /\tstop \sim \cs /\tstop$. 
The ratio of these two thus gives
\beq
\frac{|\nabla\Phi|} {|\ux-\vx|/\tstop} \sim \frac{1}{Q} \frac{\delta\Sigma}{\Sigma} \tau_s \,,
\label{eq:ratio}
\enq
the gravitational stirring becomes more important than aerodynamical drag when
$\tau_s > Q \sim 1$ in gravito-turbulent disks. However, for non-self-gravitating turbulent disks
such as MRI-driven turbulent disks, $\delta\Sigma/\Sigma$ is one order of magnitude smaller, and
$Q\gg 1$. As a results, the gravitational stirring only affect particles with $\tau_s > 10 Q \gg 1$. 
\begin{figure}
\includegraphics[width=\columnwidth]{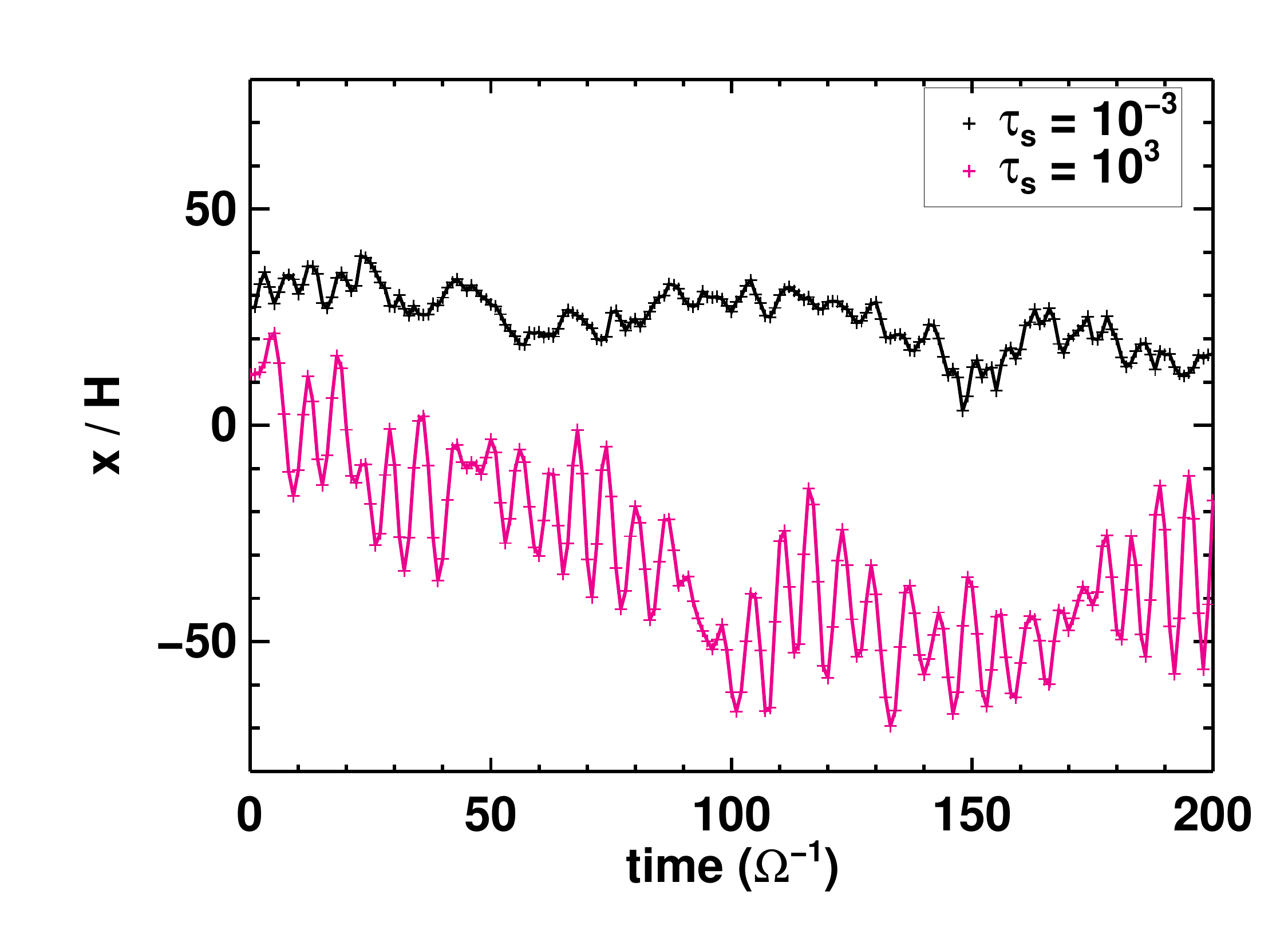}\\
\includegraphics[width=\columnwidth]{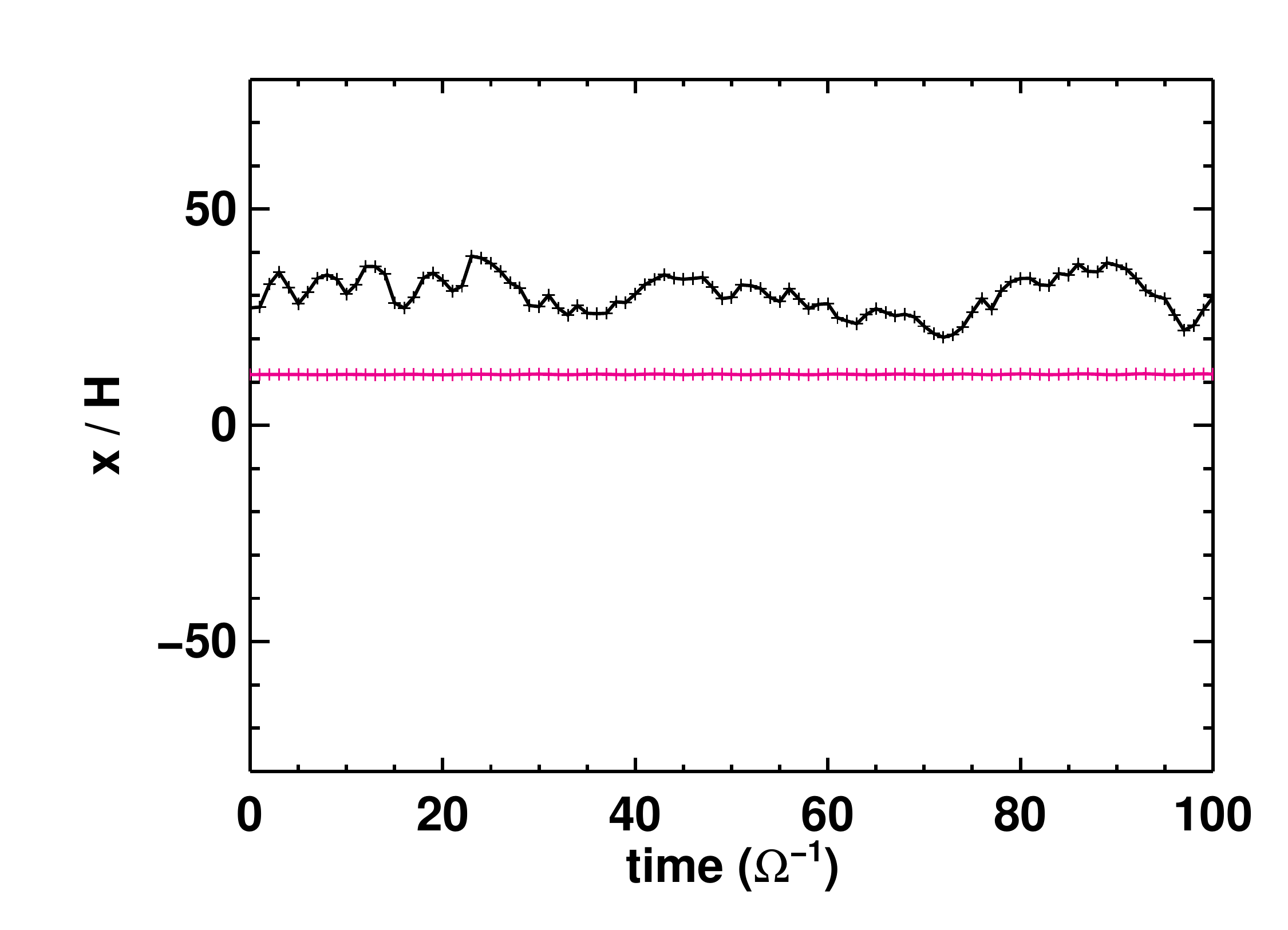}
\caption{\small{Examples of radial displacement for two particles with fixed stopping 
        time $\tau_s = 10^{-3}$ (black) and $10^3$ (magenta) in run with (top panel) 
        and without (bottom panel) gravity pull from the gas in the calculation. The
        gravitational stirring of the $\tau_s=10^3$ particle causes large amplitude oscillations which are absent in the bottom panel when gravity from the gas 
        is removed. }}
\label{fig:disp}
\end{figure}

\begin{figure}
	\includegraphics[width=\columnwidth]{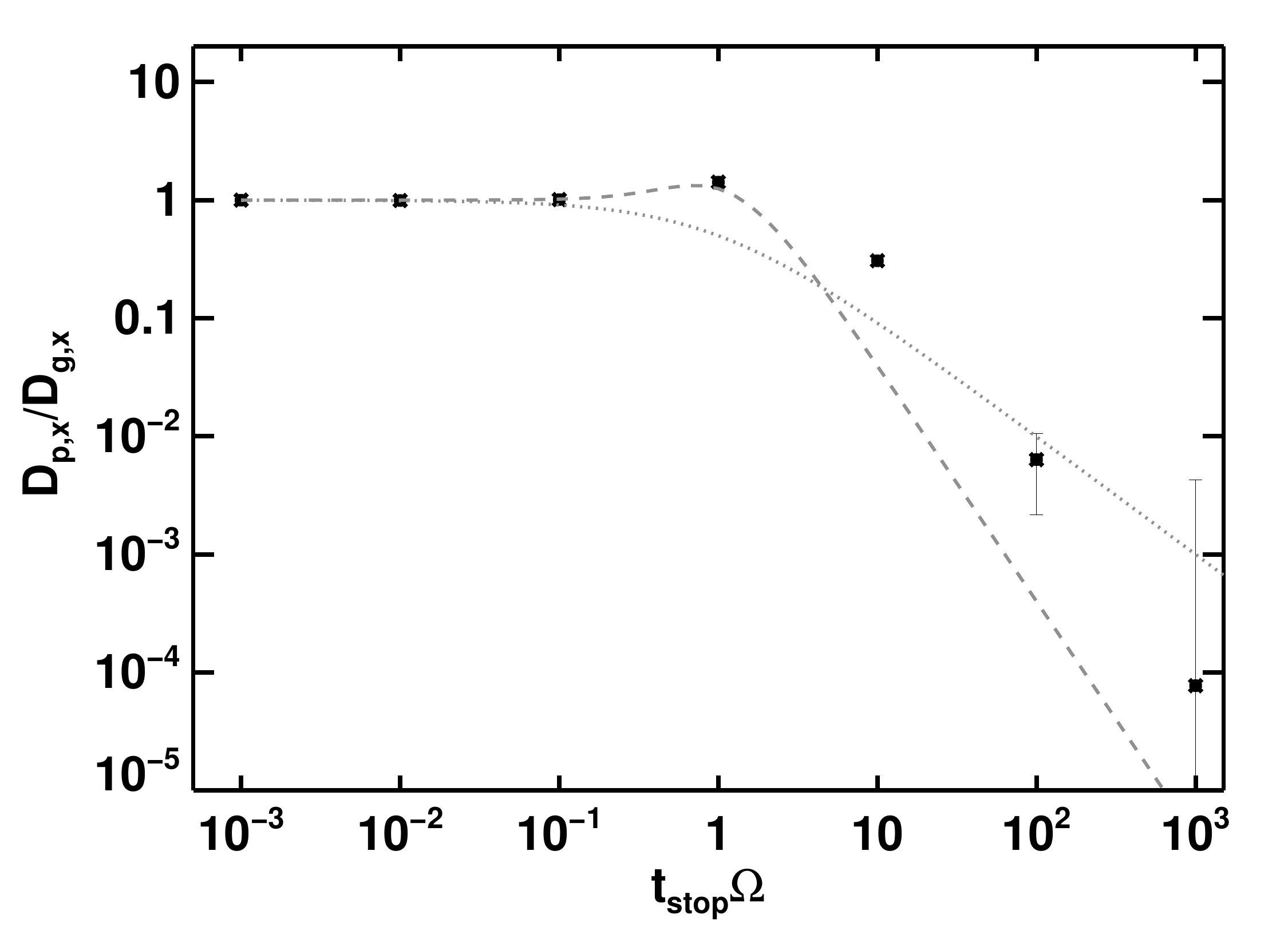} \\
	\includegraphics[width=\columnwidth]{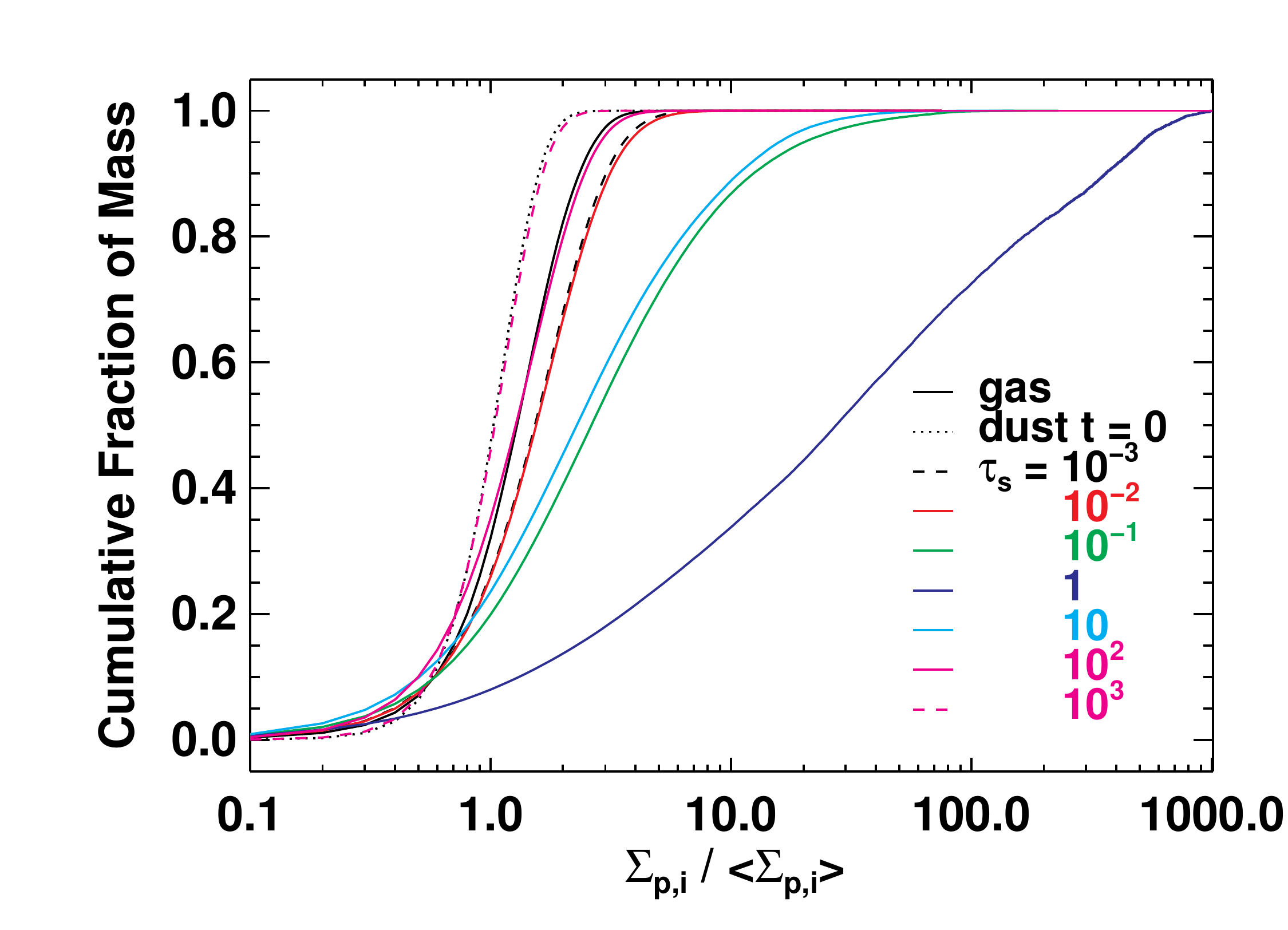}
\caption{\small{Similar to Figure~\ref{fig:fixed_ts} and \ref{fig:mass}, we show radial 
         diffusion coefficients (top panel) and cumulative fraction of dust mass (bottom 
         panel) for run tc=10.wosg, in which the gravitational forcing from the gas is
         artificially removed. At larger stopping time, the measured $\Dp$ without
         gravitational stirring are now closer to models of \citet{YL2007} (dashed gray 
         curve) or \citet{cuzzietal1993} (dotted). The clustering effect for particles of $\tau_s=0.1$ observed in Figure~\ref{fig:mass} is absent when gravitational stirring is removed.}}
\label{fig:wosg}
\end{figure}
In top panel of Figure~\ref{fig:disp}, we show the time variation of radial positions of two representative
particles in the standard simulation. The particle with larger $\tau_s=10^3$ oscillates with a
peak-to-peak amplitude of $\sim 40 G\Sigma_0/\Omega^2$ at a frequency of the orbital
frequency $\Omega$ after it is released in the turbulent disk. However, the
small particle, $\tau_s=10^{-3}$, scatters randomly over time and does not show strong
periodic variations. 

To reveal the self-gravity effect to the dust dynamics, we perform a rerun (tc=10.wosg) of the
standard simulation but suppress the gravitational forcing manually, i.e. removing
$-\nabla\Phi$ term in Equation~\ref{eq:eom_par}. The radially projected
trajectories of the same two representative particles are plotted in the bottom panel of
Figure~\ref{fig:disp} for comparison. Clearly, the small particle is not affected by the
missing of
gravitational acceleration, and appears to behave similarly as in the standard run. 
However, the
larger particle in tc=10.wosg barely moves radially when the only acceleration/deceleration
is friction.   

Without the gravitational stirring, the diffusion of particle also look significantly 
different than
in Figure~\ref{fig:fixed_ts}. In Figure~\ref{fig:wosg}, we present the diffusion coefficients
measured in run tc=10.wosg using the
same method as in Figure~\ref{fig:fixed_ts}. Again, the small $\tstop$ particles are unaffected
and therefore give nearly the same $\Dp$ as in the standard run. $\Dp$ of particles with
$\tau_s > 1$ drops rapidly as $\Dp \propto (\tau_s)^{-2}$ which manifests the lack of
gravitational forcing. We note the exact relation between diffusion coefficient and stopping
time slightly deviates from models of \citet{cuzzietal1993, YL2007}. We speculate that is a
result of different power spectra in gravito-turbulent disks than normal homogeneous turbulence model adopted in these models. 

In Figure~\ref{fig:wosg}, we also compute the cumulative mass fraction for run tc=10.wosg. As
expected, the $\Omega\tstop = 1$ particles show the strongest concentration and more
contribution towards the higher concentration ($\gtrsim 100$) than the standard tc=10 run. Both
$\Omega\tstop = 0.1$ and $\Omega\tstop = 10.0$ show similar secondary clustering but are much
weaker than the concentration of $\Omega\tstop = 0.1$ in the case where gravity is present.

Without the stochastic gravitational stirring, the averaged particle eccentricity for 
all types of
particles never exceed
$e\sim 0.3 (H/R)$ as shown in Figure~\ref{fig:ecc_ts} (green squares). 
The small particles ($\tau_s\leq 1$) have similar eccentricities as in the
standard run; however, those larger particles are dynamically unimportant, and $e$ 
approaches nearly zero (initial value) as $\tau_s$ increases.
The gravitational stirring is also responsible for the increasing relative velocities for
particles of $\tau_s\ge 1$.
As shown in top row of Figure~\ref{fig:vrel} in green squares, instead of rising up, 
$v_{\rm rel}$ decreases while $v^{\prime}_{\rm rel}$ stays constant with increasing $\tau_s$
when gravitational forcing is not included, which is consistent with the results from 
previous study \citep{OC2007,carb2010}.

\subsection{Fixed particle size\label{sec:fixed_size}}
\begin{figure*}
\includegraphics[width=8cm]{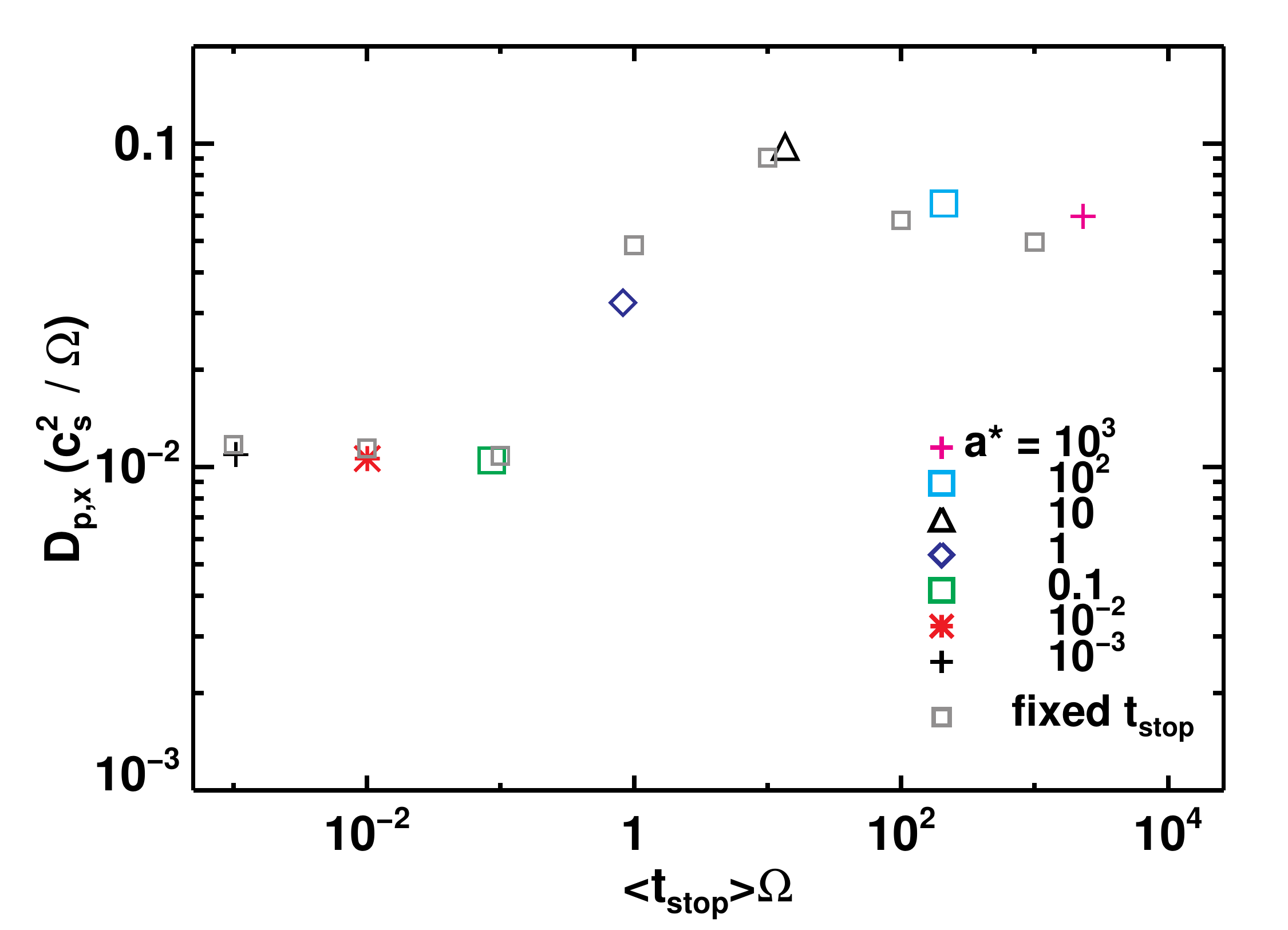} \hfill
\includegraphics[width=8cm]{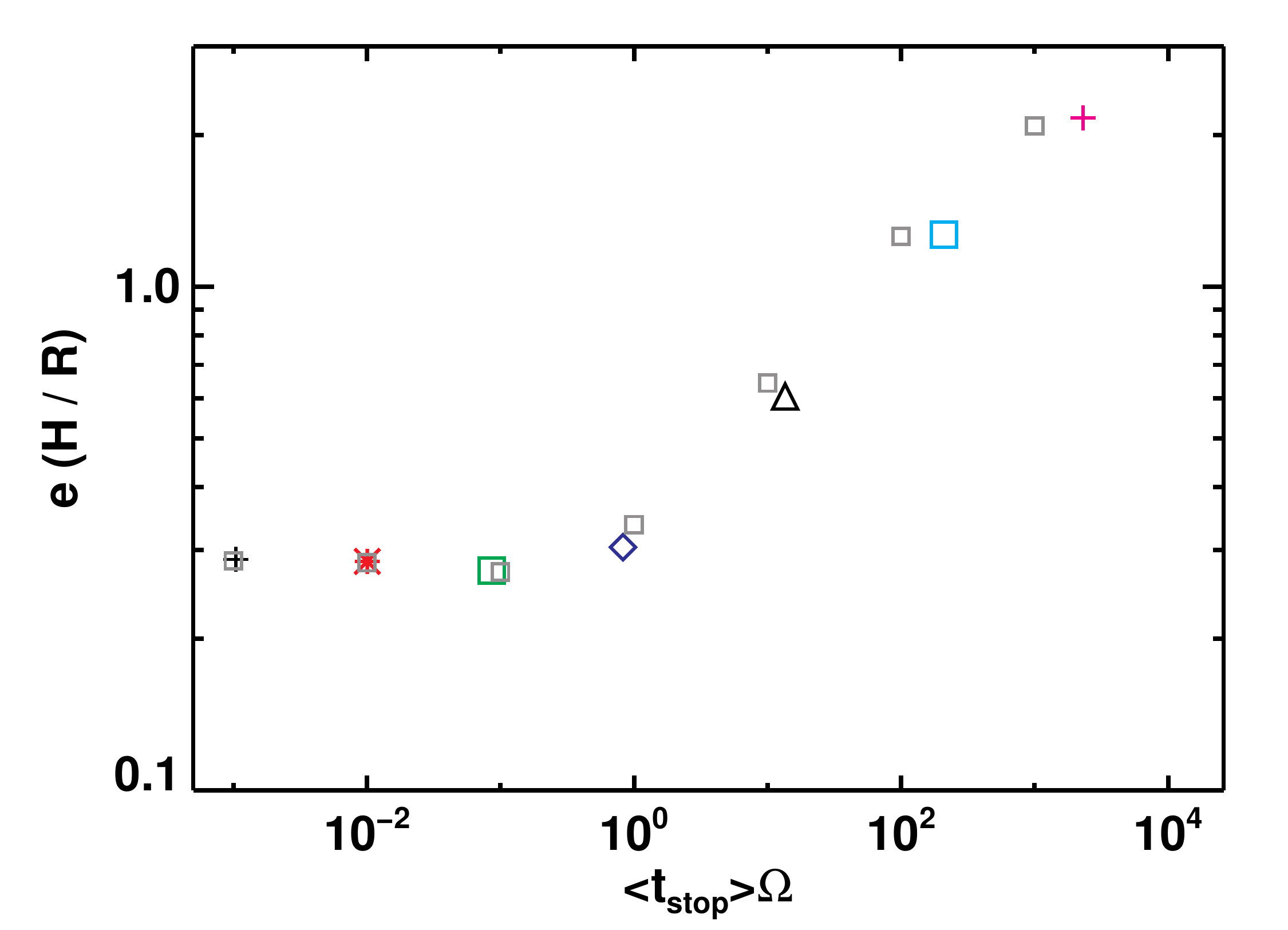} \hfill
\includegraphics[width=8cm]{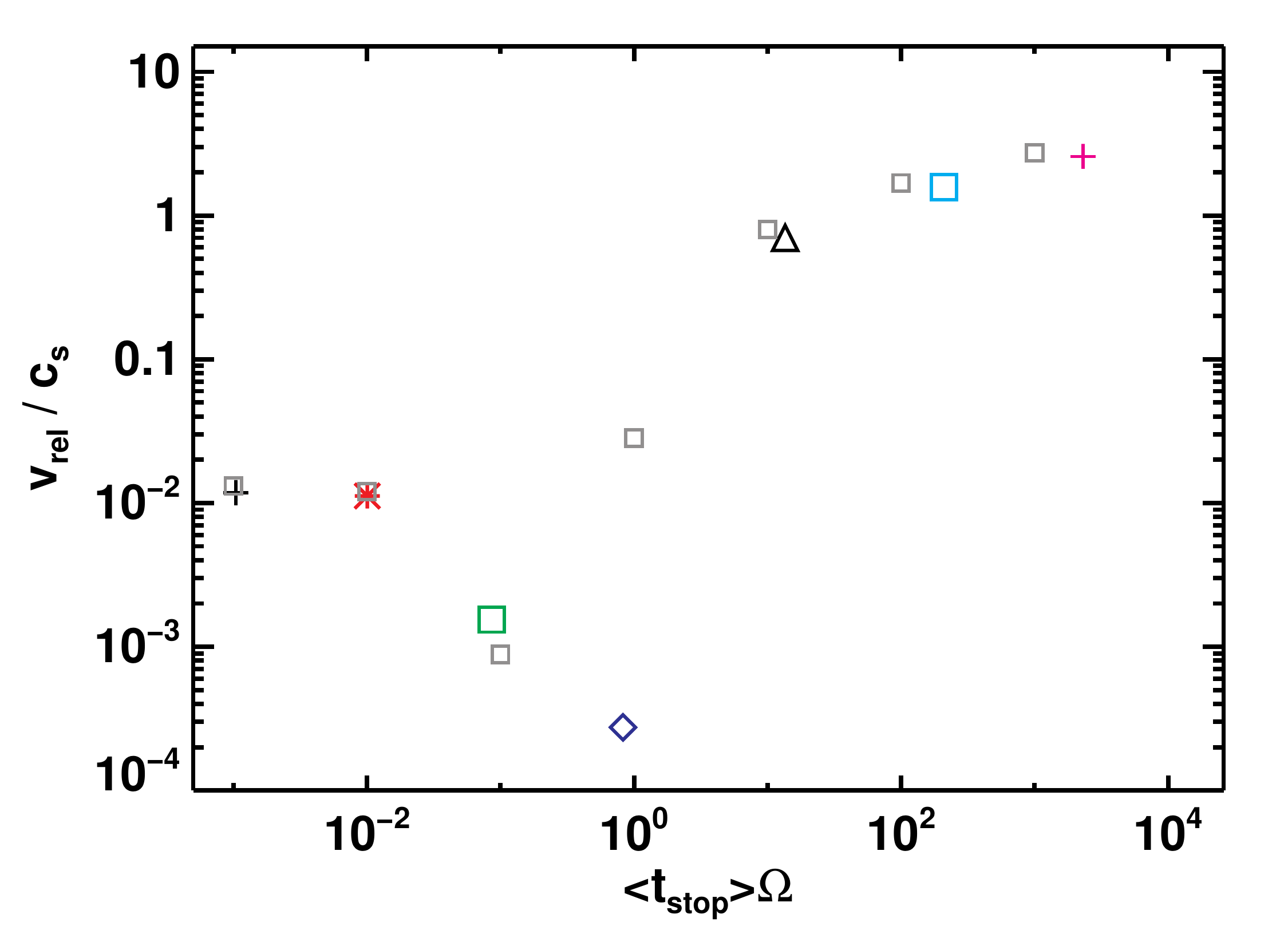}\hfill
\includegraphics[width=8cm]{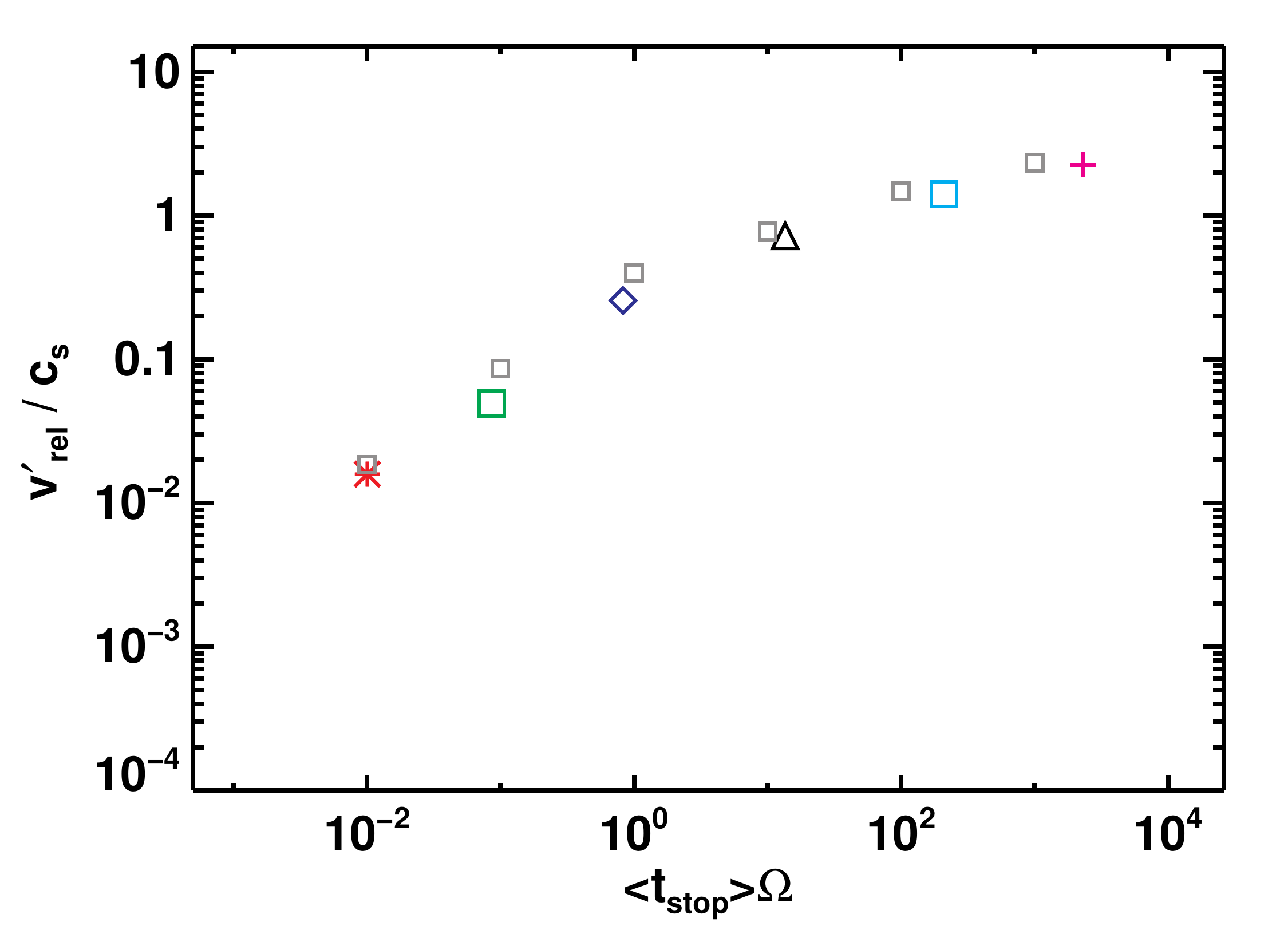}
\caption{\small{
Diffusion coefficients (top left), particle eccentricity (top right), relative velocities $v_{\rm rel}$ (bottom left) and $v^{\prime}_{\rm rel}$ (bottom right) versus the effective stopping time for fixed size simulation tc=10.size. The effective stopping times are calculated by time and particle averaging of $\tstop$ of individual particle.  Results of tc=10 from Figure~\ref{fig:fixed_ts}, which are shown with smaller grey squares, almost match fixed size results and therefore validate the usage of fixed stopping time particles in this work. We note the constant $\tstop$ approach could overestimate the relative speed for intermediate size particles with $a^*=1$ or effective $\tau_s\sim 1$. 
}}
\label{fig:fixed_size}
\end{figure*}
\begin{figure}
	\includegraphics[width=\columnwidth]{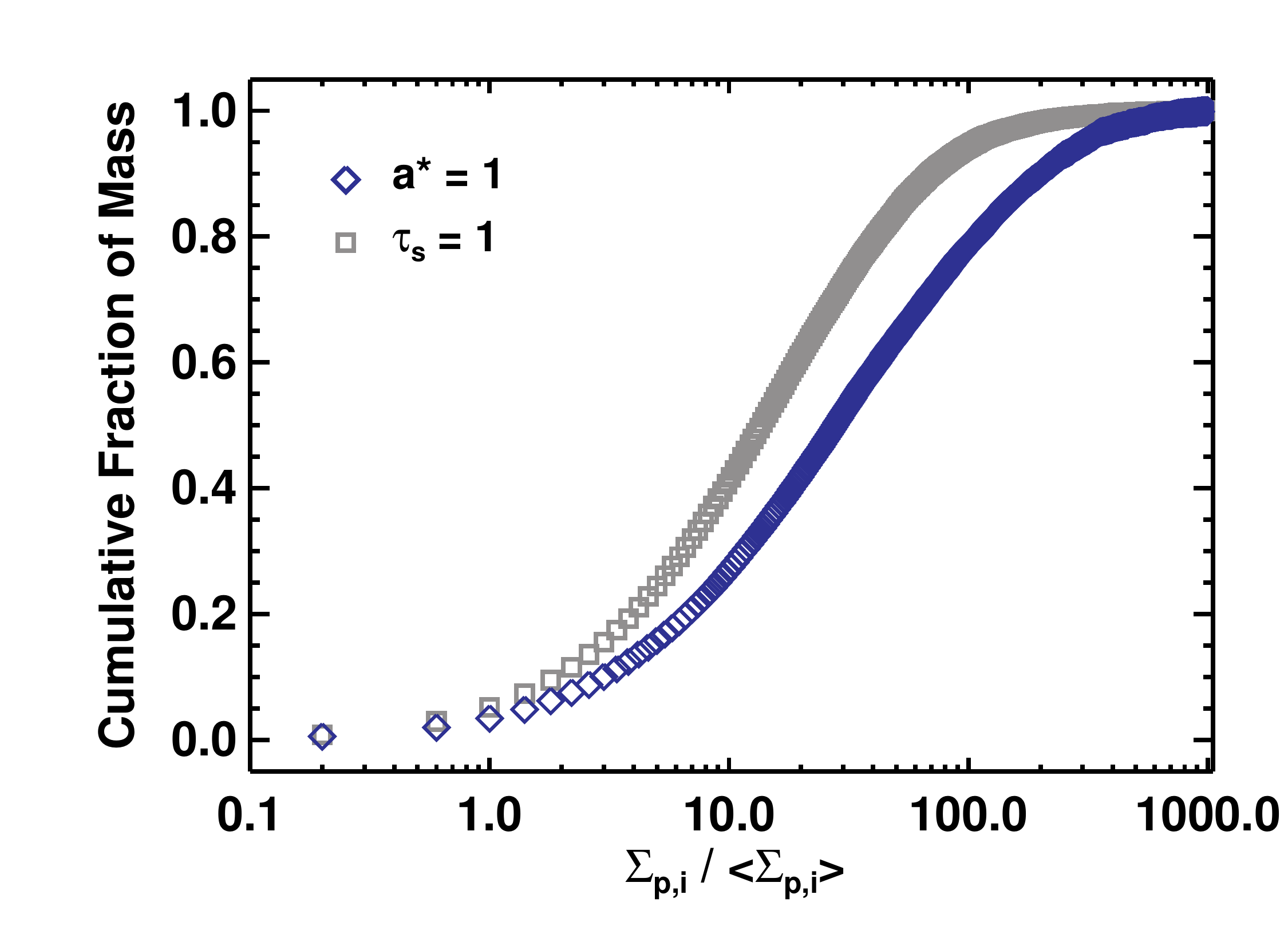}
\caption{\small{How particles concentrate
in a run with fixed stopping time
$\tau_s = 1$ (from run tc=10)
versus~a run with fixed particle size
$a^*=1$ (from run tc=10.size). The latter
case, having an effective $\tau_s \simeq 0.83$, shows stronger concentration. }}
\label{fig:cluster_fixsize}
\end{figure}
In case that the fluctuations of the density and sound speed are large ($\lesssim O(1)$), as is the case in
gravito-turbulent disks, the constant stopping time assumption for individual particles might not be
valid as the particles would have varying $\tstop$ as they travel around different regions of the
disk. We therefore, in this section, test if our results still hold statistically when we fix particle size
instead of the stopping time.

We set up seven types of particles with fixed size, dimensionless size $a_i^* = 10^{i-4}$ for
$i=1$-$7$, as described in Equation~(\ref{eq:epstein}) and section~\ref{sec:ic}. Everything else is
kept the same as the standard run tc=10. 
For each type of particles, we define an effective stopping time
$\langle\langle \tstop{_{,i}}\drangle\rangle_t$ by
first measuring the stopping time of individual particle according to
Equation~\ref{eq:epstein}, and then averaging over all particles
of the same type and time. We find the averaged stopping time
$\langle\langle\tau_s{_{,i}}\rangle\rangle_t\Omega\simeq 
[1.04\times 10^{-3},1.01\times10^{-2}, 0.086, 0.827, 13.5, 2.10\times 10^2, 2.31\times 10^3]$ for $a_i^* = 10^{-3}$--$10^3$. For small particles ($a_i^* < 0.1$),
$\langle\langle \tau_s{_{,i}}\rangle\rangle_t \simeq a_i^* \Omega^{-1}$ as 
designed because we choose the converting factor$f$ in Equation~\ref{eq:epstein} 
to match $a^*$ with $\tau_s$ of the fixed stopping time case. For intermediate 
size, the averaged stopping time is only $\lesssim 20\%$ smaller than the 
designed values. For larger particle with $a_i^* > 1$, the effective stopping 
time is roughly $\sim 2 a_i^* \Omega^{-1}$.

After converting $a^*$ to the effective stopping time, we find the results stay the same as
the simulations using particles of constant stopping times. In Figure~\ref{fig:fixed_size}, 
we show the diffusion coefficient, particle eccentricity, and relative velocities in large
colored symbols. They almost follow the results using fixed $\tstop$ (gray squares). Therefore, statistically, these results validate the usage of fixed $\tstop$ particles even in the highly
fluctuated
density field of a gravito-turbulent disk. In general, the results of fixed $\tstop$ can 
approximate the results of fixed size particles.
But this approach (constant $\tstop$) might underestimate the
clustering effect for the intermediate size particle $\tau_s = 1$ as shown in
Figure~\ref{fig:cluster_fixsize}. This also leads to an overestimate of the relative speed $v_{\rm rel}$ for intermediate size particles of $\tau_s \sim 1$. The actual relative speed for intermediate 
size particles would become even smaller than what we have obtained.

\subsection{Effects of cooling time\label{sec:tcool}}
\begin{figure*}
\includegraphics[width=8cm]{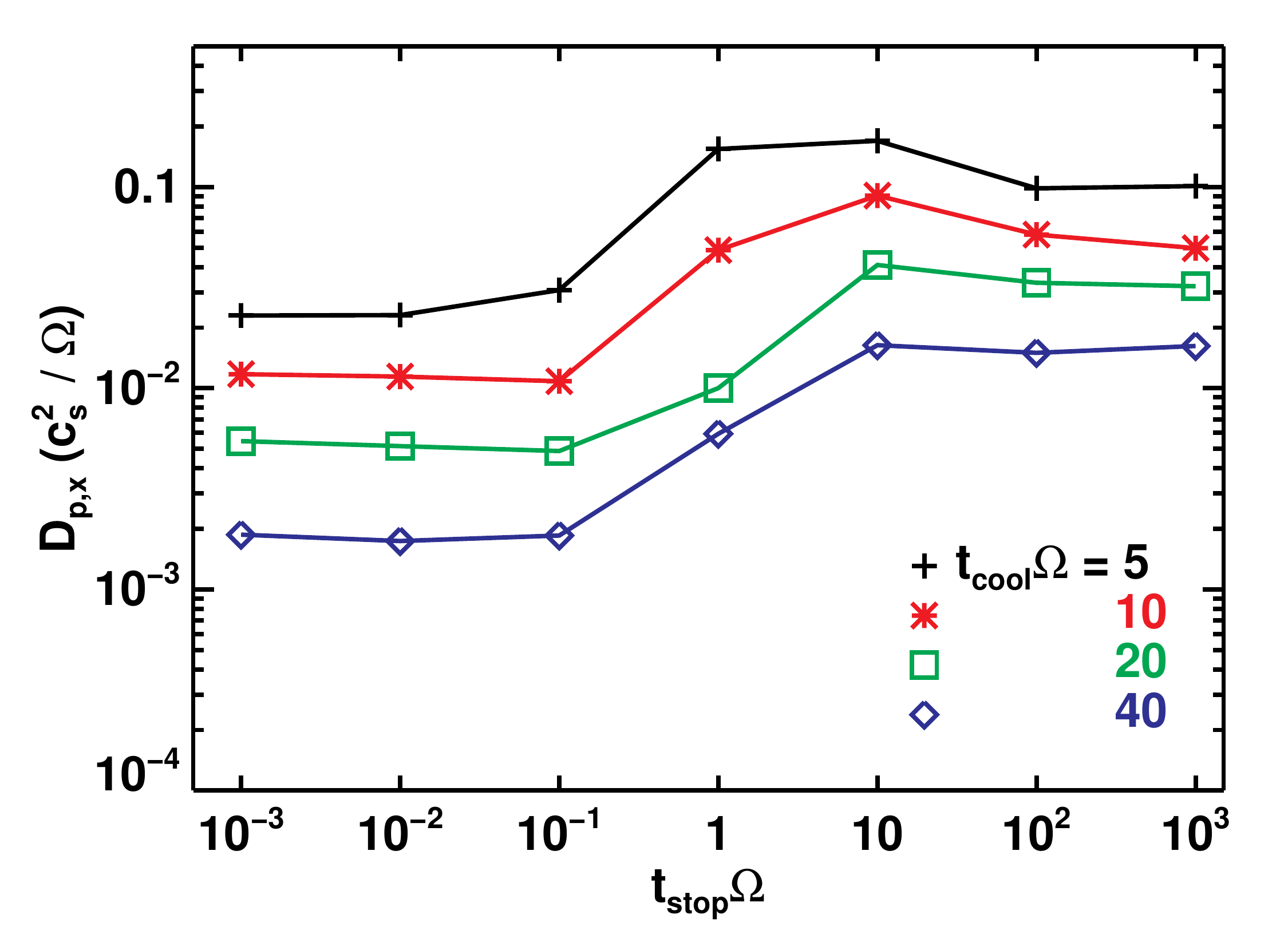} \hfill
\includegraphics[width=8cm]{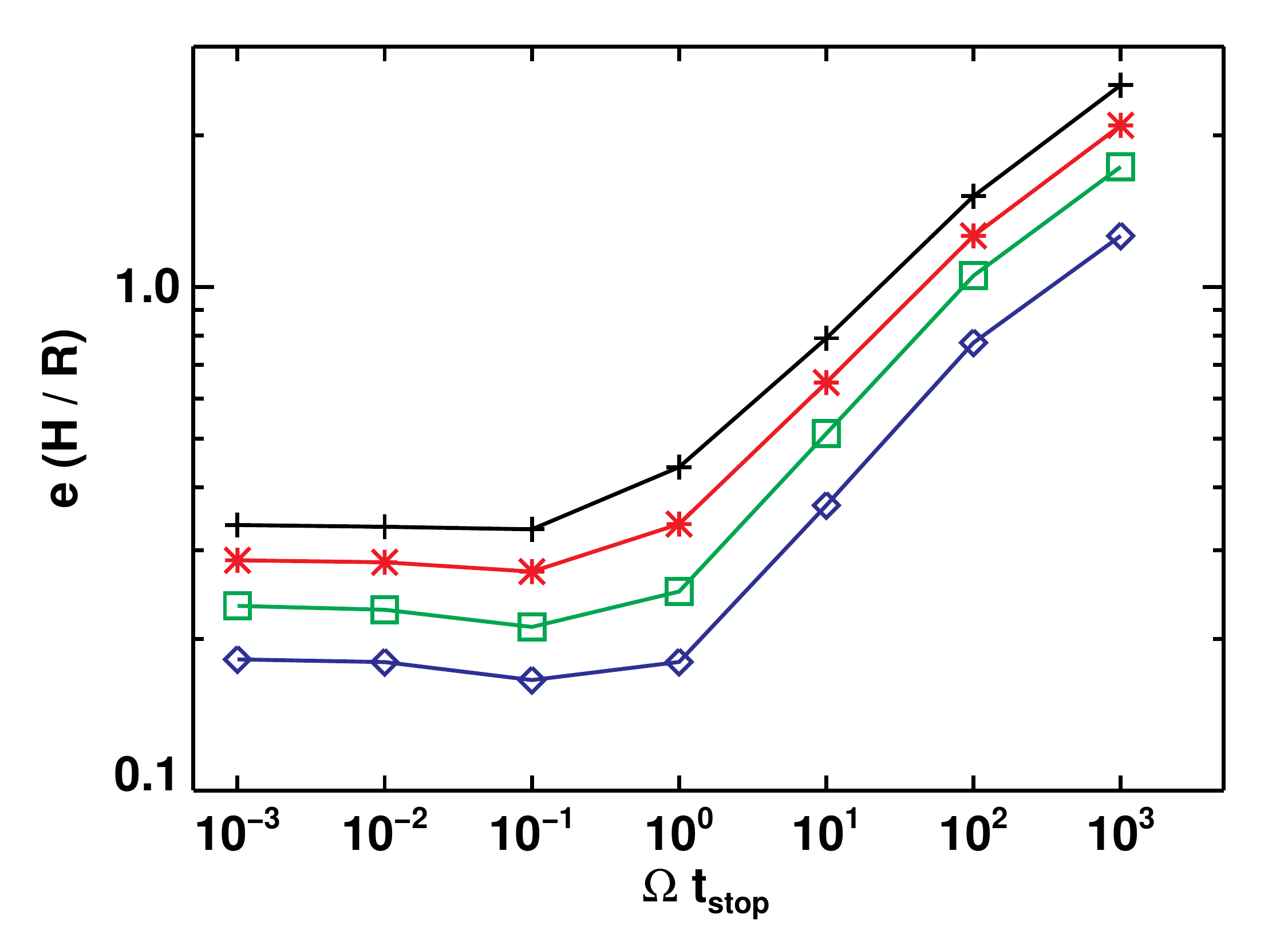} \hfill
\includegraphics[width=8cm]{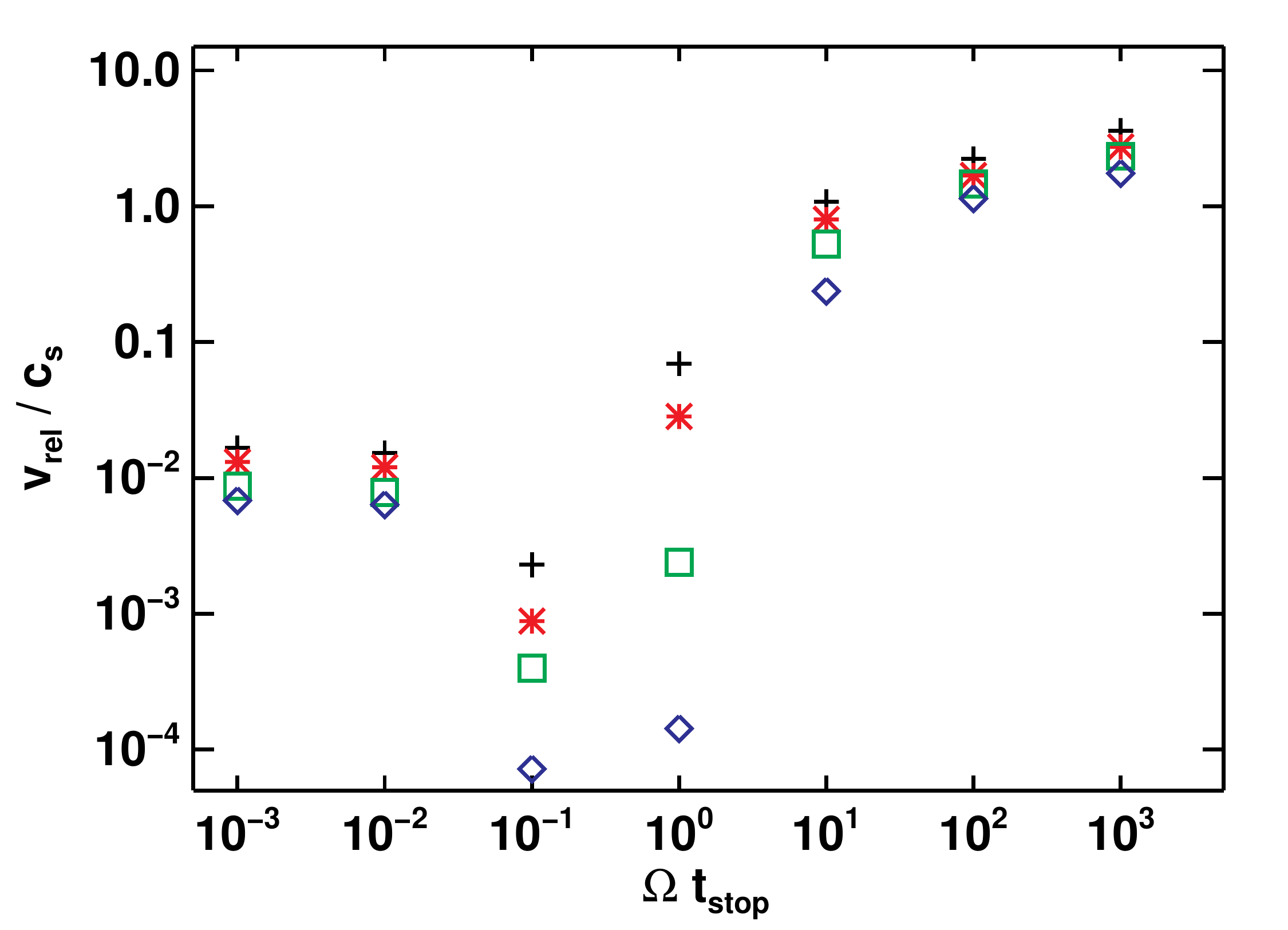}\hfill
\includegraphics[width=8cm]{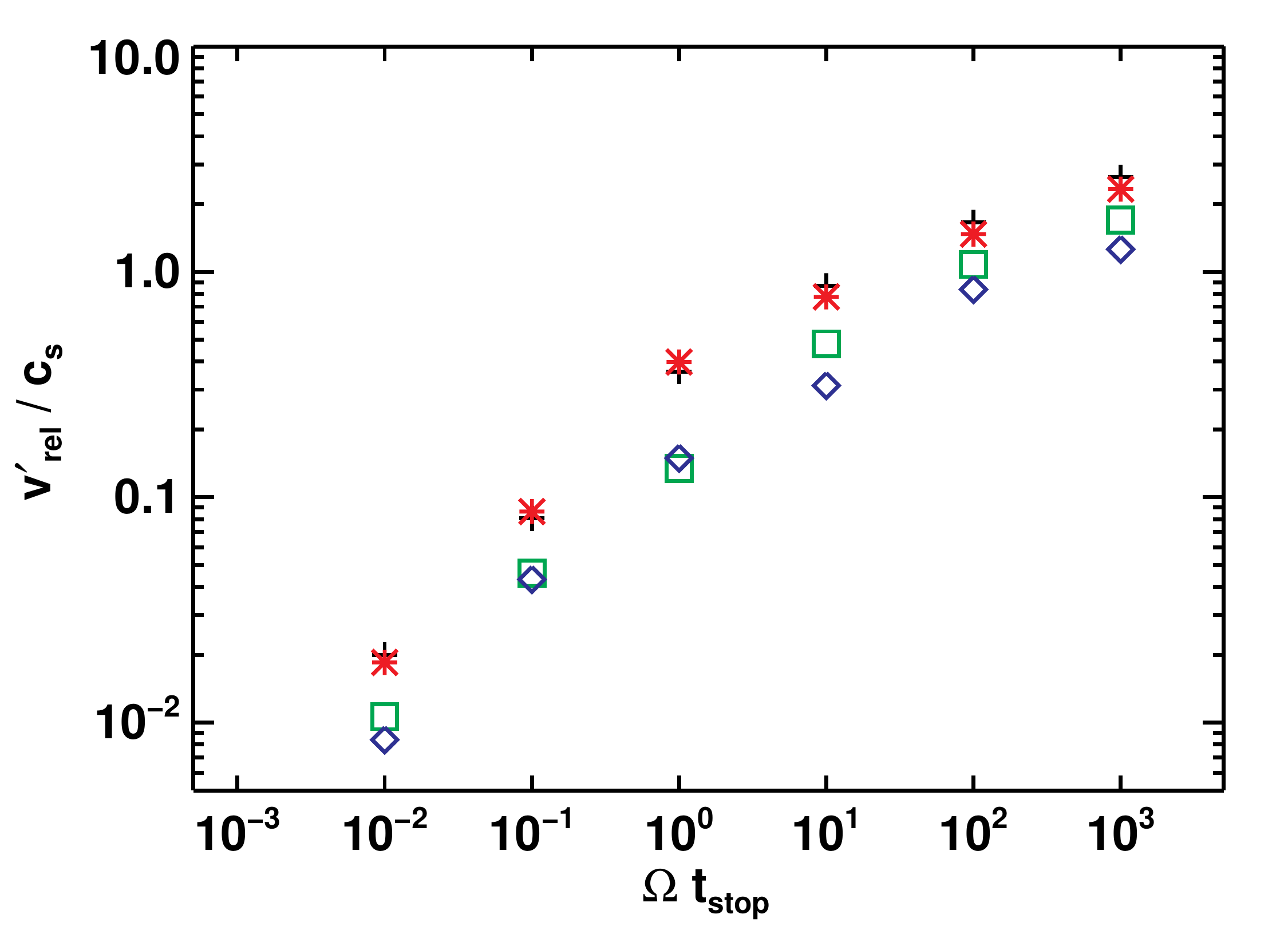}
\caption{\small{The dependence on cooling
time (different symbols) of the radial diffusion coefficient (top left), saturated
particle eccentricity (top right), relative velocity between particles of the same
type (bottom left), and relative
velocity with respect to $\tau_s=10^{-3}$ particles (bottom right).
The measured $\Dp$ is nearly inversely proportional to $\tcool$, while particle eccentricity and relative velocities roughly follow $\tcool^{-1/2}$.  }}
\label{fig:tcool}
\end{figure*}
The dust dynamics would also be affected by the strength of the turbulence. In a
gravito-turblent disk, the strength of the turbulence, e.g., in term of $\alpha$, is inversely
proportional to the cooling time \citep{gammie01,ShiChiang2014}. We can therefore 
study how the dust dynamics changes with the turbulent strength by varying $\tcool$.

We perform simulations with $\tcool\Omega = 5$ (run tc=5), $20$ (run tc=20), and $40$ (run tc=40)
using the same particle distribution as in
the $\tcool\Omega = 10$ standard run. We then measure the diffusion coefficient in each $\tcool$
case and present the results in Figure~\ref{fig:tcool}. We find $\Dp$ of various cooling time share
rather similar profile. It is flat at $\tau_s < 0.1$ end when aerodynamical
coupling is strong. $\Dp$ traces $\Dg$ in this regime, and the Schmidt number, which measures the ratio of angular momentum transport and mass diffusion,  
${\rm Sc} \equiv \alpha \cs^2\Omega^{-1}/\Dg  \sim 1.4$--$2.4$, stays roughly constant.
The diffusion coefficient rises up for $ 0.1 \leq \tau_s \leq 1$, indicating
the increasing effect of the gravitational stirring. 
In the very long stopping time regime, $\tau_s \geq 10$, the profile levels off again as the dominant gravitational acceleration does not
depend on $\tstop$. The magnitude of $\Dp$ is usually $5$--$9$ times of the diffusion of small particle or gas. 
For given particle size,  we find $\Dp$ scales roughly as
$\propto 1/\tcool$ which indicates stronger diffusion in stronger turbulence.

\begin{figure*}
	\includegraphics[width=16cm]{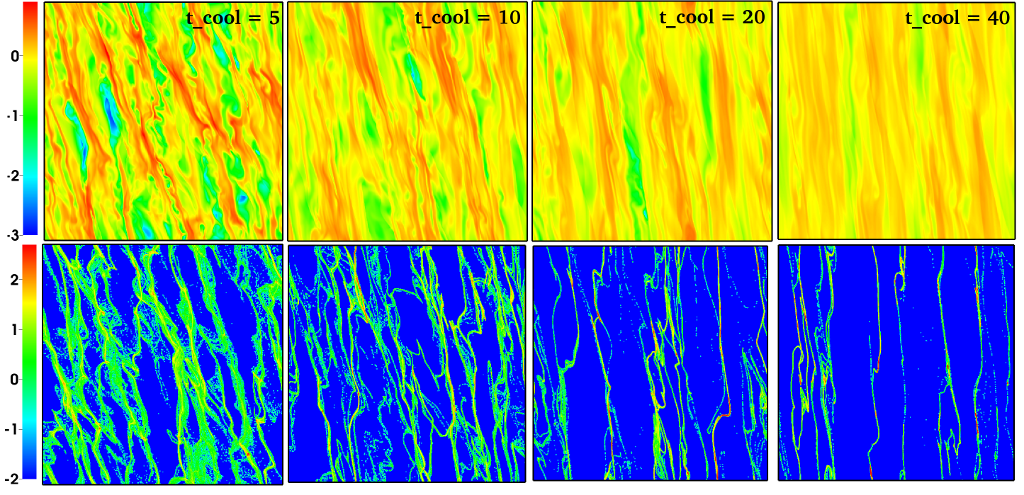}
\caption{\small{The logarithmic values of the gas (top) and $\Omega\tstop = 1$ particle density
		(bottom) at the end of simulations with various cooling time as shown in the top
		panels of each
		column. Color bars are on the left of each row. As $\tcool$ increases, the
		fluctuation of gas density becomes weaker, the tilt angle of shearing waves also
		gets smaller, both indicates relatively weaker angular momentum transport as the
		cooling time scale gets longer. However, stronger clustering effects are found for
		longer cooling cases as the velocity dispersion from the gas gets weaker. 
}}
\label{fig:cluster_tcool}
\end{figure*}

\begin{figure}
	\includegraphics[width=\columnwidth]{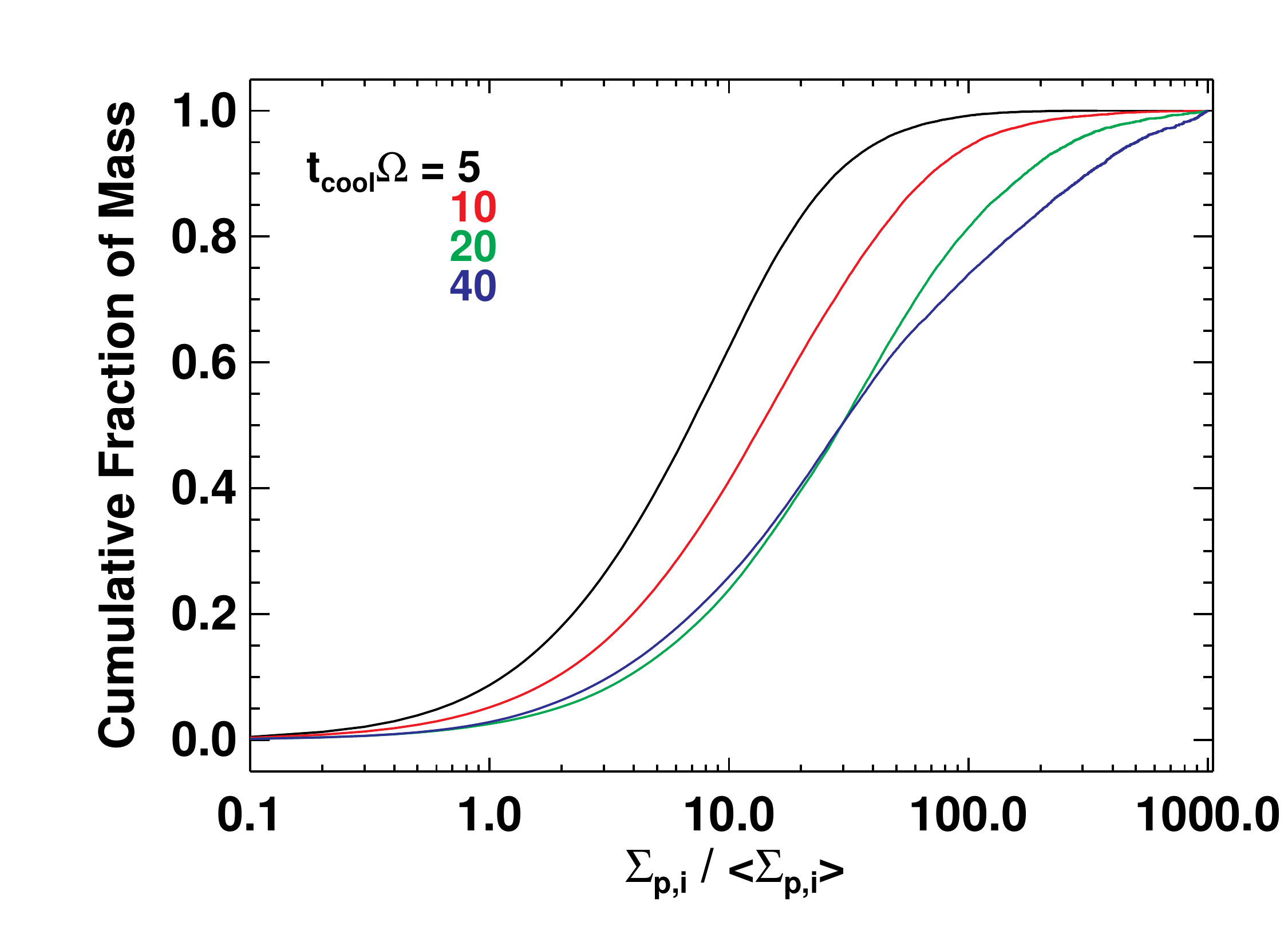}
\caption{\small{Similar to Figure~\ref{fig:mass}, the cumulative mass fraction of the particles with
		$\Omega\tstop = 1$ for various cooling times. Longer $\tcool$ results in stronger
		concentration and clustering. 
}}
\label{fig:cdf_tcool}
\end{figure}

Varying the cooling time also affects the particle concentration. In
Figure~\ref{fig:cluster_tcool}, we show the density snapshots of simulations using different cooling
times at the top row. As $\tcool$ gets longer, we find the turbulent amplitude diminishes, so does the tilt angle
of the shearing wave features.  This leads to less frequent interactions between separated shearing
waves and longer life time for existing density features. 
In the bottom row of Figure~\ref{fig:cluster_tcool}, we find strong interactions of shearing waves
in $\Omega\tcool = 5$ run leads to plume-like structure of $\tau_s = 1$ particle with typical
concentration of $\sim 10$-$100$. The concentration is greater, $\gtrsim 100-10^3$ for $\Omega\tcool =
40$ case, and particles are more spatially confined in very narrow streams. In
Figure~\ref{fig:cdf_tcool}, we quantitatively confirm this result by showing the cumulative mass
fraction of the $\tau_s=1$ particles for various cooling times. As $\tcool$ rises, more and
more mass actually concentrate towards larger dust density, although the density dispersion of gas
decreases.

The particle relative velocity and eccentricity drops as the cooling time gets longer (see Figure~\ref{fig:tcool}). In general, from $\Omega\tcool=10$ to
$40$, $v_{\rm rel}$, $v^{\prime}_{\rm rel}$ and $e$ diminish by less than a factor of $\sim 2$-$3$,
roughly scales as $\propto \alpha^{1/2}$, similar to the way velocity/density dispersion scales. 
Exceptions occur at the intermediate size particles (as found in section~\ref{sec:vel}),
$\tau_s=0.1$ and $1$, where we find the relative velocity of the same kind $v_{\rm rel}$ drops
rapidly with increasing cooling time. It diminishes $\sim 30$ times from $\Omega\tcool=5$ to $40$
for $\tau_s=0.1$ particles and $\lesssim 10^3$ times for particles with $\tau_s=1$. As
the intermediate-sized particles show the strongest clustering, the rather slow relative velocity
would favor gravitational collapsing.

Based on our results in
Table~\ref{tab:tab1}, the density dispersion usually follows
$\delta\Sigma/\Sigma \simeq 2(\Omega\tcool)^{-1/2} \simeq 5 \alpha^{1/2}$ in
gravito-turbulent disks. As already mentioned in section~\ref{sec:selfg}, this is much stronger
fluctuation than found in MRI disks ($\sim \lesssim \alpha^{1/2}$). Similar as in
section~\ref{sec:ecc}, we also measure the dimensionless turbulent strength $\gamma$ based on the
eccentricity growth for different $\tcool$ runs and list them in Table~\ref{tab:tab1}. The strength
parameter $\gamma\simeq 0.07 \alpha^{1/2}$ seems about one order of magnitude greater than those found in MRI-driven
turbulent disks \citep{NG2010,yangetal2012} which also explains why we observe much larger
eccentricity and relative speed for large particles ($\tau_s > 1$) in our gravito-turbulent disks. 

\subsection{Domain size and resolution effects \label{sec:box}}
In this section, we investigate if our results are affected by the size of the computational domain
we adopt. By doubling the size of the shearing sheet and meanwhile keeping the numerical resolution
the same, run tc=10.dble quadruples the numbers of grid cells and
dust particles. 
Following the discussion in section~\ref{sec:ic}, we first start with a pure gas simulation with
$\Omega\tcool=10$ using now an extended
domain. This simulation runs for $\sim 200\Omega^{-1}$ and quasi-steady gravito-turbulent stage is
confirmed after $\lesssim 20\Omega^{-1}$. We then inject the particles at $t=50\Omega^{-1}$ similar
to our standard run, and
rerun for another $200\Omega^{-1}$ with now both gas and particles on the grids. 
The results are summarized in Table~\ref{tab:tab1} (for the gas) and \ref{tab:tab2} (for the dust
diffusion). 

We find the gas properties of tc=10.dble, do not deviate much
from those of the
standard run tc=10. The diffusion coefficients for different type of dust particles are also very
similar, $\sim 20\%$ of difference at most. No significant variations are observed comparing the
particle eccentricity/relative velocities between the standard (black asterisk) and the double-sized (cyan triangles) simulations
in Figure~\ref{fig:ecc_ts} and \ref{fig:vrel} either. We find similar clustering for the intermediate size
particles as well. All these suggest converging results are obtained by adopting domain size of our standard run.

In another run tc=10.hires, we double the number of grid cells in
$x$ and $y$ directions and {quadruple} the total numbers of particles for each species to study the effects of numerical resolution. 
The gas properties and the radial diffusion coefficients in this
higher resolution run are listed in Table~\ref{tab:tab2}. We find doubling the
resolution would slightly increase $\alpha_{\rm tot}$ and $\Dp$ by
$\lesssim 20\%$. The particle eccentricity and relative velocities (see red diamond symbols in
Figure~\ref{fig:ecc_ts} and top two panels of
Figure~\ref{fig:vrel}) are also very similar to those from the standard run. We therefore confirm
good numerical convergence in our results. 

\section{DISCUSSION\label{sec:disc}}

In this section, we try to construct a model of gravito-turbulent disk, and convert our
scale free results in previous sections into a more physical and realistic format. 
An optically thin turbulent flow driven by gravity of its own can be easily found on the outskirts of 
protoplanetary disks. At an orbital distance of $R\sim 100$AU from a solar-mass star, a disk mass
of $\Sigma R^2\sim 0.01 M_{\odot}$ or surface density of $\Sigma\sim 10 \rm{g}\,\rm{cm}^{-2}$, and a
temperature of $T\sim 10\rm{K}$ would lead to a Toomre Q about unity and a cooling time scale much
longer than the local dynamical time. Depending on the dust opacity of the disk, the cooling time
can vary from several times to orders of magnitude longer than the local $\Omega^{-1}$
\citep{ShiChiang2014}. In this disk, $H/R \sim 0.1$ and gas density $\rho_{\rm g}\sim \Sigma/H\sim
10^{-13}\rm{g}\,\rm{cm^{-3}}$.

\subsection{Particle sizes\label{sec:disc_size}}
The mean free path of the gas at radius $R\sim 100$AU is $\lambda \sim 10^4\rm{cm}$, which is typically much
greater than the dust particle size we have simulated. Therefore the particle's stopping time
can be characterized as 
\begin{eqnarray}
	\tau_s  \simeq 0.1 \left(\frac{s}{\rm{cm}}\right)
	\left(\frac{R}{100\,\rm{AU}}\right)^{-3/2}
	\left(\frac{T}{10\,\rm{K}}\right)^{-1/2}\nonumber  \\
\times\left(\frac{\rho_s}{1\,\rm{g\,cm^{-3}}}\right)
\left(\frac{\rho_g}{10^{-13}\,\rm{g\,cm^{-3}}}\right)^{-1} \,,
\end{eqnarray}
using the Epstein's law, where $s$ is the physical size of the dust particle. 
Specifically in our simulations, particles with $\tau_s = 10^{-3}$
would be translated to $0.1$~mm in particle size, and $\tau_s = 10^3$ corresponds to $0.1$~km
planetesimals.

For particles even smaller than $0.1$~mm, they are perfectly coupled with gas, and
will behave very similarly as the sub-mm-sized particles we explored using our simulations. For planetesimals even bigger than
$0.1$~km in size, the stopping time starts to depend on the relative velocity between the dust and
gas.
However, similar to Equation~\ref{eq:ratio}, we can show that in the Stokes regime, the gravitational stirring always
dominates the gas drag in our gravito-turbulent disk. The general empirical formula for
the drag force reads $F_{\rm D} = 0.5 C_{\rm D}({\rm{Re}})\pi s^2\rho_g {\rm v_{gs}^2}$, where $\rm{Re}\sim s
\rm{v_{gs}}/\nu\sim s {\rm v_{gs}}/\lambda \cs$ is the fluid Reynolds number, the relative velocity between gas
and dust ${\rm v_{gs}}\sim v^{\prime}_{\rm rel} \gtrsim \cs$ for large particles as implied by our simulations, and the
dimensionless coefficient $C_{\rm D}(\rm{Re})\sim 0.1$-$10$ for typical Re of
$s > 0.1$~km planetesimals in our gravito-turbulent disk described above
\citep{Cheng2009,PMC2011}. The ratio between the specific drag force and the
gravitational acceleration is therefore
\begin{eqnarray}
\label{eq:ratio2}
\frac{F_{\rm D}/m_s}{|\nabla\Phi|} \sim  7\times 10^{-3} \left(\frac{C_{\rm d}}{10}\right)
\left(\frac{s}{10^4\,\rm{cm}}\right)^{-1}  \nonumber \\
\times
\left(\frac{\rho_s}{1\,\rm{g}\,\rm{cm^{-3}}}\right)^{-1}
\left(\frac{H/R}{0.1}\right)^{-1} \nonumber \\ 
\times
\left(\frac{R}{100\,{\rm{AU}}}\right)^{-1}
\left(\frac{{\rm v_{gs}}}{0.1\,\rm{km}\,\rm{s^{-1}}}\right)^2 \,,
\end{eqnarray}
where $|\nabla\Phi|\sim G\delta\Sigma\sim 5\sqrt{\alpha}\,G\Sigma$ (see section~\ref{sec:tcool}) is
the gravitational force density, and
$m_s$ is the mass of the assumed spherical planetesimal of radius $s$. Given
Equation~\ref{eq:ratio2}, we can see that the dynamics of the km- or larger
planetesimals is mostly determined by the gravity of the gas in gravito-turbulent disks.
Therefore, our results for $\tau_s=10^3$, or $s=0.1$\,km, particles could be easily extrapolated
to even bigger planetesimals.

\subsection{Implications\label{sec:disc_imply}}
\subsubsection{Fast mass transport via turbulent diffusion\label{sec:diffusion}}
\begin{figure}
\includegraphics[width=\columnwidth]{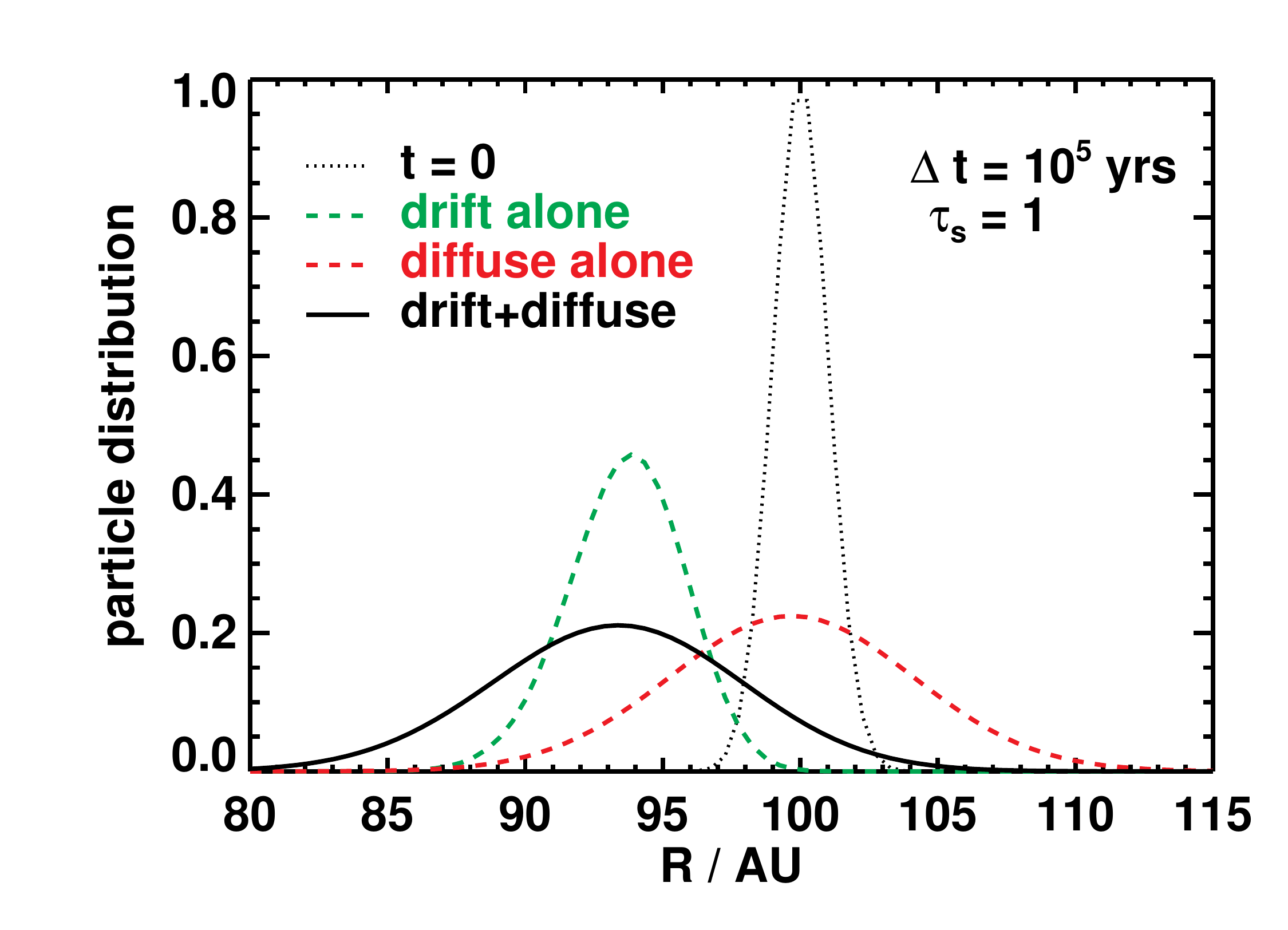}
\caption{\small{The particle ($\tau_s=1$) distribution after $10^5$ years due to the effects of
		radial drift and diffusion (using $\Dp = 0.1\cs^2/\Omega$). An 1D
		advection-diffusion model \citep{AdamsBloch2009} is used to evolve the distribution. 
		About $10\%$ of the initial values can still be found at $R=100$\,AU (black solid) in contrast to
		the radial drift alone case (green dashed) where nearly all particles drift out of
		$R=100$\,AU.
}}
\label{fig:dd}
\end{figure}

Radial diffusion due to turbulent and gravitational stirring can contribute to the mass transport of
the solids. 
The time for particles across a radial distance $\Delta R$ solely due to the diffusion process is 
$t_{\rm diff} \sim \Delta R^2/\Dp$. If we take
$\Omega\tcool = 10$ run as an example,
$t_{\rm diff} \sim 10^5 \left(\Delta R/10\rm{AU}\right)^2$\,yr for cm or smaller particles ($\tau_s < 1$) or 
$\sim 10^4\left(\Delta R/10\rm{AU}\right)^2$\,yr for particles of $10$\,cm or bigger ($\tau_s
\geq 1$). Both suggests the radial diffusion in gravito-turbulent disks might play a role in the
radial transport of the solids. For comparison, the drift time scale due to the radial
pressure gradient of the gas is $t_{\rm drift}\sim \Delta R/v_{\rm drift}\gtrsim 10^4 \left(\Delta
R/10\rm{AU}\right)$\,yr, with radial drift
velocity $v_{\rm drift}\sim \tau_s (1+\tau_s^2)^{-1}\eta v_{\rm K}$ and $\eta\sim \left(\cs/v_{\rm
K}\right)^2$. 
We thus find 
\begin{eqnarray}
\frac{t_{\rm diff}}{t_{\rm drift}} \simeq
\begin{cases} 
10\tau_s \left(\frac{\Delta
R}{10\rm{AU}}\right) \left(\frac{\alpha}{0.02}\right)^{-1} & \rm{if~~} \tau_s < 1 \\
\tau_s^{-1} \left(\frac{\Delta
R}{10\rm{AU}}\right) \left(\frac{\alpha}{0.02}\right)^{-1} & \rm{if~~} \tau_s \geq 1 \,,
\end{cases}
\end{eqnarray}
in which we put back in the cooling time dependence of the diffusion coefficient (see
section~\ref{sec:tcool}) as $\Dp \propto \alpha \propto (\Omega\tcool)^{-1}$ roughly.
The radial transport due to turbulent diffusion and gravitational stirring really becomes comparable to the radial drift
effect especially for the large particles with $\tau_s \geq 1$. Outward
diffusive transport counters, to some extent, inward drift of large particles and {partially alleviates} the radial drift barrier problem \citep{weidenschilling77,boss2015}. {In Figure~\ref{fig:dd}, we show this effect for particles with $\tau_s=1$ --- those that drift fastest --- using a diffusion coefficient $\Dp=0.1 \cs^2\Omega^{-1}$. We
evolve a narrow Gaussian distribution of particles at $100$\,AU for $10^5$ years by solving the Fokker-Planck
equation, including both radial drift and turbulent diffusion \citep{AdamsBloch2009}. The
finite volume algorithm \texttt{FiPy} \citep{FiPy:2009} is used to solve the partial differential
equation.\footnote{Downloadable at \url{http://www.ctcms.nist.gov/fipy/}}
About 10\% of the original set of particles (black solid curve in Figure~\ref{fig:dd}) can still be found at $R=100$\,AU; by contrast, if radial
diffusion is turned off (green dashed curve), practically none remain at the original location.}
On the other hand, as we discuss in
the next section,
inward diffusion might also help particles move out of turbulent regions and avoid 
disruption induced by gravitational torques. It is clear that 
radial diffusion of dust in gravito-turbulent disks is significant.

\subsubsection{Fragmentation and run-away accretion barriers\label{sec:fragment}}
Large relative velocity between particle pairs might lead to catastrophic disruptions.
For micron-to-meter-sized particles, the critical collisional speeds which would results in
fragmentation is $v_{\rm f}\gtrsim 1\,\rm{m}\,\rm{s^{-1}}$
\citep{blumwurm08,SL2009,carb2010}, which is $\gtrsim 0.01 \cs$ in our disk. Based on the
probability distribution function of 
relative velocity for the equal-sized particles (bottom left panel of Figure~\ref{fig:vrel}),
particles of decimeter or smaller ($\tau_s \leq 1 $) are mostly below this fragmentation barrier.

As particle grows even bigger, its self-gravity would strengthen the particle against catastrophic disruption. 
The criteria of fragmentation is such that the specific kinetic energy of a collision ($v_{\rm
rel}^2/2$) exceeds the critical disruption energy 
\beq
Q_{\rm D} \simeq \left[Q_0\left(\frac{s}{1\rm{cm}}\right)^a\!\! +
B\left(\frac{\rho_s}{1 \rm{g}\,\rm{cm^{-3}}}\right)
\left(\frac{s}{1\rm{cm}}\right)^b \right]\,\rm{erg}\,\rm{g^{-1}}\,,
\label{eq:Qd}
\enq
where $Q_0\sim 10^7\rm{-}10^8$ is the material strength, $B\simeq
0.3\rm{-}2.1$ parameterize the self-gravity effect, $a\simeq -0.4$, and $b\simeq
1.3$ for pair-collisions between equal-sized particles \citep{BA1999,IGM2008}. We note that
different material properties, projectile-to-target mass ratio
and impact velocity could all cause differences in this criteria \citep{SL2009}.
Therefore, we should take the above energy criteria as a rough order-of-magnitude estimation. 

Following \citet{IGM2008} and using Equation~\ref{eq:Qd}, we find that a collision results in
destruction if the relative velocity exceeds the fragmentation velocity 
\beq
v_{\rm f} \simeq 0.13 \cs \left(\frac{s}{1\,\rm{km}}\right)^{0.65}
                          \left(\frac{\rho_s}{3\,\rm{g}\,\rm{cm^{-3}}}\right)^{1/2}
		\left(\frac{R}{100\,\rm{AU}}\right)^{1/2}\,.
\label{eq:vf}
\enq
Thus particles with ${\rm{m}} \leq s \leq {\rm{km}}$ ($10\leq\tau_s\leq10^4$) would have 
fragmentation velocity  $\sim (0.01{\rm{-}} 0.1) \cs$, much lower than the most probable relative
velocity of those particles found in our simulations (see again the bottom left panel of
Figure~\ref{fig:vrel}). 
The high relative velocity found in our
gravito-turbulent disks would therefore limit the size that the majority of planetesimal formation
could reach. 

If the main channel of planetesimal formation is through gravitational run-away accretion, it would
require the relative velocity to be even lower than the escape velocity of the collisional outcome
\citep{IGM2008,OO2013a,OO2013b},
\beq
v_{\rm esc} \simeq 4\times 10^{-3} \cs 
\left(\frac{s}{1\,\rm{km}}\right) 
\left(\frac{\rho_s}{3\,\rm{g}\,\rm{cm^{-3}}}\right)^{1/2}
\left(\frac{R}{100\,\rm{AU}}\right) \,.
\label{eq:vesc}
\enq
The planetesimal formation from collisional accretion is strongly disfavored in our
gravito-turbulent disks. 
\begin{figure*}
	\includegraphics[width=16cm]{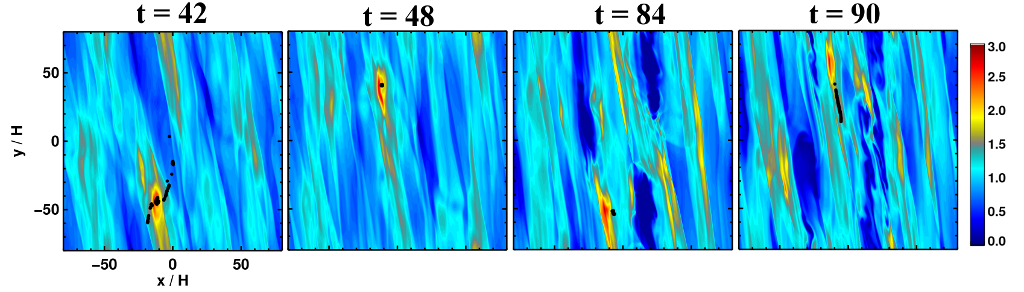}
\caption{\small{Illustration of the time evolution of the dust concentration for $\Omega\tcool=40$
		run, in which a selection of particles ($\sim 20,000$ in number) with $\tau_s =1$
		are marked as black dots, and the color contour at the background shows the density
		fluctuation in linear scale. We first identify a high density ($\Sigma_p\gtrsim
		1000$) particle cluster at
		$t=48\Omega^{-1}$, as show in the second panel with black dots. We then track each
		individual particles backward/forward in time. The compact cluster disappears only after
		$t=84\Omega^{-1}$, which lasts nearly $6$ orbits before being sheared away.  
}}
\label{fig:cluster_time}
\end{figure*}

There are a few conditions that might allay the difficulties: 
(1)~When taking into account of $\tcool$, a longer
cooling time scale would reduce the relative velocity (only weakly depends on $\tcool$ though; see
Figure~\ref{fig:tcool}) and thus lower the possibility of fragmentation. 
(2)~We note the distribution functions show a sizeable spread toward smaller relative velocity,
therefor a small fraction of planetesimals  might still survive the collision or even accrete
although the mean relative velocity is high \citep{windmark2012b}. 
(3)~The strong diffusion may bring planetesimals outside the gravito-turbulent zone, and proceed
further growth in a less violent environment.
(4) If the pre-existing sizes of the planetesimals are big enough,
they might avoid fragmentation and/or run-away barrier since $v_{\rm rel}$ scales with $\tau_s$
differently than $v_{\rm esc}$ and $v_{\rm f}$ do. We find $v_{\rm rel}/\cs \simeq (\tau_s/10)^{0.16}$ after
extrapolating our relative speed results in the bottom left panel of Figure~\ref{fig:tcool} toward
high $\tau_s$ end. We thus find only planetesimals $\gtrsim 100\,{\rm km}$ could pass the fragmentation
barrier; only $\gtrsim 1000\,{\rm km}$-sized planetesimals would result in collisional accretion.
Such large planetesimals might form via gravitational instability as we will discuss next.

\subsubsection{Gravitational collapse\label{sec:collapse}}
Despite the problems caused by the big relative velocity for particles bigger than meter-size, those
cm-to-decimeter-sized ($\tau_s=0.1\rm{-}1$) pebbles, find themselves mostly in very dense coherent
structures and possessing very low relative velocities. As shown in Figure~\ref{fig:8panel} and
\ref{fig:cluster_tcool}, the concentration factor $\Sigma_{\rm p}/\langle\Sigma_{\rm p}\rangle$ is
typically $\gtrsim 10$-$100$ in those filament-like structures. Concentrations can go even higher
than $\gtrsim 100$-$10^3$ times the original particle density for longer cooling time $\tcool =
40\Omega^{-1}$. 
This density enhancement of the dust particles would push the dust-to-gas ratio from $\sim 1/100$ to 
$\Sigma_{\rm  p}/\Sigma_g \gtrsim 10$. More importantly, the resulting local dust density
\beq
\rho_{\rm p} \gtrsim 10^{-12} \left(\frac{\Sigma_{\rm p}/\Sigma_g}{10}\right) {\rm g}\,{\rm cm^{-3}}\,,
\enq
will exceed the Roche density
\beq
\rho_{\rm Roche} \sim \frac{M_*}{R^3}\sim 10^{-12}\left(\frac{M_*}{M_\odot}\right)
\left(\frac{R}{100\,\rm{AU}}\right)^{-3}\,,
\enq
where $M_*$ is the stellar mass, and therefore trigger gravitational collapse.  
This opens up a potential channel of planetesimal formation via gravitational collapse
\citep{boss2000,britschetal2008,gibbonsetal2012,gibbonsetal2014,sc2013}.
In Figure~\ref{fig:cluster_time}, we show a clump of $\tau_s = 1$ particles which possess over-density
$\Sigma_{\rm p}/\langle\Sigma_{\rm p}\rangle \gtrsim 10^3$ could survive for $\simeq 6 \,{\rm
orbits}$, or $6,000 {\rm yr}$ at $R=100\,\rm{AU}$, before getting disrupted in a gravito-turbulent
disk with $\Omega\tcool = 40$. This is significantly longer than the dynamical time that is necessary for
the process of gravitational collapse.
Even if the dust density does not cross the Roche density, the $\rho_{\rm p}\sim \rho_{g}$ in our
	gravito-turbulent disks could still trigger other instabilities such as streaming
	instability \citep{goodmanpindor00,youdingoodman05} for those intermediate size particles. As a result, the dust density increases
further and graviational collapse would still occur eventually.
The resulting planetesimals would then diffuse out the gravito-turbulent region as discussed in
section~\ref{sec:diffusion} and avoid being destroyed (if $s\lesssim 100\,\rm{km}$) due to collision
with planetesimals of the same size.

\subsubsection{Comparison with MRI-driven turbulent disks\label{sec:mri_disk}}
In general, we find relatively stronger radial diffusion and eccentricity growth in our
gravito-turbulent disks than in MRI-driven turbulent disks of similar $\alpha$.

According to \citet{yangetal2009,yangetal2012}, the standard deviation of radial drift can be
described as 
\beq
\sigma(\Delta x) = C_x\,\xi \,H\left( \frac{t\Omega}{2\pi} \right)^{1/2}\,,
\label{eq:dx_yang}
\enq
where $C_x$ is a dimensionless coefficient and $\xi \equiv 4\pi G\rho_0(2\pi/\Omega)^2=4(2\pi)^{3/2} / Q$
assuming $\rho_0= \Sigma_0/2H$. Comparing it with Equation~(\ref{eq:def_dp}), we can convert our
diffusion coefficient to 
\beq
C_x \simeq  0.056\, Q \left(\frac{\Dp}{\cs^2/\Omega}\right)^{1/2}\,.
\label{eq:dp2cx}
\enq
Take our standard tc=10 run ($\alpha\simeq 0.02$) as an example, 
$\Dp\geq 0.05\cs^2/\Omega$ for large particles with $\tau_s > 1$, we 
therefore obtain a dimensionless $C_x \geq 0.038$, more than one order of magnitude larger than $C_x$
measured in MRI disks (cf. Table~1 in \citet{NG2010} and Table~2 in \citet{yangetal2012}). 

The excitation of eccentricity can also be described as \citep{yangetal2009,yangetal2012}
\beq
\sigma(\Delta e) = C_e\,\xi \left(\frac{H}{R}\right)\left(\frac{t\Omega}{2\pi}\right)^{1/2}\,,
\label{eq:de_yang}
\enq
where $C_e$ is a dimensionless coefficient characterizes the growth rate. After comparing it with
Equation~\ref{eq:ecc}, we have
\beq
C_e \simeq 0.040\, Q C \,. 
\label{eq:c2ce}
\enq
Recalling $C \simeq 0.15$ for $\tau_s=10^2$ when $\Omega\tcool=10$ in section~\ref{sec:ecc}, we find
the typical $C_e \simeq 0.018$, much greater than $10^{-3}$--$10^{-4}$ found in \citet{yangetal2012}
using local shearing boxes, and one order of magnitude greater than reported in \citet{NG2010} with
global simulations.

The increasing eccentricity and diffusion is a result of stronger gravitational forcing in
gravito-turbulent disk than in MRI disk. 
Quantitatively, the parameter $\gamma$ (closely related to $\delta\Sigma/\Sigma$) reflects the
strength of this stirring. As discussed in
section~\ref{sec:tcool} and also shown in Table~\ref{tab:tab1}, this dimensionless parameter
$\gamma$ is at least one order of magnitude larger than found in MRI disks \citep[e.g.,
][]{yangetal2012}. 
As
discussed in section~\ref{sec:selfg}, the stronger forcing also pushes more types of particles
($\tau_s >1$) into the gravity dominated regime; while
in MRI disk, as shown in Figure 15 and 23 of \citet{NG2010}, this occurs only for particles with
$\tau_s > 100$.

\section{SUMMARY AND CONCLUSION\label{sec:conclusion}}
We have studied the dynamics of dust in gravito-turbulent disks, i.e., gaseous 
disks whose turbulence is driven by self-gravity, using 2D 
hybrid (particle and gas) simulations in a local shearing sheet approximation. For dust particles, we
included the aerodynamic drag and gravitational pull from self-gravitating gas, and neglected
particle self-gravity and feedback. We obtained the density distribution, radial diffusion
coefficient and relative velocities for dust particles with stopping times distributed
from $10^{-3}\Omega^{-1}$ to
$10^{3}\Omega^{-1}$.
We summarize our main results as follows:
\begin{enumerate}
\item{Particles with small stopping times ($\tau_s < 0.1$) are aerodynamically well-coupled
to gas and therefore trace the gas distribution. Diffusion coefficients for
small particles are
close to those for gas. The gas diffusion
coefficient $\Dg$ is related to the angular momentum transport parameter $\alpha$
via a
roughly constant Schmidt number ${\rm Sc} = \alpha/\Dg = 1.4{\rm -}2.4$. Small particles also have low
eccentricities ($e \sim 0.1 (H/R) (\alpha/0.01)^{1/2}$) 
and low relative velocities ($\lesssim 0.01{\rm -} 0.1\,(\alpha/0.01)^{1/2}\cs$). }

\item{Particles with larger stopping times ($\tau_s \gtrsim 1$) are more strongly gravitationally
forced from self-gravitating gas than by aerodynamic drag. The stronger forcing results in diffusion
coefficients that are $\gtrsim 5$ times greater than those of smaller particles.
Turbulent diffusion in gravito-turbulent disks therefore plays an important role in radial transport of large bodies.
Strong stochastic gravitational stirring generates large eccentricities ($e \gtrsim 0.5{\rm -}1 (H/R) (\alpha/0.01)^{1/2}$)
and large relative velocity ($\gtrsim (\alpha/0.01)^{1/2} \cs$). These disfavor planetesimal
formation by pairwise collision as the typical collisional speed exceeds both the escape
and fragmentation velocities.}

\item{Particles of intermediate size ($\tau_s= 0.1 {\rm -}1$) are marginally coupled to gas, and
are collected by both gas drag and gravitational torques (see equation~\ref{eq:ratio}) 
into filament-like structures having large
overdensities ($\gtrsim 10{\rm-} 10^3$ times the background) and low relative velocities
($\lesssim 0.01 \cs$) between like-sized particles. The density concentration is high enough to
trigger direct gravitational collapse. Nascent planetesimals can avoid collisional disruption by 
diffusing to less turbulent regions of the disk.}

\item{Longer cooling times result in weaker turbulence ($\alpha\propto (\Omega\tcool)^{-1}$),
		weaker particle diffusion ($\propto\alpha$), lower relative velocities and
		eccentricities
		($\propto \sqrt{\alpha}$), and stronger clustering for
	intermediate-size ($\tau_s=0.1{\rm -}1$) particles. }

\item{Compared to MRI-turbulent disks, gravito-turbulent disks show almost one order-of-magnitude
		stronger density fluctuations for similar $\alpha$ ($\delta\Sigma/\Sigma \simeq 5
		\sqrt{\alpha}$ versus
		$\sqrt{0.5\alpha}$). Stronger gravitational stirring leads to
	higher eccentricities and relative velocities between particles.}

\end{enumerate}

{In a recent paper that parallels ours, \citet{BoothClarke2016} study the relative velocities of dust particles
in global smoothed-particle hydrodynamics (SPH) simulations of 
gravito-turbulent protoplanetary disks. Like us, they find large relative velocities for particles with $\tau_s \gtrsim 3$ (compare our Figure~\ref{fig:vrel} with their
Figures~6 and 7).}

Future investigations could
improve upon our work in any number 
of ways. 
As we use massless test particles in 2D hydrodynamic local
simulations with a simplistic cooling prescription, the effects of 
vertical sedimentation, particle feedback, particle self-gravity {\citep{gibbonsetal2014}}, and self-consistent heating and
cooling {on dust dynamics could be further explored}. Local and eventually global 3D simulations which account for these
physical effects would be clear next steps. 

\section*{Acknowledgements}
We thank Xuening Bai for fixing a bug related to the orbital advection scheme for particles. 
We also thank the anonymous referee for stimulating comments which led to improvement of this
paper.
This work was supported in part by the
National Science Foundation under grant PHY-1144374, "A Max-Planck/Princeton Research Center for
Plasma Physics" and grant PHY-0821899, "Center for Magnetic Self-Organization". ZZ acknowledges support by NASA through Hubble 
Fellowship grant HST-HF-51333.01-A awarded by the Space Telescope
Science Institute, which is operated by the Association of
Universities for Research in Astronomy, Inc., for NASA, under 
contract NAS 5-26555.
Financial support for EC was provided
by the NSF and NASA Origins.
Resources supporting this work were provided by the Princeton Institute of Computational Science and Engineering (PICSciE) and Stampede at Texas Advanced Computing Center (TACC), the University of Texas at
Austin through XSEDE grant TG-AST130002.

\clearpage

{\renewcommand{\arraystretch}{1.4}
\begin{deluxetable}{ccccccccc}
\tabletypesize{\footnotesize}
\tablecolumns{9} \tablewidth{0pc}
\tablecaption{Gas Properties in Gravito-Turbulent Disks \label{tab:tab1}}
\setlength{\tabcolsep}{0.1in}
\tablehead{\colhead{Name} & 
\colhead{$\Omega\tcool$}   &
\colhead{$Q$} &
\colhead{$\delta\Sigma_g/\Sigma_g$}\tablenotemark{(a)} &
\colhead{$\gamma$}\tablenotemark{(b)} &
\colhead{$\delta\ux/\cs$}\tablenotemark{(c)} &
\colhead{$\atot$}\tablenotemark{(d)} &
\colhead{$D_{g,x}\,(\cs^2\Omega^{-1})$\tablenotemark{(e)}  } & 
\colhead{${\rm Sc}$\tablenotemark{(f)} }}
\startdata
tc=5       
&  $5$   & $3.35$ & $0.82$ & $0.013$ & $0.38\, (0.91)$    & $0.035$ & $0.022$ & $1.6\, (1.7)$ \\
tc=10  
&  $10$  & $3.19$ & $0.63$ & $9.4\times 10^{-3}$ & $0.31\, (0.61)$ & $0.020$ & $0.014$ & $1.4\,
(1.7)$\\
tc=10.dble \tablenotemark{(g)}         
&  $10$  & $3.15$ & $0.61$ & $9.0\times 10^{-3}$ & $0.34\, (0.64)$ & $0.022$ & $0.016$ & $1.4\,
(1.5)$\\
tc=10.hires \tablenotemark{(h)}
&  $10$  & $3.09$ & $0.59$ & $9.5\times 10^{-3}$ & $0.33\, (0.59)$ & $0.024$ & $0.015$ & $1.6\, (1.7)$\\
tc=20
&  $20$  & $3.16$ & $0.52$ & $8.3\times 10^{-3}$ & $0.26\, (0.38)$ & $0.011$ & $0.0061$ & $1.8\,
(2.0)$\\
tc=40         
& $40$   & $2.99$ & $0.32$ & $5.2\times 10^{-3}$ & $0.20\, (0.23)$ & $0.006$ & $0.0025$ & $2.4\,
(2.9)$\\
\enddata
\tablenotetext{(a)}{Time- and spatial averaged density dispersion, i.e., $\langle\!\langle\delta\Sigma_g^2\rangle\!\rangle_{t}^{1/2}/\Sigma_g$.}
\tablenotetext{(b)}{Dimensionless parameter used in \citet{IGM2008} which characterize the amplitude
of the random gravity field.} 
\tablenotetext{(c)}{Velocity dispersion with density weight, i.e., $\langle\!\langle
\ux^2\drangle\rangle_{t}^{1/2}/\cs$, where $\cs$ is the density weighted sound speed with
$\Gamma=2$. The dispersions calculated without density weight are in the parentheses.}
\tablenotetext{(d)}{$\atot =\ar + \ag$, the total internal stress normalized with averaged
pressure.}
\tablenotetext{(e)}{Gas diffusion coefficient calculated using the autocorrelation function of Equation~\ref{eq:def_dg}.}
\tablenotetext{(f)}{Schmidt number ${\rm Sc}=\alpha \cs^2\Omega^{-1}/\Dg$. The values in parentheses are calculated
assuming $\Dg\simeq \Dp$ for $\tau_s=10^{-3}$ particles (see Table~\ref{tab:tab2}). }
\tablenotetext{(g)}{Similar to run tc=10 but using doubled domain sized and fixed grid resolution
and particle number.}
\tablenotetext{(h)}{Similar to run tc=10 but using doubled grid resolution and particle
number.} 
\end{deluxetable}
}

{\renewcommand{\arraystretch}{1.4}
\begin{deluxetable}{cccccccc}
\tabletypesize{\footnotesize}
\tablecolumns{8} \tablewidth{0pc}
\tablecaption{ $\Dp /(\cs^2\Omega^{-1})$: Dust Radial Diffusion coefficient\label{tab:tab2}}
\setlength{\tabcolsep}{0.1in}
\tablehead{\colhead{$$} & \colhead{$\tau_s = 10^{-3}$} & \colhead{$10^{-2}$} & \colhead{$0.1$} &
\colhead{$1$} & \colhead{$10$} & \colhead{$10^2$} & \colhead{$10^3$} }

\startdata
tc=5       
&  $0.021$   & $0.021$ & $0.028$ & $0.14$ & $0.15$    & $0.089$ & $0.092$ \\
tc=10  
&  $0.012$  & $0.011$ & $0.011$ & $0.049$ & $0.091$ & $0.058$ & $0.050$ \\
tc=10.wosg \tablenotemark{(a)}
&  $0.011$  & $0.011$ & $0.011$ & $0.016$ & $0.003$ & $7\times 10^{-5}$ & $9\times 10^{-7}$\\
tc=10.dble        
&  $0.015$  & $0.015$ & $0.015$ & $0.050$ & $0.085$ & $0.056$ & $0.051$\\
tc=10.hires 
&  $0.014$  & $0.014$ & $0.016$ & $0.058$ & $0.096$ & $0.063$ & $0.057$\\
tc=20
&  $0.0055$  & $0.0052$ & $0.0049$ & $0.010$ & $0.042$ & $0.034$ & $0.0033$\\
tc=40         
& $0.0021$   & $0.0020$ & $0.0021$ & $0.0067$ & $0.019$ & $0.017$ & $0.018$\\
\enddata
\tablenotetext{(a)}{Similar to run tc=10 but the gravitational forcing from the gas is artificially
	removed.}
\end{deluxetable}
}

\clearpage

\bibliographystyle{mnras}

\begin{thebibliography}{}
\makeatletter
\relax
\def\mn@urlcharsother{\let\do\@makeother \do\$\do\&\do\#\do\^\do\_\do\%\do\~}
\def\mn@doi{\begingroup\mn@urlcharsother \@ifnextchar [ {\mn@doi@}
  {\mn@doi@[]}}
\def\mn@doi@[#1]#2{\def\@tempa{#1}\ifx\@tempa\@empty \href
  {http://dx.doi.org/#2} {doi:#2}\else \href {http://dx.doi.org/#2} {#1}\fi
  \endgroup}
\def\mn@eprint#1#2{\mn@eprint@#1:#2::\@nil}
\def\mn@eprint@arXiv#1{\href {http://arxiv.org/abs/#1} {{\tt arXiv:#1}}}
\def\mn@eprint@dblp#1{\href {http://dblp.uni-trier.de/rec/bibtex/#1.xml}
  {dblp:#1}}
\def\mn@eprint@#1:#2:#3:#4\@nil{\def\@tempa {#1}\def\@tempb {#2}\def\@tempc
  {#3}\ifx \@tempc \@empty \let \@tempc \@tempb \let \@tempb \@tempa \fi \ifx
  \@tempb \@empty \def\@tempb {arXiv}\fi \@ifundefined
  {mn@eprint@\@tempb}{\@tempb:\@tempc}{\expandafter \expandafter \csname
  mn@eprint@\@tempb\endcsname \expandafter{\@tempc}}}

\bibitem[\protect\citeauthoryear{{Adams} \& {Bloch}}{{Adams} \&
  {Bloch}}{2009}]{AdamsBloch2009}
{Adams} F.~C.,  {Bloch} A.~M.,  2009, \mn@doi [\apj]
  {10.1088/0004-637X/701/2/1381}, \href
  {http://adsabs.harvard.edu/abs/2009ApJ...701.1381A} {701, 1381}

\bibitem[\protect\citeauthoryear{{Alexander}, {Grossman}, {Ebel}  \&
  {Ciesla}}{{Alexander} et~al.}{2008}]{alexanderetal2008sci}
{Alexander} C.~M.~O.~.,  {Grossman} J.~N.,  {Ebel} D.~S.,   {Ciesla} F.~J.,
  2008, \mn@doi [Science] {10.1126/science.1156561}, \href
  {http://adsabs.harvard.edu/abs/2008Sci...320.1617A} {320, 1617}

\bibitem[\protect\citeauthoryear{{Andrews}}{{Andrews}}{2015}]{andrews2015PASP}
{Andrews} S.~M.,  2015, \mn@doi [\pasp] {10.1086/683178}, \href
  {http://adsabs.harvard.edu/abs/2015PASP..127..961A} {127, 961}

\bibitem[\protect\citeauthoryear{{Andrews} \& {Williams}}{{Andrews} \&
  {Williams}}{2007}]{andrews_williams2007}
{Andrews} S.~M.,  {Williams} J.~P.,  2007, \mn@doi [\apj] {10.1086/522885},
  \href {http://adsabs.harvard.edu/abs/2007ApJ...671.1800A} {671, 1800}

\bibitem[\protect\citeauthoryear{{Armitage}}{{Armitage}}{2011}]{armitage2011araa}
{Armitage} P.~J.,  2011, \mn@doi [\araa] {10.1146/annurev-astro-081710-102521},
  \href {http://adsabs.harvard.edu/abs/2011ARA%26A..49..195A} {49, 195}


\bibitem[\protect\citeauthoryear{{Bai} \& {Stone}}{{Bai} \&
  {Stone}}{2010}]{baistone10a}
{Bai} X.-N.,  {Stone} J.~M.,  2010, \mn@doi [\apjs]
  {10.1088/0067-0049/190/2/297}, \href
  {http://adsabs.harvard.edu/abs/2010ApJS..190..297B} {190, 297}

\bibitem[\protect\citeauthoryear{{Balbus} \& {Hawley}}{{Balbus} \&
  {Hawley}}{1991}]{BH91}
{Balbus} S.~A.,  {Hawley} J.~F.,  1991, \mn@doi [\apj] {10.1086/170270}, \href
  {http://adsabs.harvard.edu/abs/1991ApJ...376..214B} {376, 214}

\bibitem[\protect\citeauthoryear{{Balbus} \& {Hawley}}{{Balbus} \&
  {Hawley}}{1998}]{BH98}
{Balbus} S.~A.,  {Hawley} J.~F.,  1998, \mn@doi [Reviews of Modern Physics]
  {10.1103/RevModPhys.70.1}, \href
  {http://adsabs.harvard.edu/abs/1998RvMP...70....1B} {70, 1}

\bibitem[\protect\citeauthoryear{{Barranco} \& {Marcus}}{{Barranco} \&
  {Marcus}}{2005}]{barrancomarcus05}
{Barranco} J.~A.,  {Marcus} P.~S.,  2005, \mn@doi [\apj] {10.1086/428639},
  \href {http://adsabs.harvard.edu/abs/2005ApJ...623.1157B} {623, 1157}

\bibitem[\protect\citeauthoryear{{Benz} \& {Asphaug}}{{Benz} \&
  {Asphaug}}{1999}]{BA1999}
{Benz} W.,  {Asphaug} E.,  1999, \mn@doi [\icarus] {10.1006/icar.1999.6204},
  \href {http://adsabs.harvard.edu/abs/1999Icar..142....5B} {142, 5}

\bibitem[\protect\citeauthoryear{{Blum} \& {Wurm}}{{Blum} \&
  {Wurm}}{2008}]{blumwurm08}
{Blum} J.,  {Wurm} G.,  2008, \mn@doi [\araa]
  {10.1146/annurev.astro.46.060407.145152}, \href
  {http://adsabs.harvard.edu/abs/2008ARA%26A..46...21B} {46, 21}


\bibitem[\protect\citeauthoryear{{Booth} \& {Clarke}}{{Booth} \&
  {Clarke}}{2016}]{BoothClarke2016}
{Booth} R.~A.,  {Clarke} C.~J.,  2016, preprint, \href
  {http://adsabs.harvard.edu/abs/2016arXiv160300029B} {} (\mn@eprint {arXiv}
  {1603.00029})

\bibitem[\protect\citeauthoryear{{Boss}}{{Boss}}{2000}]{boss2000}
{Boss} A.~P.,  2000, \mn@doi [\apjl] {10.1086/312737}, \href
  {http://adsabs.harvard.edu/abs/2000ApJ...536L.101B} {536, L101}

\bibitem[\protect\citeauthoryear{{Boss}}{{Boss}}{2015}]{boss2015}
{Boss} A.~P.,  2015, \mn@doi [\apj] {10.1088/0004-637X/807/1/10}, \href
  {http://adsabs.harvard.edu/abs/2015ApJ...807...10B} {807, 10}

\bibitem[\protect\citeauthoryear{{Britsch}, {Clarke}  \& {Lodato}}{{Britsch}
  et~al.}{2008}]{britschetal2008}
{Britsch} M.,  {Clarke} C.~J.,   {Lodato} G.,  2008, \mn@doi [\mnras]
  {10.1111/j.1365-2966.2008.12910.x}, \href
  {http://adsabs.harvard.edu/abs/2008MNRAS.385.1067B} {385, 1067}

\bibitem[\protect\citeauthoryear{{Carballido}, {Cuzzi}  \&
  {Hogan}}{{Carballido} et~al.}{2010}]{carb2010}
{Carballido} A.,  {Cuzzi} J.~N.,   {Hogan} R.~C.,  2010, \mn@doi [\mnras]
  {10.1111/j.1365-2966.2010.16653.x}, \href
  {http://adsabs.harvard.edu/abs/2010MNRAS.405.2339C} {405, 2339}

\bibitem[\protect\citeauthoryear{{Carballido}, {Bai}  \& {Cuzzi}}{{Carballido}
  et~al.}{2011}]{carb2011}
{Carballido} A.,  {Bai} X.-N.,   {Cuzzi} J.~N.,  2011, \mn@doi [\mnras]
  {10.1111/j.1365-2966.2011.18661.x}, \href
  {http://adsabs.harvard.edu/abs/2011MNRAS.415...93C} {415, 93}

\bibitem[\protect\citeauthoryear{Cheng}{Cheng}{2009}]{Cheng2009}
Cheng N.-S.,  2009, \mn@doi [Powder Technology]
  {http://dx.doi.org/10.1016/j.powtec.2008.07.006}, 189, 395

\bibitem[\protect\citeauthoryear{{Chiang} \& {Youdin}}{{Chiang} \&
  {Youdin}}{2010}]{chiangyoudin10}
{Chiang} E.,  {Youdin} A.,  2010, Annual Reviews of Earth and Planetary
  Science, 38

\bibitem[\protect\citeauthoryear{{Cossins}, {Lodato}  \& {Clarke}}{{Cossins}
  et~al.}{2009}]{Cossins2009}
{Cossins} P.,  {Lodato} G.,   {Clarke} C.~J.,  2009, \mn@doi [\mnras]
  {10.1111/j.1365-2966.2008.14275.x}, \href
  {http://adsabs.harvard.edu/abs/2009MNRAS.393.1157C} {393, 1157}

\bibitem[\protect\citeauthoryear{{Cuzzi} \& {Alexander}}{{Cuzzi} \&
  {Alexander}}{2006}]{cuzzi2006nature}
{Cuzzi} J.~N.,  {Alexander} C.~M.~O.,  2006, \mn@doi [\nat]
  {10.1038/nature04834}, \href
  {http://adsabs.harvard.edu/abs/2006Natur.441..483C} {441, 483}

\bibitem[\protect\citeauthoryear{{Cuzzi}, {Dobrovolskis}  \&
  {Champney}}{{Cuzzi} et~al.}{1993}]{cuzzietal1993}
{Cuzzi} J.~N.,  {Dobrovolskis} A.~R.,   {Champney} J.~M.,  1993, \mn@doi
  [\icarus] {10.1006/icar.1993.1161}, \href
  {http://adsabs.harvard.edu/abs/1993Icar..106..102C} {106, 102}

\bibitem[\protect\citeauthoryear{{Eisner}, {Hillenbrand}, {Carpenter}  \&
  {Wolf}}{{Eisner} et~al.}{2005}]{eisneretal2005}
{Eisner} J.~A.,  {Hillenbrand} L.~A.,  {Carpenter} J.~M.,   {Wolf} S.,  2005,
  \mn@doi [\apj] {10.1086/497161}, \href
  {http://adsabs.harvard.edu/abs/2005ApJ...635..396E} {635, 396}

\bibitem[\protect\citeauthoryear{{Flaherty}, {Hughes}, {Rosenfeld}, {Andrews},
  {Chiang}, {Simon}, {Kerzner}  \& {Wilner}}{{Flaherty}
  et~al.}{2015}]{flahertyetal2015}
{Flaherty} K.~M.,  {Hughes} A.~M.,  {Rosenfeld} K.~A.,  {Andrews} S.~M.,
  {Chiang} E.,  {Simon} J.~B.,  {Kerzner} S.,   {Wilner} D.~J.,  2015, \mn@doi
  [\apj] {10.1088/0004-637X/813/2/99}, \href
  {http://adsabs.harvard.edu/abs/2015ApJ...813...99F} {813, 99}

\bibitem[\protect\citeauthoryear{{Forgan}, {Armitage}  \& {Simon}}{{Forgan}
  et~al.}{2012}]{Forgan2012}
{Forgan} D.,  {Armitage} P.~J.,   {Simon} J.~B.,  2012, \mn@doi [\mnras]
  {10.1111/j.1365-2966.2012.21962.x}, \href
  {http://adsabs.harvard.edu/abs/2012MNRAS.426.2419F} {426, 2419}

\bibitem[\protect\citeauthoryear{{Gammie}}{{Gammie}}{2001}]{gammie01}
{Gammie} C.~F.,  2001, \mn@doi [\apj] {10.1086/320631}, \href
  {http://adsabs.harvard.edu/abs/2001ApJ...553..174G} {553, 174}

\bibitem[\protect\citeauthoryear{{Garaud}, {Meru}, {Galvagni}  \&
  {Olczak}}{{Garaud} et~al.}{2013}]{garaudetal2013}
{Garaud} P.,  {Meru} F.,  {Galvagni} M.,   {Olczak} C.,  2013, \mn@doi [\apj]
  {10.1088/0004-637X/764/2/146}, \href
  {http://adsabs.harvard.edu/abs/2013ApJ...764..146G} {764, 146}

\bibitem[\protect\citeauthoryear{{Gibbons}, {Rice}  \&
  {Mamatsashvili}}{{Gibbons} et~al.}{2012}]{gibbonsetal2012}
{Gibbons} P.~G.,  {Rice} W.~K.~M.,   {Mamatsashvili} G.~R.,  2012, \mn@doi
  [\mnras] {10.1111/j.1365-2966.2012.21731.x}, \href
  {http://adsabs.harvard.edu/abs/2012MNRAS.426.1444G} {426, 1444}

\bibitem[\protect\citeauthoryear{{Gibbons}, {Mamatsashvili}  \&
  {Rice}}{{Gibbons} et~al.}{2014}]{gibbonsetal2014}
{Gibbons} P.~G.,  {Mamatsashvili} G.~R.,   {Rice} W.~K.~M.,  2014, \mn@doi
  [\mnras] {10.1093/mnras/stu809}, \href
  {http://adsabs.harvard.edu/abs/2014MNRAS.442..361G} {442, 361}

\bibitem[\protect\citeauthoryear{{Gibbons}, {Mamatsashvili}  \&
  {Rice}}{{Gibbons} et~al.}{2015}]{gibbonsetal2015}
{Gibbons} P.~G.,  {Mamatsashvili} G.~R.,   {Rice} W.~K.~M.,  2015, \mn@doi
  [\mnras] {10.1093/mnras/stv1766}, \href
  {http://adsabs.harvard.edu/abs/2015MNRAS.453.4232G} {453, 4232}

\bibitem[\protect\citeauthoryear{{Goldreich} \& {Lynden-Bell}}{{Goldreich} \&
  {Lynden-Bell}}{1965}]{goldreichbell65}
{Goldreich} P.,  {Lynden-Bell} D.,  1965, \mnras, 130, 125

\bibitem[\protect\citeauthoryear{{Goodman} \& {Pindor}}{{Goodman} \&
  {Pindor}}{2000}]{goodmanpindor00}
{Goodman} J.,  {Pindor} B.,  2000, \mn@doi [Icarus] {10.1006/icar.2000.6467},
  \href {http://adsabs.harvard.edu/abs/2000Icar..148..537G} {148, 537}

\bibitem[\protect\citeauthoryear{{Gressel}, {Nelson}  \& {Turner}}{{Gressel}
  et~al.}{2012}]{gnt2012mnras}
{Gressel} O.,  {Nelson} R.~P.,   {Turner} N.~J.,  2012, \mn@doi [\mnras]
  {10.1111/j.1365-2966.2012.20701.x}, \href
  {http://adsabs.harvard.edu/abs/2012MNRAS.422.1140G} {422, 1140}

\bibitem[\protect\citeauthoryear{Guyer, Wheeler  \& Warren}{Guyer
  et~al.}{2009}]{FiPy:2009}
Guyer J.~E.,  Wheeler D.,   Warren J.~A.,  2009, \mn@doi [Computing in Science
  & Engineering] {10.1109/MCSE.2009.52}, 11, 6

\bibitem[\protect\citeauthoryear{{Ida}, {Guillot}  \& {Morbidelli}}{{Ida}
  et~al.}{2008}]{IGM2008}
{Ida} S.,  {Guillot} T.,   {Morbidelli} A.,  2008, \mn@doi [\apj]
  {10.1086/591903}, \href {http://adsabs.harvard.edu/abs/2008ApJ...686.1292I}
  {686, 1292}

\bibitem[\protect\citeauthoryear{{Jacquet} \& {Thompson}}{{Jacquet} \&
  {Thompson}}{2014}]{JT2014}
{Jacquet} E.,  {Thompson} C.,  2014, \mn@doi [\apj]
  {10.1088/0004-637X/797/1/30}, \href
  {http://adsabs.harvard.edu/abs/2014ApJ...797...30J} {797, 30}

\bibitem[\protect\citeauthoryear{{Johansen}, {Youdin}  \& {Klahr}}{{Johansen}
  et~al.}{2009}]{Johansen2009}
{Johansen} A.,  {Youdin} A.,   {Klahr} H.,  2009, \mn@doi [\apj]
  {10.1088/0004-637X/697/2/1269}, \href
  {http://adsabs.harvard.edu/abs/2009ApJ...697.1269J} {697, 1269}

\bibitem[\protect\citeauthoryear{{Johansen}, {Klahr}  \& {Henning}}{{Johansen}
  et~al.}{2011}]{johansenetal2011}
{Johansen} A.,  {Klahr} H.,   {Henning} T.,  2011, \mn@doi [\aap]
  {10.1051/0004-6361/201015979}, \href
  {http://adsabs.harvard.edu/abs/2011A%26A...529A..62J} {529, A62}


\bibitem[\protect\citeauthoryear{{Johansen}, {Blum}, {Tanaka}, {Ormel},
  {Bizzarro}  \& {Rickman}}{{Johansen} et~al.}{2014}]{johansenetal2014prpl}
{Johansen} A.,  {Blum} J.,  {Tanaka} H.,  {Ormel} C.,  {Bizzarro} M.,
  {Rickman} H.,  2014, \mn@doi [Protostars and Planets VI]
  {10.2458/azu_uapress_9780816531240-ch024}, \href
  {http://adsabs.harvard.edu/abs/2014prpl.conf..547J} {pp 547--570}

\bibitem[\protect\citeauthoryear{{Johnson} \& {Gammie}}{{Johnson} \&
  {Gammie}}{2003}]{JG2003}
{Johnson} B.~M.,  {Gammie} C.~F.,  2003, \mn@doi [\apj] {10.1086/378392}, \href
  {http://adsabs.harvard.edu/abs/2003ApJ...597..131J} {597, 131}

\bibitem[\protect\citeauthoryear{{Johnson}, {Goodman}  \& {Menou}}{{Johnson}
  et~al.}{2006}]{johnsonetal2006}
{Johnson} E.~T.,  {Goodman} J.,   {Menou} K.,  2006, \mn@doi [\apj]
  {10.1086/505462}, \href {http://adsabs.harvard.edu/abs/2006ApJ...647.1413J}
  {647, 1413}

\bibitem[\protect\citeauthoryear{{Johnson}, {Guan}  \& {Gammie}}{{Johnson}
  et~al.}{2008}]{Johnson2008}
{Johnson} B.~M.,  {Guan} X.,   {Gammie} C.~F.,  2008, \mn@doi [\apjs]
  {10.1086/586707}, \href {http://adsabs.harvard.edu/abs/2008ApJS..177..373J}
  {177, 373}

\bibitem[\protect\citeauthoryear{{Klahr} \& {Bodenheimer}}{{Klahr} \&
  {Bodenheimer}}{2003}]{Klahr2003}
{Klahr} H.~H.,  {Bodenheimer} P.,  2003, \mn@doi [\apj] {10.1086/344743}, \href
  {http://adsabs.harvard.edu/abs/2003ApJ...582..869K} {582, 869}

\bibitem[\protect\citeauthoryear{{Laughlin}, {Steinacker}  \&
  {Adams}}{{Laughlin} et~al.}{2004}]{laughlinetal2004}
{Laughlin} G.,  {Steinacker} A.,   {Adams} F.~C.,  2004, \mn@doi [\apj]
  {10.1086/386316}, \href {http://adsabs.harvard.edu/abs/2004ApJ...608..489L}
  {608, 489}

\bibitem[\protect\citeauthoryear{{Lodato} \& {Rice}}{{Lodato} \&
  {Rice}}{2004}]{LR2004}
{Lodato} G.,  {Rice} W.~K.~M.,  2004, \mn@doi [\mnras]
  {10.1111/j.1365-2966.2004.07811.x}, \href
  {http://adsabs.harvard.edu/abs/2004MNRAS.351..630L} {351, 630}

\bibitem[\protect\citeauthoryear{{Lodato} \& {Rice}}{{Lodato} \&
  {Rice}}{2005}]{LR2005}
{Lodato} G.,  {Rice} W.~K.~M.,  2005, \mn@doi [\mnras]
  {10.1111/j.1365-2966.2005.08875.x}, \href
  {http://adsabs.harvard.edu/abs/2005MNRAS.358.1489L} {358, 1489}

\bibitem[\protect\citeauthoryear{{Mamatsashvili} \& {Rice}}{{Mamatsashvili} \&
  {Rice}}{2009}]{mamarice2009}
{Mamatsashvili} G.~R.,  {Rice} W.~K.~M.,  2009, \mn@doi [\mnras]
  {10.1111/j.1365-2966.2009.14481.x}, \href
  {http://adsabs.harvard.edu/abs/2009MNRAS.394.2153M} {394, 2153}

\bibitem[\protect\citeauthoryear{{Masset}}{{Masset}}{2000}]{Masset2000}
{Masset} F.,  2000, \mn@doi [\aaps] {10.1051/aas:2000116}, \href
  {http://adsabs.harvard.edu/abs/2000A%26AS..141..165M} {141, 165}


\bibitem[\protect\citeauthoryear{{Maxey}}{{Maxey}}{1987}]{maxey87}
{Maxey} M.~R.,  1987, J. Fluid Mech., 174, 441

\bibitem[\protect\citeauthoryear{{Mej{\'{\i}}a}, {Durisen}, {Pickett}  \&
  {Cai}}{{Mej{\'{\i}}a} et~al.}{2005}]{Mejia2005}
{Mej{\'{\i}}a} A.~C.,  {Durisen} R.~H.,  {Pickett} M.~K.,   {Cai} K.,  2005,
  \mn@doi [\apj] {10.1086/426707}, \href
  {http://adsabs.harvard.edu/abs/2005ApJ...619.1098M} {619, 1098}

\bibitem[\protect\citeauthoryear{{Meru} \& {Bate}}{{Meru} \&
  {Bate}}{2011}]{MB2011}
{Meru} F.,  {Bate} M.~R.,  2011, \mn@doi [\mnras]
  {10.1111/j.1745-3933.2010.00978.x}, \href
  {http://adsabs.harvard.edu/abs/2011MNRAS.411L...1M} {411, L1}

\bibitem[\protect\citeauthoryear{{Mitra}, {Wettlaufer}  \&
  {Brandenburg}}{{Mitra} et~al.}{2013}]{mitraetal2013}
{Mitra} D.,  {Wettlaufer} J.~S.,   {Brandenburg} A.,  2013, \mn@doi [\apj]
  {10.1088/0004-637X/773/2/120}, \href
  {http://adsabs.harvard.edu/abs/2013ApJ...773..120M} {773, 120}

\bibitem[\protect\citeauthoryear{{Nakamoto} \& {Miura}}{{Nakamoto} \&
  {Miura}}{2004}]{NM2004}
{Nakamoto} T.,  {Miura} H.,  2004, in {Mackwell} S.,  {Stansbery} E.,  eds,
  Lunar and Planetary Science Conference Vol. 35, Lunar and Planetary Science
  Conference.

\bibitem[\protect\citeauthoryear{{Nelson}}{{Nelson}}{2005}]{nelson2005}
{Nelson} R.~P.,  2005, \mn@doi [\aap] {10.1051/0004-6361:20042605}, \href
  {http://adsabs.harvard.edu/abs/2005A%26A...443.1067N} {443, 1067}


\bibitem[\protect\citeauthoryear{{Nelson} \& {Gressel}}{{Nelson} \&
  {Gressel}}{2010}]{NG2010}
{Nelson} R.~P.,  {Gressel} O.,  2010, \mn@doi [\mnras]
  {10.1111/j.1365-2966.2010.17327.x}, \href
  {http://adsabs.harvard.edu/abs/2010MNRAS.409..639N} {409, 639}

\bibitem[\protect\citeauthoryear{{Ogihara}, {Ida}  \& {Morbidelli}}{{Ogihara}
  et~al.}{2007}]{ogiharaetal2007}
{Ogihara} M.,  {Ida} S.,   {Morbidelli} A.,  2007, \mn@doi [\icarus]
  {10.1016/j.icarus.2006.12.006}, \href
  {http://adsabs.harvard.edu/abs/2007Icar..188..522O} {188, 522}

\bibitem[\protect\citeauthoryear{{Okuzumi} \& {Hirose}}{{Okuzumi} \&
  {Hirose}}{2011}]{OH2011}
{Okuzumi} S.,  {Hirose} S.,  2011, \mn@doi [\apj] {10.1088/0004-637X/742/2/65},
  \href {http://adsabs.harvard.edu/abs/2011ApJ...742...65O} {742, 65}

\bibitem[\protect\citeauthoryear{{Okuzumi} \& {Ormel}}{{Okuzumi} \&
  {Ormel}}{2013}]{OO2013a}
{Okuzumi} S.,  {Ormel} C.~W.,  2013, \mn@doi [\apj]
  {10.1088/0004-637X/771/1/43}, \href
  {http://adsabs.harvard.edu/abs/2013ApJ...771...43O} {771, 43}

\bibitem[\protect\citeauthoryear{{Ormel} \& {Cuzzi}}{{Ormel} \&
  {Cuzzi}}{2007}]{OC2007}
{Ormel} C.~W.,  {Cuzzi} J.~N.,  2007, \mn@doi [\aap]
  {10.1051/0004-6361:20066899}, \href
  {http://adsabs.harvard.edu/abs/2007A%26A...466..413O} {466, 413}


\bibitem[\protect\citeauthoryear{{Ormel} \& {Okuzumi}}{{Ormel} \&
  {Okuzumi}}{2013}]{OO2013b}
{Ormel} C.~W.,  {Okuzumi} S.,  2013, \mn@doi [\apj]
  {10.1088/0004-637X/771/1/44}, \href
  {http://adsabs.harvard.edu/abs/2013ApJ...771...44O} {771, 44}

\bibitem[\protect\citeauthoryear{{Paardekooper}}{{Paardekooper}}{2012}]{Paardekooper2012}
{Paardekooper} S.-J.,  2012, \mn@doi [\mnras]
  {10.1111/j.1365-2966.2012.20553.x}, \href
  {http://adsabs.harvard.edu/abs/2012MNRAS.421.3286P} {421, 3286}

\bibitem[\protect\citeauthoryear{{Paczynski}}{{Paczynski}}{1978}]{Paczynski78}
{Paczynski} B.,  1978, \actaa, \href
  {http://adsabs.harvard.edu/abs/1978AcA....28...91P} {28, 91}

\bibitem[\protect\citeauthoryear{{Pan}, {Padoan}, {Scalo}, {Kritsuk}  \&
  {Norman}}{{Pan} et~al.}{2011}]{panetal2011}
{Pan} L.,  {Padoan} P.,  {Scalo} J.,  {Kritsuk} A.~G.,   {Norman} M.~L.,  2011,
  \mn@doi [\apj] {10.1088/0004-637X/740/1/6}, \href
  {http://adsabs.harvard.edu/abs/2011ApJ...740....6P} {740, 6}

\bibitem[\protect\citeauthoryear{{Pan}, {Padoan}  \& {Scalo}}{{Pan}
  et~al.}{2014}]{PPS2014b}
{Pan} L.,  {Padoan} P.,   {Scalo} J.,  2014, \mn@doi [\apj]
  {10.1088/0004-637X/792/1/69}, \href
  {http://adsabs.harvard.edu/abs/2014ApJ...792...69P} {792, 69}

\bibitem[\protect\citeauthoryear{{Perets} \& {Murray-Clay}}{{Perets} \&
  {Murray-Clay}}{2011}]{PMC2011}
{Perets} H.~B.,  {Murray-Clay} R.~A.,  2011, \mn@doi [\apj]
  {10.1088/0004-637X/733/1/56}, \href
  {http://adsabs.harvard.edu/abs/2011ApJ...733...56P} {733, 56}

\bibitem[\protect\citeauthoryear{{Ricci}, {Testi}, {Natta}, {Neri}, {Cabrit}
  \& {Herczeg}}{{Ricci} et~al.}{2010}]{riccietal2010}
{Ricci} L.,  {Testi} L.,  {Natta} A.,  {Neri} R.,  {Cabrit} S.,   {Herczeg}
  G.~J.,  2010, \mn@doi [\aap] {10.1051/0004-6361/200913403}, \href
  {http://adsabs.harvard.edu/abs/2010A%26A...512A..15R} {512, A15}


\bibitem[\protect\citeauthoryear{{Rice}, {Armitage}, {Bate}  \&
  {Bonnell}}{{Rice} et~al.}{2003}]{Rice2003}
{Rice} W.~K.~M.,  {Armitage} P.~J.,  {Bate} M.~R.,   {Bonnell} I.~A.,  2003,
  \mn@doi [\mnras] {10.1046/j.1365-8711.2003.06253.x}, \href
  {http://adsabs.harvard.edu/abs/2003MNRAS.339.1025R} {339, 1025}

\bibitem[\protect\citeauthoryear{{Rice}, {Lodato}, {Pringle}, {Armitage}  \&
  {Bonnell}}{{Rice} et~al.}{2004}]{riceetal2004}
{Rice} W.~K.~M.,  {Lodato} G.,  {Pringle} J.~E.,  {Armitage} P.~J.,   {Bonnell}
  I.~A.,  2004, \mn@doi [\mnras] {10.1111/j.1365-2966.2004.08339.x}, \href
  {http://adsabs.harvard.edu/abs/2004MNRAS.355..543R} {355, 543}

\bibitem[\protect\citeauthoryear{{Rice}, {Lodato}, {Pringle}, {Armitage}  \&
  {Bonnell}}{{Rice} et~al.}{2006}]{riceetal2006}
{Rice} W.~K.~M.,  {Lodato} G.,  {Pringle} J.~E.,  {Armitage} P.~J.,   {Bonnell}
  I.~A.,  2006, \mn@doi [\mnras] {10.1111/j.1745-3933.2006.00215.x}, \href
  {http://adsabs.harvard.edu/abs/2006MNRAS.372L...9R} {372, L9}

\bibitem[\protect\citeauthoryear{{Rodr{\'{\i}}guez}, {Loinard}, {D'Alessio},
  {Wilner}  \& {Ho}}{{Rodr{\'{\i}}guez} et~al.}{2005}]{rodriguez2005}
{Rodr{\'{\i}}guez} L.~F.,  {Loinard} L.,  {D'Alessio} P.,  {Wilner} D.~J.,
  {Ho} P.~T.~P.,  2005, \mn@doi [\apjl] {10.1086/429223}, \href
  {http://adsabs.harvard.edu/abs/2005ApJ...621L.133R} {621, L133}

\bibitem[\protect\citeauthoryear{{Shi} \& {Chiang}}{{Shi} \&
  {Chiang}}{2013}]{sc2013}
{Shi} J.-M.,  {Chiang} E.,  2013, \mn@doi [\apj] {10.1088/0004-637X/764/1/20},
  \href {http://adsabs.harvard.edu/abs/2013ApJ...764...20S} {764, 20}

\bibitem[\protect\citeauthoryear{{Shi} \& {Chiang}}{{Shi} \&
  {Chiang}}{2014}]{ShiChiang2014}
{Shi} J.-M.,  {Chiang} E.,  2014, \mn@doi [\apj] {10.1088/0004-637X/789/1/34},
  \href {http://adsabs.harvard.edu/abs/2014ApJ...789...34S} {789, 34}

\bibitem[\protect\citeauthoryear{{Shi}, {Stone}  \& {Huang}}{{Shi}
  et~al.}{2016}]{ssh2016}
{Shi} J.-M.,  {Stone} J.~M.,   {Huang} C.~X.,  2016, \mn@doi [\mnras]
  {10.1093/mnras/stv2815}, \href
  {http://adsabs.harvard.edu/abs/2016MNRAS.456.2273S} {456, 2273}

\bibitem[\protect\citeauthoryear{{Stewart} \& {Leinhardt}}{{Stewart} \&
  {Leinhardt}}{2009}]{SL2009}
{Stewart} S.~T.,  {Leinhardt} Z.~M.,  2009, \mn@doi [\apjl]
  {10.1088/0004-637X/691/2/L133}, \href
  {http://adsabs.harvard.edu/abs/2009ApJ...691L.133S} {691, L133}

\bibitem[\protect\citeauthoryear{{Stone} \& {Gardiner}}{{Stone} \&
  {Gardiner}}{2009}]{sg09}
{Stone} J.~M.,  {Gardiner} T.,  2009, \mn@doi [\na]
  {10.1016/j.newast.2008.06.003}, \href
  {http://adsabs.harvard.edu/abs/2009NewA...14..139S} {14, 139}

\bibitem[\protect\citeauthoryear{{Stone} \& {Gardiner}}{{Stone} \&
  {Gardiner}}{2010}]{sg10}
{Stone} J.~M.,  {Gardiner} T.~A.,  2010, \mn@doi [\apjs]
  {10.1088/0067-0049/189/1/142}, \href
  {http://adsabs.harvard.edu/abs/2010ApJS..189..142S} {189, 142}

\bibitem[\protect\citeauthoryear{{Stone}, {Gardiner}, {Teuben}, {Hawley}  \&
  {Simon}}{{Stone} et~al.}{2008}]{stoneetal08}
{Stone} J.~M.,  {Gardiner} T.~A.,  {Teuben} P.,  {Hawley} J.~F.,   {Simon}
  J.~B.,  2008, \mn@doi [\apjs] {10.1086/588755}, \href
  {http://adsabs.harvard.edu/abs/2008ApJS..178..137S} {178, 137}

\bibitem[\protect\citeauthoryear{{Testi} et~al.,}{{Testi}
  et~al.}{2014}]{testietal2014prpl}
{Testi} L.,  et~al., 2014, \mn@doi [Protostars and Planets VI]
  {10.2458/azu_uapress_9780816531240-ch015}, \href
  {http://adsabs.harvard.edu/abs/2014prpl.conf..339T} {pp 339--361}

\bibitem[\protect\citeauthoryear{{Tobin}, {Hartmann}, {Chiang}, {Wilner},
  {Looney}, {Loinard}, {Calvet}  \& {D'Alessio}}{{Tobin}
  et~al.}{2013}]{tobinetal2013}
{Tobin} J.~J.,  {Hartmann} L.,  {Chiang} H.-F.,  {Wilner} D.~J.,  {Looney}
  L.~W.,  {Loinard} L.,  {Calvet} N.,   {D'Alessio} P.,  2013, \mn@doi [\apj]
  {10.1088/0004-637X/771/1/48}, \href
  {http://adsabs.harvard.edu/abs/2013ApJ...771...48T} {771, 48}

\bibitem[\protect\citeauthoryear{{Toomre}}{{Toomre}}{1964}]{toomre64}
{Toomre} A.,  1964, \mn@doi [\apj] {10.1086/147861}, \href
  {http://adsabs.harvard.edu/abs/1964ApJ...139.1217T} {139, 1217}

\bibitem[\protect\citeauthoryear{{Voelk}, {Jones}, {Morfill}  \&
  {Roeser}}{{Voelk} et~al.}{1980}]{voelketal1980}
{Voelk} H.~J.,  {Jones} F.~C.,  {Morfill} G.~E.,   {Roeser} S.,  1980, \aap,
  \href {http://adsabs.harvard.edu/abs/1980A%26A....85..316V} {85, 316}


\bibitem[\protect\citeauthoryear{{Walmswell}, {Clarke}  \&
  {Cossins}}{{Walmswell} et~al.}{2013}]{walmswelletal2013}
{Walmswell} J.,  {Clarke} C.,   {Cossins} P.,  2013, \mn@doi [\mnras]
  {10.1093/mnras/stt314}, \href
  {http://adsabs.harvard.edu/abs/2013MNRAS.431.1903W} {431, 1903}

\bibitem[\protect\citeauthoryear{{Weidenschilling}}{{Weidenschilling}}{1977}]{weidenschilling77}
{Weidenschilling} S.~J.,  1977, \mnras, \href
  {http://adsabs.harvard.edu/abs/1977MNRAS.180...57W} {180, 57}

\bibitem[\protect\citeauthoryear{{Weidenschilling} \&
  {Cuzzi}}{{Weidenschilling} \& {Cuzzi}}{1993}]{weidenschillingcuzzi93}
{Weidenschilling} S.~J.,  {Cuzzi} J.~N.,  1993, in {E.~H.~Levy \& J.~I.~Lunine}
  ed., Protostars and Planets III. pp 1031--1060

\bibitem[\protect\citeauthoryear{{Williams} \& {Cieza}}{{Williams} \&
  {Cieza}}{2011}]{william_cieza2011araa}
{Williams} J.~P.,  {Cieza} L.~A.,  2011, \mn@doi [\araa]
  {10.1146/annurev-astro-081710-102548}, \href
  {http://adsabs.harvard.edu/abs/2011ARA%26A..49...67W} {49, 67}


\bibitem[\protect\citeauthoryear{{Windmark}, {Birnstiel}, {Ormel}  \&
  {Dullemond}}{{Windmark} et~al.}{2012}]{windmark2012b}
{Windmark} F.,  {Birnstiel} T.,  {Ormel} C.~W.,   {Dullemond} C.~P.,  2012,
  \mn@doi [\aap] {10.1051/0004-6361/201220004}, \href
  {http://adsabs.harvard.edu/abs/2012A%26A...544L..16W} {544, L16}


\bibitem[\protect\citeauthoryear{{Yang}, {Mac Low}  \& {Menou}}{{Yang}
  et~al.}{2009}]{yangetal2009}
{Yang} C.-C.,  {Mac Low} M.-M.,   {Menou} K.,  2009, \mn@doi [\apj]
  {10.1088/0004-637X/707/2/1233}, \href
  {http://adsabs.harvard.edu/abs/2009ApJ...707.1233Y} {707, 1233}

\bibitem[\protect\citeauthoryear{{Yang}, {Mac Low}  \& {Menou}}{{Yang}
  et~al.}{2012}]{yangetal2012}
{Yang} C.-C.,  {Mac Low} M.-M.,   {Menou} K.,  2012, \mn@doi [\apj]
  {10.1088/0004-637X/748/2/79}, \href
  {http://adsabs.harvard.edu/abs/2012ApJ...748...79Y} {748, 79}

\bibitem[\protect\citeauthoryear{{Youdin} \& {Goodman}}{{Youdin} \&
  {Goodman}}{2005}]{youdingoodman05}
{Youdin} A.~N.,  {Goodman} J.,  2005, \apj, 620, 459

\bibitem[\protect\citeauthoryear{{Youdin} \& {Lithwick}}{{Youdin} \&
  {Lithwick}}{2007}]{YL2007}
{Youdin} A.~N.,  {Lithwick} Y.,  2007, \mn@doi [\icarus]
  {10.1016/j.icarus.2007.07.012}, \href
  {http://adsabs.harvard.edu/abs/2007Icar..192..588Y} {192, 588}

\bibitem[\protect\citeauthoryear{{Zhu}, {Stone}  \& {Bai}}{{Zhu}
  et~al.}{2015}]{zhuetal2015}
{Zhu} Z.,  {Stone} J.~M.,   {Bai} X.-N.,  2015, \mn@doi [\apj]
  {10.1088/0004-637X/801/2/81}, \href
  {http://adsabs.harvard.edu/abs/2015ApJ...801...81Z} {801, 81}

\bibitem[\protect\citeauthoryear{van Leer}{van Leer}{2006}]{vl06}
van Leer B.,  2006, Comm. Comput. Phys., 1, 192

\makeatother
\end{thebibliography}

\bsp	

\label{lastpage}
\end{document}